\documentclass[12pt,12pt,sort&compress]{article}
\usepackage[T1]{fontenc}
\usepackage[latin9]{inputenc}
\usepackage{color}
\usepackage{array}
\usepackage{multirow}
\usepackage{amsmath}
\usepackage{amssymb}
\usepackage{graphicx}
\usepackage{esint}
\usepackage[numbers]{natbib}
\PassOptionsToPackage{normalem}{ulem}
\usepackage{ulem}

\makeatletter

\providecommand{\tabularnewline}{\\}

\@ifundefined{date}{}{\date{}}

\usepackage{indentfirst}
\usepackage{amsfonts}
\usepackage[T1]{fontenc}
\usepackage{ae,aecompl}
\usepackage{sidecap}
\usepackage[section]{placeins}
\usepackage{epsf}

\usepackage{fancyhdr}

\makeatother

\begin{document}

\title{Thermodynamic theory of equilibrium fluctuations}

\author{\noindent Y.~Mishin\\
{\normalsize{} Department of Physics and Astronomy, George Mason University,
MSN 3F3,}\\
 {\normalsize{}Fairfax, VA 22030, USA; }\\
{\normalsize{}Email: ymishin@gmu.edu} }

\maketitle
\noindent \begin{center}
\today
\par\end{center}
\begin{abstract}
The postulational basis of classical thermodynamics has been expanded
to incorporate equilibrium fluctuations. The main additional elements
of the proposed thermodynamic theory are the concept of quasi-equilibrium
states, a definition of non-equilibrium entropy, a fundamental equation
of state in the entropy representation, and a fluctuation postulate
describing the probability distribution of macroscopic parameters
of an isolated system. Although these elements introduce a statistical
component that does not exist in classical thermodynamics, the logical
structure of the theory is different from that of statistical mechanics
and represents an expanded version of thermodynamics. Based on this
theory, we present a regular procedure for calculations of equilibrium
fluctuations of extensive parameters, intensive parameters and densities
in systems with any number of fluctuating parameters. The proposed
fluctuation formalism is demonstrated by four applications: (1) derivation
of the complete set of fluctuation relations for a simple fluid in
three different ensembles; (2) fluctuations in finite-reservoir systems
interpolating between the canonical and micro-canonical ensembles;
(3) derivation of fluctuation relations for excess properties of grain
boundaries in binary solid solutions, and (4) derivation of the grain
boundary width distribution for pre-melted grain boundaries in alloys.
The last two applications offer an efficient fluctuation-based approach
to calculations of interface excess properties and extraction of the
disjoining potential in pre-melted grain boundaries. Possible future
extensions of the theory are outlined.
\end{abstract}
\noindent \textbf{Keywords:} Thermodynamics, fluctuation, entropy,
grain boundary, pre-melting

\section{Introduction}

\subsection{Historical background and goal of the work}

Fluctuations of thermodynamic properties play a crucial role in many
physical phenomena and diverse applications. Fluctuations are especially
important in nanometer-scale systems where they can lead to a large
variability of mechanical and functional properties and create noises
affecting performance of devices. Fluctuations are unavoidable in
molecular dynamics and Monte Carlo simulations of materials, where
the system dimensions rarely exceed a few nanometers. In fact, in
many atomistic calculations, equilibrium properties of interest are
extracted by analyzing statistical fluctuations of other properties
\citep{FrenkelS02,LandauBinder09,HoytFluct10}. Examples include the
calculations of elastic coefficients of solid materials from strain
and/or stress fluctuations in molecular dynamics \citep{Parrinello1982,Gusev1996,Zhou2002,Cui2007}
or Monte Carlo \citep{Meyers2005,Mishin:2015aa} simulations; calculation
of partial molar properties of solutions from concentration fluctuations
\citep{Debenedetti:1987aa,Debenedetti:1987bb}; and calculation of
the interface free energy of solid-liquid \citep{Davidchack06,Amini08}
and solid-sold \citep{Hoyt01,Foiles06a,Hoyt09a,Song10,Fensin2010,Karma-2012a,Adland2013a,Mishin2014,Schmitz_2014,Hickman_to_be_published_2015}
interfaces from capillary fluctuations or fluctuations of the interface
width.

Presently, fluctuations are primarily discussed in the statistical-mechanical
literature and typically in the context of specific (usually, very
simple) models. At the same time, the thermodynamics community traditionally
relies on classical thermodynamics \citep{Willard_Gibbs,Munster:1970aa,Callen_book_1960,Callen_book_1985,Tisza:1961aa}
which, by the macroscopic and equilibrium nature of this discipline,
disregards fluctuations and operates solely in terms of static properties.
In fact, the very term ``fluctuation'' has a temporal connotation
incompatible with the time-independent character of classical thermodynamics. 

There is, however, a third direction in thermal physics that pursues
a generalized form of thermodynamics that incorporates fluctuations
of thermodynamic parameters around equilibrium. By contrast to the
statistical-mechanical approach, such theories seek to introduce fluctuations
\emph{directly} into the thermodynamic framework via additional assumptions,
postulates or similar elements of the logical structure. The goal
of such theories is to expand the scope of thermodynamics by introducing
statistical elements while preserving the traditional axiomatic approach
that distinguishes this discipline. It is this direction in thermal
physics that constitutes the subject of the present paper.\footnote{Although we do not wish to enter into terminological discussions,
we point out that an expanded thermodynamic theory that incorporates
statistical elements such as fluctuations could be called statistical
thermodynamics. This term would distinguish it from both statistical
mechanics and classical thermodynamics. Unfortunately, the term statistical
thermodynamics is already used with several different meanings, most
notably as a collective reference to statistical mechanics and thermodynamics
(which we refer to as thermal physics).}

The logical foundations of classical thermodynamics have been the
subject of research over the past hundred or more years, starting
with the works of Carathéodory \citep{Caratheodory_1909} and Ehrenfest
\citep{Ehrenfest:1959aa} and continuing in modern times \citep{Callen_book_1960,Tisza:1961aa,Landsberg:1970,Kestin:1970,Gurney:1970aa,Masavetas:1988aa,Jongschaap:2001aa,Callen_book_1985,Weihold_book}.
In spite of the fundamental importance of the traditional three laws
of thermodynamics, they are essentially a reflection of the historical
development of the discipline and do not constitute an autonomous
and logically complete structure. The four-postulate structure proposed
by Callen \citep{Callen_book_1960,Callen_book_1985} is detached from
the historical context, deeply thought-through, but still far from
complete. The most rigorous and complete postulational basis of classical
thermodynamics has been formulated by Tisza \citep{Tisza:1961aa}.
His theory, called the Macroscopic Thermodynamics of Equilibrium (MTE),
presents an elegant logical structure comprising a set of interconnected
definitions, postulates and corollaries. This rigor comes at a price:
Tisza's MTE is restricted to a certain class of rather simple thermodynamic
systems. Nevertheless, it demonstrates an approach that can be applied
for the construction of similar postulational structures for other
classes of systems. 

Unfortunately, attempts to create a similarly rigorous thermodynamic
formalism that would include fluctuations have not been very successful.
The thermodynamic fluctuation theory by Greene and Callen \citep{Green:1951aa,Callen:1965aa}
and Callen's Postulate II$^{\prime}$ \citep{Callen_book_1960}\footnote{Callen's Postulate II$^{\prime}$ \citep{Callen_book_1960} was abandoned
in a later edition of his book \citep{Callen_book_1985}.} (which generalizes his entropy Postulate II to include fluctuations)
turned out to be insufficient. They required additional assumptions,
such as the approximation of average values of thermodynamic properties
by their most probable values, and relied on the assumption that there
is no distinction between the ``canonical thermodynamics'' and ``microcanonical
thermodynamics''. The latter assumption was later criticized by Tisza
and Quay \citep{Tisza:1963} from the standpoint of statistical mechanics.
Tisza and Quay's \citep{Tisza:1963} own treatment of fluctuations
was an amalgamation of statistical mechanics and thermodynamics, and
in this sense, did not achieve the goal either. In fact, the authors
\citep{Tisza:1963} admit that their theory, called the Statistical
Thermodynamic of Equilibrium, is only a stepping stone toward a future
more general theory whose development encounters major difficulties. 

There are two major challenges in the development of a thermodynamic
theory of fluctuations: (1) appropriate definition of non-equilibrium
entropy that would seamlessly connect with the system of other definitions
and postulates of thermodynamics, and (2) formulation of a general
fluctuation law of thermodynamic parameters. Historically, the first
work addressing these issues was Einstein's 1910 paper on critical
opalescence \citep{Einstein_1910}. In his theory, the point of departure
was Boltzmann's equation
\begin{equation}
S=k\ln W+\textrm{const},\label{eq:300}
\end{equation}
where $k$ is Boltzmann's constant and $W$ is the probability associated
with the entropy $S$ (our notations are slightly different from Einstein's).
Einstein considered a completely isolated thermodynamic system with
a fixed energy $E$. He assumed that all equilibrium and non-equilibrium
(fluctuated) thermodynamic states of the system can be described by
a set of macroscopic parameters $\lambda_{1},...,\lambda_{n}$. For
a discrete set of such parameters, he proposed to interpreted $W$
as the fractions of time spent by the fluctuating system in each of
the states compatible with the given energy $E$. Based on this interpretation,
Einstein inverted Eq.(\ref{eq:300}) to obtain the probability of
a macroscopic state and thus a given set of parameters $\lambda_{1},...,\lambda_{n}$:
\begin{equation}
W=\textrm{const}\cdot\exp\left(\dfrac{S}{k}\right).\label{eq:301}
\end{equation}
For continuos state variables, the probability that the parameters
$\lambda_{1},...,\lambda_{n}$ be found between $\lambda_{1}$ and
$\lambda_{1}+d\lambda_{1}$ ,..., $\lambda_{n}$ and $\lambda_{n}+d\lambda_{n}$
is
\begin{equation}
dW=\textrm{const}\cdot\exp\left(\dfrac{S-S_{0}}{k}\right)d\lambda_{1}...d\lambda_{n},\label{eq:302}
\end{equation}
where $S_{0}$ is the maximum entropy. Einstein's theory does not
make any reference to micro-states of the system. Instead, it relates
the entropy \emph{directly} to the probabilities of macroscopic states.
As such, this theory belongs more to the realm of thermodynamics than
statistical mechanics.\footnote{The first statistical-mechanical theory of fluctuations was developed
by Gibbs \citep{Willard_Gibbs_2}, whose work was apparently unknown
to Einstein at the time. Gibbs' treatment of fluctuations was based
\emph{explicitly} on classical dynamics of particles, with micro-states
defined in the momentum-coordinate phase space. Einstein's theory
\citep{Einstein_1910} is obviously more general, even though Eqs.(\ref{eq:301})
and (\ref{eq:302}) \emph{per se} do not suggest any means of calculating
the probabilities or the entropy for any particular system. For a
more detailed comparison of the Gibbs and Einstein approaches see
Tisza and Quay \citep{Tisza:1963} and a more recent paper by Rudoi
and Sukhanov \citep{Rudoi:2000aa}. } 

Equations (\ref{eq:301}) and (\ref{eq:302}) essentially constitute
a fluctuation law that can be built into the formalism of thermodynamics.
The fluctuations described by these equations occur in an isolated
system, thus the respective entropy has a micro-canonical meaning.
For canonical systems, the distribution function of macroscopic parameters
can be derived by considering a relatively small (but macroscopic)
subsystem of an isolated system and treating the remaining part (complementary
system) as a reservoir. 

The situation is complicated by the fact that the most advanced thermodynamic
theories of fluctuations \citep{Callen_book_1960,Tisza:1963} postulate
the fluctuation law directly in a canonical form and derive properties
of isolated systems as a liming case of canonical. This approach is
different from Einstein's \citep{Einstein_1910} and is rooted in
statistical-mechanical arguments. The treatment of a canonical system
as an asymptotic case when the complementary system tends to infinity
has certain conceptual disadvantages in rigorous formulations of statistical
mechanics. Alternative theories \citep{Szilard:1925aa,Tisza:1963,Khinchin_1949}
postulate the canonical distribution for an infinite reservoir from
the outset and derive properties of isolated and finite systems only
later. Thus, postulating thermodynamic fluctuations in a canonical
form is more consistent with the aforementioned statistical-mechanical
theories. 

In the present paper, we take the view that, if a thermodynamic fluctuation
theory is to be relatively autonomous, then the simplicity of its
internal logical structure is more important than consistency with
all aspects the formal structure of statistical mechanics. We thus
propose to build the theory on a fluctuation law postulated for an
isolated system. The law which will be adopted in this work will look
similar to Eq.(\ref{eq:302}), except for certain refinements in justification
and interpretation. 

We emphasize that this paper is focused on \emph{equilibrium} fluctuations
and will not discuss driven or other strongly non-equilibrium systems.
Furthermore, given that our approach is purely thermodynamic, we are
only interested in statistical distributions of fluctuating parameters
without attempting to describe the actual dynamics of the fluctuations
\citep{Evans:1993aa,Bertini:2015aa}.

\subsection{Organization of the paper}

The outline of the paper is a follows. In Secs.~\ref{sec:Definition}-\ref{sec:Stat-mech}
we formulate the basic assumptions of the theory describing equilibrium
fluctuations in an isolated system. We then consider the usual quadratic
expansion of the entropy around equilibrium and arrive at the Gaussian
law of fluctuations and a set of relations for mean-square fluctuations
and covariances of thermodynamic properties (Sec.~\ref{sec:Gaussian-law}).
In Sec.~\ref{sec:Canonical} we derive the fluctuation law for generalized
canonical systems, treating the system of interest and the reservoir
as a combined isolated system. We derive the Gaussian law of canonical
fluctuations in two forms, called the entropy scheme and the energy
scheme, and present the fluctuation formalism in both schemes. To
illustrate the calculation techniques, we compute fluctuations of
all properties of a simple fluid in three different ensembles. In
Sec.~\ref{sec:Finite-reservoir} we analyze systems connected to
a finite-size reservoir. The finite-reservoir ensemble interpolates
between the micro-canonical and canonical ensembles and has been implemented
in recent computer simulations \citep{Challa88,Huller1994b,Huller1995,Johal2003,Velazquez2009,Velazquez2010a,Velazquez2010b,Sadigh2012,Sadigh2012a,Mishin2014}.
This section demonstrates the ease with which the theory can handle
finite-size systems as opposed to alternate theories \citep{Callen_book_1960,Tisza:1963}
built directly on the canonical formalism. 

Sections \ref{sec:Fluctuations-in-GBs} and \ref{sec:Pre-melting}
are devoted to applications of the theory to fluctuations in a single-phase
interface, called a grain boundary (GB), in a binary solid solution.
This is a non-trivial problem, which is addressed by conceptually
partitioning the system into an imaginary perfect crystal (grain)
and an excess system representing the GB, each described by their
own fundamental equation. The goal is to describe fluctuations in
the excess system. In Sec.~\ref{sec:Fluctuations-in-GBs} we develop
a set of equations linking fluctuations of some interface properties
to equilibrium values of other properties, such as the excess heat
capacity, excess elastic moduli and others. An interesting result
here is the prediction of fluctuations of the interface free energy,
an important excess quantity that is usually treated as a static property.
Next, we address the problem of width fluctuations of a pre-melted
GB represented as a thin liquid layer subject to a disjoining pressure.
We derive the distribution function of the GB width expressed through
the disjoining potential describing interactions between the two solid-liquid
interfaces bounding the liquid layer. While the functional form of
the width distribution is known for single-component systems \citep{Hoyt09a,Fensin2010,Song10,Adland2013a},
it has not been previously derived for binary mixtures. Finally, in
Sec.~\ref{sec:Conclusions} we summarize the paper and outline future
work. 

Although the paper is focused on thermodynamics, in Secs.~\ref{sec:Stat-mech}
and \ref{sub:Generalized-canonical-ensembles} and a few other parts
of the text we do discuss statistical distributions of micro-states
and other aspects of statistical mechanics. It should be emphasized
that these discussions are \emph{not} part of the proposed fluctuation
theory. They are only included here to provide a statistical-mechanical
interpretation of certain thermodynamic statements or demonstrate
their consistency with principles of statistical mechanics. In addition,
we slightly deviate from the traditional macroscopic terminology of
classical thermodynamics and talk about the number of particles (instead
of moles) and ensembles (instead of types of imposed constraints).
Obviously, this is only a matter of terminology that does not compromise
the thermodynamic nature of the theory. 

Finally, we generally prefer a smooth development of the logic and
do not follow the mathematical style of presentation \citep{Caratheodory_1909,Tisza:1961aa,Tisza:1963}
in which all statements are broken into definitions, postulates, corollaries
and other deductive steps.

\section{Definition of equilibrium fluctuations\label{sec:Definition}}

Consider an isolated thermodynamic system, i.e., a system incapable
of exchanging heat or matter with its environment or performing work
on the environment. Some physical properties of an isolated system
are strictly fixed by conservation laws (e.g., the total energy, total
momentum, total amounts of chemical elements\footnote{In a system without nuclear reactions.}),
whereas other properties can vary. Thermodynamics postulates that
all non-conserved properties averaged over a certain time scale $t_{TD}$,
which we call the thermodynamic time scale, eventually stop varying
and remain constant for as long as the system remains isolated. The
quiescent state of an isolated system in which all properties remain
constant on the $t_{TD}$ time scale is called the state of thermodynamic
equilibrium. 

Even after thermodynamic equilibrium has been reached, non-conserved
properties continue to vary on a shorter time scale, $t_{f}$$\ll t_{TD}$,
called the fluctuation time scale. Such continual variations of properties
of an equilibrium system are called \emph{equilibrium fluctuations}. 

For any physical property $X$, let $\overline{X}$ be its value averaged
over the thermodynamic time scale $t_{TD}$. During the equilibrium
fluctuations, instantaneous values of $X$ randomly deviate from $\overline{X}$.
The goal of the fluctuation theory is to predict the probability distribution
of any property $X$ around its average $\overline{X}$ and compute
the moments and other statistical characteristics of this distribution.

\section{The fundamental equation}

Consider an isolated thermodynamic system whose equilibrium properties
averaged over the $t_{TD}$ time scale are fully defined by a set
of $n$ \emph{conserved} extensive parameters $X_{1},...,X_{n}$.
(Tisza \citep{Tisza:1961aa} refers to such conserved extensive properties
as ``additive invariants''.) For example, all equilibrium properties
of a simple fluid enclosed in an isolated rigid box are fully defined
by its energy $E$, volume $V$, and the number of particles $N$.\footnote{This can be considered a definition of the simple fluid. Physically,
it is a homogeneous single-component fluid without electric, magnetic
or other contributions to its thermodynamic properties.} As long as the parameters $X_{1},...,X_{n}$ remain fixed, the system
remains in a state of equilibrium. One can temporarily break the isolation,
change all or some of the parameters $X_{1},...,X_{n}$, and then
isolate and re-equilibrate the system to a new state. By repeating
such perturbation/re-equilibration steps, one can vary the parameters
$X_{1},...,X_{n}$ and measure all equilibrium thermodynamic properties
of the system as functions of $X_{1},...,X_{n}$. 

In particular, the described procedure can be applied to measure the
system entropy $S$ as a function of the conserved parameters $X_{1},...,X_{n}$.
Here, the entropy is treated as a property defined in statistical
mechanics. Namely, for an equilibrium isolated system, $S$ is identified
with the micro-canonical entropy $S=k\ln\Omega_{max}$, where $\Omega_{max}$
is the maximum number of micro-states compatible with the given set
of parameters $X_{1},...,X_{n}$ (see Sec.~\ref{sec:Stat-mech} for
a more detailed statistical-mechanical interpretation of entropy).
The entropy $S$ so defined will be referred to as the ``equilibrium
entropy'' in order to distinguish it from the ``non-equilibrium entropy''
defined later.

The function 
\begin{equation}
S=S\left(X_{1},...,X_{n}\right)\label{eq:1}
\end{equation}
is called the \emph{fundamental equation} of the system in the entropy
representation \citep{Willard_Gibbs,Callen_book_1960,Tisza:1961aa,Callen_book_1985}.
More accurately, Eq.(\ref{eq:1}) is the fundamental equation of a
particular phase of the system. Different phases are specified by
different fundamental equations (\ref{eq:1}). For a simple fluid,
the fundamental equation has the form $S=S(E,V,N)$.

It is important to recognize that, at this point of the development,
the entropy $S$ has been defined only for an equilibrium isolated
system. Furthermore, the arguments $X_{1},...,X_{n}$ of the fundamental
equation (\ref{eq:1}) are understood as conserved properties that
remain fixed (no fluctuations!) in every given state of the isolated
system.

\section{The non-equilibrium entropy\label{sec:Non-equilibrium-entropy}}

When a non-equilibrium isolated system approaches equilibrium, it
is often possible to define a property called the \emph{non-equilibrium
entropy}. Consider two examples. 

\emph{\uline{Example 1.}} When the system is close enough to equilibrium,
it can be mentally partitioned into small but macroscopic subsystems
that can be considered as isolated and equilibrium for a period of
time on the order of $t_{q}$. Such subsystems are called quasi-equilibrium
and the entire isolated system is said to be in a quasi-equilibrium
state.\footnote{On the time scale $t_{q}$, the quasi-equilibrium system can be treated
as if it was an equilibrium system with subsystems separated by isolating
walls.} On the quasi-equilibrium time scale $t_{q}$, each subsystem $\alpha$
can be described by a fundamental equation $S_{\alpha}=S_{\alpha}\left(X_{1}^{\alpha},...,X_{n}^{\alpha}\right)$,
where $X_{1}^{\alpha},...,X_{n}^{\alpha}$ are extensive properties
assumed to be fixed. The non-equilibrium entropy $\hat{S}$ is \emph{defined}
as the sum of the entropies of all quasi-equilibrium subsystems:
\begin{equation}
\hat{S}\left(\lambda_{1},...,\lambda_{m},X_{1},...,X_{n}\right)\equiv\sum_{\alpha}S_{\alpha}\left(X_{1}^{\alpha},...,X_{n}^{\alpha}\right).\label{eq:2}
\end{equation}
Here, $\lambda_{1},...,\lambda_{m}$ are so-called \emph{internal
parameters} describing the distribution of the conserved extensive
properties $X_{1},...,X_{n}$ of the entire isolated system over its
quasi-equilibrium subsystems.\footnote{To avoid notational confusion, we note that our parameters $\lambda_{1},...,\lambda_{m}$
are \emph{independent} variables describing the macroscopic states
subject to isolation constraints. In Einstein's work \citep{Einstein_1910},
$\lambda_{1},...,\lambda_{n}$ are not independent variables. His
Eq.(\ref{eq:302}) is only valid on the $(n-1)$-dimensional constant-energy
surface $E(\lambda_{1},...,\lambda_{n})=\textrm{const}$ in the $n$-dimensional
parameter space of $\lambda_{1},...,\lambda_{n}$.} As the system evolves towards equilibrium, these internal parameters
vary on the time scale $\gg t_{q}$, as do the local parameters $X_{1}^{\alpha},...,X_{n}^{\alpha}$
of the subsystems and thus the non-equilibrium entropy $\hat{S}$.
Note that the quasi-equilibrium system as a whole does not have a
fundamental equation; its properties depend not only on the total
amounts of the extensive quantities $X_{1},...,X_{n}$ but also on
their distribution among the subsystems. 

\begin{figure}
\noindent \begin{centering}
\includegraphics[clip,scale=0.5]{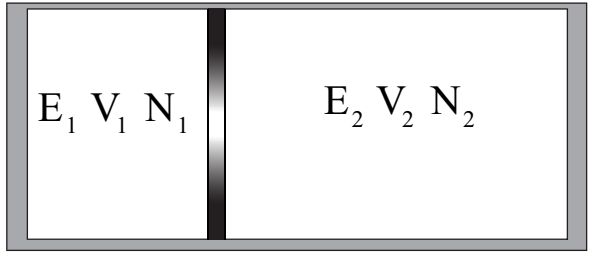}
\par\end{centering}

\protect\caption{A model of an isolated system: a rigid cylinder divided in two compartments
containing different amounts of the same substance separated by a
movable piston. \label{fig:cylinder}}

\end{figure}

As a specific example, consider an isolated cylinder with rigid side
walls, rigid top and bottom, and a movable piston dividing the cylinder
in two compartments 1 and 2 (Fig.~\ref{fig:cylinder}). The compartments
are filled with different amounts of the same simple fluid obeying
a fundamental equation $S=S(E,V,N)$. The system is initially not
in equilibrium. Heat and particles can diffuse through the piston,
but so slowly that each compartment maintains its own thermodynamic
equilibrium at all times. Thus, as the system drifts toward equilibrium,
it always remains in quasi-equilibrium. During this quasi-equilibrium
process, the total energy $E=E_{1}+E_{2}$, total volume $V=V_{1}+V_{2}$
and the total number of particles $N=N_{1}+N_{2}$ remain fixed. These
properties are the additive invariants denoted as $X_{i}$. The non-equilibrium
entropy of the system equals
\begin{equation}
\hat{S}(\underbrace{E_{1},V_{1},N_{1}}_{\lambda},E,V,N)=S\left(E_{1},V_{1},N_{1}\right)+S\left(E-E_{1},V-V_{1},N-N_{1}\right),\label{eq:3}
\end{equation}
where $E_{1}$, $V_{1}$ and $N_{1}$ play the role of the internal
$\lambda$-parameters describing the distribution of the fixed quantities
$E$, $V$ and $N$ between the compartments.

\emph{\uline{Example 2.}} Many systems reach equilibrium with respect
to some thermodynamic properties while other properties are still
in the process of equilibration. In such cases, we can talk about
quasi-equilibrium as equilibrium with respect to some properties without
equilibrium with respect to other properties. For example, chemical
reactions usually occur much slower in comparison with thermal and
mechanical equilibration. There is a certain time scale $t_{q}$ on
which chemical reactions can be considered as ``frozen'' and the
system can be treated \emph{as if} it were in full equilibrium. Accordingly,
the system can be assigned a non-equilibrium entropy $\hat{S}$ computed
by ignoring the reactions and treating the amounts of all chemical
components as fixed parameters. Of course, on the time scale much
longer than $t_{q}$, the entropy slowly drifts as the system evolves
towards chemical equilibrium. The non-equilibrium entropy so defined
is a function of not only the conserved parameters $X_{1},...,X_{n}$
but also some progress variables $\lambda_{1},...,\lambda_{m}$ describing
the extents of the chemical reactions. The latter thus play the role
of the internal parameters specifying the quasi-equilibrium states.

As a simple illustration, consider an isolated box containing a mixture
of three chemically different gases A, B and C. In the absence of
chemical reactions, this mixture is described by a fundamental equation
$S=S(E,V,N_{A},N_{B},N_{C})$. Let the system initially contain the
amounts of the components $N_{A}^{(i)}$, $N_{B}^{(i)}$ and $N_{C}^{(i)}$.
Now suppose that the components can transform to each other via a
chemical reaction $A+B\rightleftharpoons C$. Assuming that the reaction
is slow, the system always maintains mechanical and thermal equilibrium,
while its chemical composition gradually changes. Let $\Delta N_{A}$
be the change in the number of particles A relative to the initial
state. The non-equilibrium entropy is defined by 
\begin{equation}
\hat{S}(\underbrace{\Delta N_{A}}_{\lambda},E,V,N_{A}^{(i)},N_{B}^{(i)},N_{C}^{(i)})=S(E,V,N_{A}^{(i)}-\Delta N_{A},N_{B}^{(i)}-\Delta N_{A},N_{C}^{(i)}+\Delta N_{A}).\label{eq:4}
\end{equation}
Here, $\Delta N_{A}$ characterizes the progress of the chemical reaction
and plays the role of an internal parameter $\lambda$, whereas $E,V,N_{A}^{(i)},N_{B}^{(i)},N_{C}^{(i)}$
is a set of fixed extensive parameters $X_{1},...,X_{n}$.

\section{The second law of thermodynamics\label{sec:Second-law}}

In the extended thermodynamics including fluctuations, the second
law (entropy postulate) can be formulated as two statements: 
\begin{enumerate}
\item The non-equilibrium entropy $\hat{S}$ of an isolated system averaged
over the thermodynamic time scale $t_{TD}$, which we denote $\tilde{S}$,
increases with time and reaches a maximum value $\overline{S}$ when
the system arrives at equilibrium. 
\item The individual values of $\hat{S}$ fluctuate but can never exceed
the equilibrium value $S$. 
\end{enumerate}
Note that this formulation of the second law is stronger than the
frequently used weak formulation. The latter only states that entropy
should increase but does not guarantee that an equilibrium state will
be eventually reached.\footnote{The weak formulation of the second law of thermodynamics has a ``negative''
(prohibitive) character stating what \emph{cannot} happen - the entropy
cannot decrease; whereas the strong formulation makes a ``positive''
statement predicting what \emph{will} happen - the system will reach
equilibrium and the entropy a maximum.} Clearly, to talk about equilibrium fluctuations we must first ensure
that the system actually arrives at equilibrium.

Once equilibrium has been reached, the average $\overline{X}$ of
any thermodynamic property $X$ remains constant while instantaneous
values of $X$ can randomly deviate from $\overline{X}$ on the fluctuation
time scale $t_{f}$. This includes fluctuations of the non-equilibrium
entropy $\hat{S}$. However, the latter is a special case because
it can only fluctuate down from its maximum value $S$. As a result,
the average entropy $\overline{S}$ is always slightly smaller than
$S$.

\begin{figure}
\noindent \begin{centering}
\includegraphics[scale=0.72]{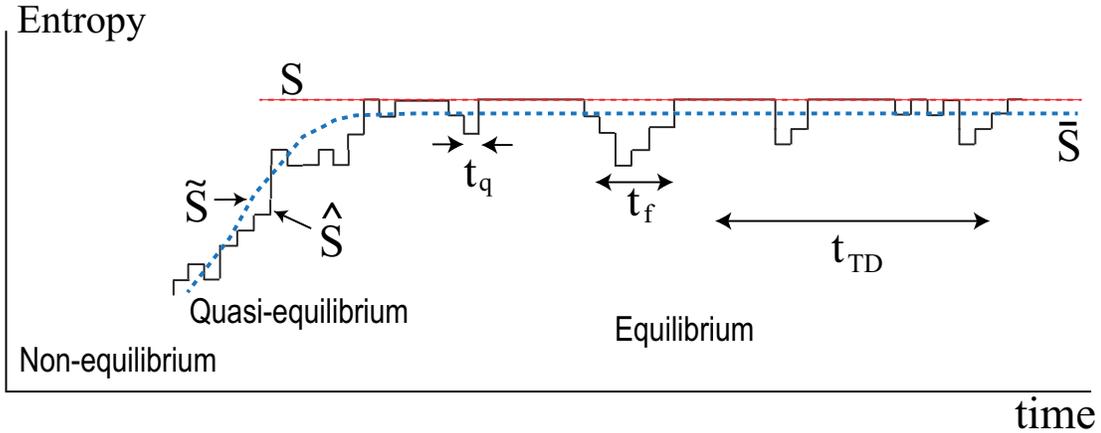}
\par\end{centering}

\protect\caption{Schematic temperature dependence of entropy of an isolated system.
The system is initially in a highly non-equilibrium state in which
the non-equilibrium entropy $\hat{S}$ is undefined. As the system
evolves towards equilibrium, it eventually reaches the quasi-equilibrium
stage in which the non-equilibrium entropy $\hat{S}$ can be defined
on a time scale $t_{q}$. The subsequent time dependence of $\hat{S}$
is shown by a broken line to emphasize that individual values of $\hat{S}$
can only be measured within time intervals on the order of $t_{q}$.
The values of $\hat{S}$ can never exceed the maximum (micro-canonical)
entropy $S$. By the second law of thermodynamics, $\hat{S}$ averaged
over the thermodynamic time scale $t_{TD}$ ($\tilde{S}$, shown by
the dashed blue line) monotonically increases with time and eventually
levels out when the system reaches equilibrium. In the equilibrium
state, $\hat{S}$ can randomly deviate down from its maximum value
$S$ on the fluctuation time scale $t_{f}$ such that $t_{q}\ll t_{f}\ll t_{TD}$,
always returning to the maximum value when the fluctuation is over.\label{fig:time-scales}}
\end{figure}

The time evolution of an isolated system approaching equilibrium is
illustrated schematically in Fig.~\ref{fig:time-scales}. The plot
assumes the existence of three different time scales\footnote{Generally, the time scales may depend on the observable. For the present
discussion, it will suffice to represent each time scale by one ``characteristic''
order of magnitude.} such that
\begin{equation}
t_{TD}\gg t_{f}\gg t_{q}.\label{eq:5}
\end{equation}
It should be remembered that $\hat{S}$ is only defined on the quasi-equilibrium
time scale $t_{q}$. The inequality (\ref{eq:5}) reflects the important
assumption of the fluctuation theory that all states arising during
the fluctuations are quasi-equilibrium states. As such, they can always
be assigned a non-equilibrium entropy $\hat{S}$. The assumption that
the approach to equilibrium (relaxation) and equilibrium fluctuations
realize the same, quasi-equilibrium, states of the system bears a
certain analogy with the fluctuation-dissipation theorem in statistical
mechanics \citep{Nyquist_1928,Callen_Welton_1951,Landau-Lifshitz-Stat-phys}.

\begin{figure}
\noindent \begin{centering}
\includegraphics[scale=0.78]{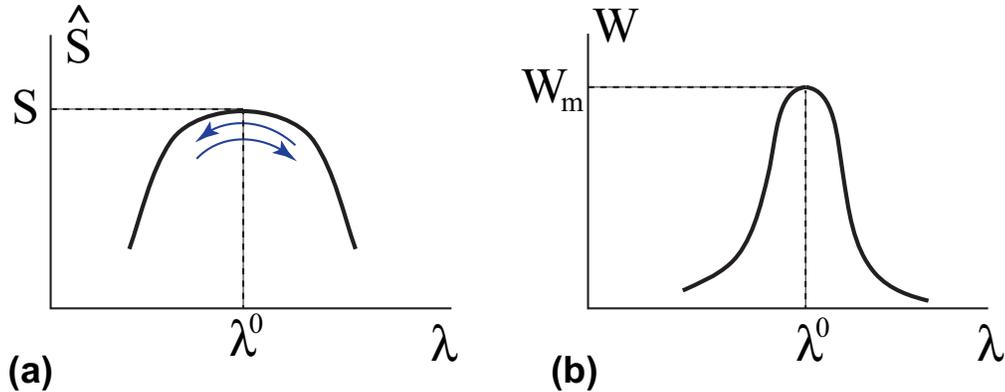}
\par\end{centering}

\protect\caption{Schematic diagram of equilibrium fluctuations in an isolated system
with one internal parameter $\lambda$. (a) Non-equilibrium entropy
$\hat{S}$ reaches its maximum $S$ at the equilibrium value of the
internal parameter $\lambda^{0}$. The arrows symbolize equilibrium
fluctuations of $\lambda$ on either side of $\lambda^{0}$ accompanied
by downward deviations of $\hat{S}$ from $S$. (b) The probability
density of $\lambda$ has a sharp maximum at $\lambda=\lambda^{0}$
of a height $W_{m}$.\label{fig:Fluctuations_1}}
\end{figure}

The non-equilibrium entropy $\hat{S}$ reaches its maximum value $S$
for some set of internal parameters $\lambda_{1}^{0},...,\lambda_{m}^{0}$
called the equilibrium internal parameters. The latter must satisfy
the necessary conditions of extremum,
\begin{equation}
\left(\dfrac{\partial\hat{S}}{\partial\lambda_{i}}\right)_{\lambda_{j\neq i},X_{1},...,X_{n}}=0,\,\,\,\,\,i=1,...,m,\label{eq:6}
\end{equation}
and the relation
\begin{equation}
\hat{S}\left(\lambda_{1}^{0},...,\lambda_{m}^{0},X_{1},...,X_{n}\right)=S\left(X_{1},...,X_{n}\right),\label{eq:7}
\end{equation}
where the right-hand side is the fundamental equation of the substance.
This is illustrated schematically in Fig.~\ref{fig:Fluctuations_1}(a)
for a single internal parameter $\lambda$.

As an example, consider again the isolated cylinder with two compartments
discussed in Sec.~\ref{sec:Non-equilibrium-entropy} (Fig.~\ref{fig:cylinder}).
We have three internal parameters: $\lambda_{1}=E_{1}$, $\lambda_{2}=V_{1}$
and $\lambda_{3}=N_{1}$. Applying Eq.(\ref{eq:6}) to $\hat{S}$
given by Eq.(\ref{eq:3}), we have
\begin{equation}
\dfrac{\partial\hat{S}}{\partial E_{1}}=\dfrac{\partial S(E_{1},V_{1},N_{1})}{\partial E_{1}}-\dfrac{\partial S(E_{2},V_{2},N_{2})}{\partial E_{2}}=\dfrac{1}{T_{1}}-\dfrac{1}{T_{2}}=0,\label{eq:8}
\end{equation}

\begin{equation}
\dfrac{\partial\hat{S}}{\partial V_{1}}=\dfrac{\partial S(E_{1},V_{1},N_{1})}{\partial V_{1}}-\dfrac{\partial S(E_{2},V_{2},N_{2})}{\partial V_{2}}=\dfrac{p_{1}}{T_{1}}-\dfrac{p_{2}}{T_{2}}=0,\label{eq:9}
\end{equation}
\begin{equation}
\dfrac{\partial\hat{S}}{\partial N_{1}}=\dfrac{\partial S(E_{1},V_{1},N_{1})}{\partial N_{1}}-\dfrac{\partial S(E_{2},V_{2},N_{2})}{\partial N_{2}}=-\dfrac{\mu_{1}}{T_{1}}+\dfrac{\mu_{2}}{T_{2}}=0,\label{eq:10}
\end{equation}
where we used the standard thermodynamic relations \citep{Callen_book_1960,Callen_book_1985,Munster:1970aa,Landau-Lifshitz-Stat-phys}
$\partial S/\partial E=1/T$, $\partial S/\partial V=p/T$ and $\partial S/\partial N=-\mu/T$
($T$ being temperature, $p$ pressure and $\mu$ chemical potential).
We thus recover the well-known thermodynamic equilibrium conditions
$T_{1}=T_{2}$, $p_{1}=p_{2}$ and $\mu_{1}=\mu_{2}$. These conditions
must be satisfied for the equilibrium parameter set $(E_{1}^{0},V_{1}^{0},N_{1}^{0})$. 

As another example, the non-equilibrium entropy of a gas mixture with
the chemical reaction $A+B\rightleftharpoons C$ is defined by Eq.(\ref{eq:4})
with a single internal parameter $\lambda=\Delta N_{A}$. The necessary
condition of equilibrium reads
\begin{eqnarray}
 &  & \dfrac{\partial S(E,V,N_{A}^{(i)}-\Delta N_{A},N_{B}^{(i)}-\Delta N_{A},N_{C}^{(i)}+\Delta N_{A})}{\partial\Delta N_{A}}\nonumber \\
 & = & -\dfrac{\partial S(E,V,N_{A},N_{B},N_{C})}{\partial N_{A}}-\dfrac{\partial S(E,V,N_{A},N_{B},N_{C})}{\partial N_{B}}+\dfrac{\partial S(E,V,N_{A},N_{B},N_{C})}{\partial N_{C}}\nonumber \\
 & = & 0.\label{eq:11}
\end{eqnarray}
We obtained the standard chemical equilibrium condition $\mu_{A}+\mu_{B}=\mu_{C}$,
which must be satisfied at some $\lambda^{0}=\Delta N_{A}^{0}$. The
equilibrium amounts of the chemical components are then $N_{A}^{0}=N_{A}^{(i)}-\Delta N_{A}^{0}$,
$N_{B}^{0}=N_{B}^{(i)}-\Delta N_{A}^{0}$ and $N_{C}^{0}=N_{C}^{(i)}+\Delta N_{A}^{0}$.

\section{The general law of fluctuations\label{sec:General-fluctations}}

During equilibrium fluctuations, the internal parameters randomly
deviate from the equilibrium values $\lambda_{1}^{0},...,\lambda_{m}^{0}$
on the time scale of $t_{f}$. These fluctuations of the internal
parameters are accompanied by deviations of the non-equilibrium entropy
$\hat{S}$ down from its maximum value $S$ {[}Fig.~\ref{fig:Fluctuations_1}(a){]}. 

The \emph{Fundamental Postulate} of the thermodynamic theory of fluctuations
states that the probability density $W(\lambda_{1},...,\lambda_{m})$
of internal parameters of an equilibrium isolated system is 
\begin{equation}
W(\lambda_{1},...,\lambda_{m})=A\exp\left[\frac{\hat{S}(\lambda_{1},...,\lambda_{m})}{k}\right],\label{eq:12}
\end{equation}
where $A$ is a constant. To simplify the notations, we do not display
the fixed parameters $X_{1},...,X_{n}$ as arguments of $W$ and $\hat{S}$. 

By definition, $W(\lambda_{1},...,\lambda_{m})d\lambda_{1}...d\lambda_{m}$
is the probability of finding the internal parameters in the infinitesimal
region $d\lambda_{1}...d\lambda_{m}$ of the parameter space. The
constant $A$ can be found from the normalization condition 
\begin{equation}
\int W(\lambda_{1},...,\lambda_{m})d\lambda_{1}...d\lambda_{m}=1,\label{eq:13}
\end{equation}
where the integration extends over the entire parameter space. Since
$\hat{S}$ reaches its maximum value
\begin{equation}
S=\hat{S}(\lambda_{1}^{0},...,\lambda_{m}^{0})\label{eq:14}
\end{equation}
 at the equilibrium parameter set $\lambda_{1}^{0},...,\lambda_{m}^{0}$
{[}cf.~Eq.(\ref{eq:7}){]}, $W$ has a peak of the height
\begin{equation}
W_{m}=A\exp\left(\frac{S}{k}\right)\label{eq:15}
\end{equation}
at $\lambda_{1}^{0},...,\lambda_{m}^{0}$ {[}Fig.~\ref{fig:Fluctuations_1}(b){]}.
In other words, $\lambda_{1}^{0},...,\lambda_{m}^{0}$ is the \emph{most
probable} parameter set. Using this fact, Eq.(\ref{eq:12}) can be
recast as
\begin{equation}
W(\lambda_{1},...,\lambda_{m})=W_{m}\exp\left[\frac{\hat{S}(\lambda_{1},...,\lambda_{m})-S}{k}\right].\label{eq:16}
\end{equation}

Equation (\ref{eq:16}) is the central equation of the thermodynamic
fluctuation theory. It is the departure point for thermodynamic analysis
of all statistical properties of equilibrium fluctuations. 

The average (expectation) value of any internal parameter $\lambda_{i}$
is obtained by 
\begin{equation}
\overline{\lambda}_{i}=W_{m}\int\lambda_{i}\exp\left[\frac{\hat{S}(\lambda_{1},...,\lambda_{m})-S}{k}\right]d\lambda_{1}...d\lambda_{m}.\label{eq:17}
\end{equation}
Note that this average does not generally coincide with the most probable
value $\lambda_{i}^{0}$. It is common to use the symbol $\Delta$
to denote deviations of fluctuating properties from their average
values. For example,
\begin{equation}
\Delta\lambda_{i}\equiv\lambda_{i}-\overline{\lambda}_{i}.\label{eq:17-1}
\end{equation}
The covariance (second correlation moment) of any pair of internal
parameters can be computed by 
\begin{equation}
\overline{\Delta\lambda_{i}\Delta\lambda_{j}}=W_{m}\int\Delta\lambda_{i}\Delta\lambda_{j}\exp\left[\frac{\hat{S}(\lambda_{1},...,\lambda_{m})-S}{k}\right]d\lambda_{1}...d\lambda_{m},\label{eq:18}
\end{equation}
while the mean-square fluctuation of $\lambda_{i}$ is 
\begin{equation}
\overline{(\Delta\lambda_{i})^{2}}=W_{m}\int(\Delta\lambda_{i})^{2}\exp\left[\frac{\hat{S}(\lambda_{1},...,\lambda_{m})-S}{k}\right]d\lambda_{1}...d\lambda_{m}.\label{eq:19}
\end{equation}

\section{Statistical-mechanical interpretation of fluctuations\label{sec:Stat-mech}}

In this section we will discuss the general law of fluctuations formulated
above from the viewpoint of statistical-mechanics.

As before, we consider an isolated system whose equilibrium state
is defined by a set of fixed parameters $X_{1},...,X_{n}$. If multiple
observations are made, the system can be found in different \emph{macro-states},
each specified by a set of internal parameters $\lambda_{1},...,\lambda_{m}$.
Let us first consider a system with discrete $\lambda$-parameters.
Each parameter set $\lambda_{1},...,\lambda_{m}$ is implemented by
a certain number $\Omega$ of \emph{micro-states}. The latter number
is sometimes called the degeneracy of the macro-state. The macro-states
are analogs of the equilibrium and quasi-equilibrium states introduced
in the thermodynamic theory (Sec.~\ref{sec:Non-equilibrium-entropy}).
The degeneracy of a given macro-state is a function of the internal
parameters $\lambda_{1},...,\lambda_{m}$ and the fixed parameters
$X_{1},...,X_{n}$:
\begin{equation}
\Omega=\Omega\left(\lambda_{1},...,\lambda_{m},X_{1},...,X_{n}\right).\label{eq:19-1}
\end{equation}

It is postulated that for an isolated equilibrium system, there is
one unique macro-state for which the degeneracy reaches a maximum
value $\Omega_{max}$. Denoting the internal parameters corresponding
to this macro-state $\lambda_{1}^{0},...,\lambda_{m}^{0}$, the maximum
degeneracy depends solely on the fixed parameters $X_{1},...,X_{n}$:
\begin{equation}
\Omega_{max}\left(X_{1},...,X_{n}\right)\equiv\Omega\left(\lambda_{1}^{0},...,\lambda_{m}^{0},X_{1},...,X_{n}\right).\label{eq:19-2}
\end{equation}

For a given set of fixed parameters $X_{1},...,X_{n}$, the degeneracy
$\Omega$ of an equilibrium isolated system fluctuates with time from
its maximum value $\Omega_{max}$ down to smaller values. These equilibrium
fluctuations follow the \emph{Fundamental Postulate of Statistical
Mechanics} stating that the probability $P$ of a single \emph{micro}-state
is the same for all micro-states and depends only on the fixed parameters
$X_{1},...,X_{n}$:
\begin{equation}
P=P\left(X_{1},...,X_{n}\right).\label{eq:19-3}
\end{equation}
This statement is often referred to as the \emph{micro-canonical distribution}.
It immediately follows that the probability of any macro-state is
simply $P\Omega$. Thus, macro-states with a larger degeneracy have
a proportionately higher probability to occur. The macro-state with
the largest degeneracy $\Omega_{max}$ is the most probable one. 

Analysis of fluctuations relies solely on the fact that $P$ is the
same for all micro-states and does not require a specific knowledge
of $P$. We note, however, that $P$ can be found from the normalization
condition stating that the sum of probabilities of all possible macro-states
of the system be unity. In the approximation where fluctuations are
neglected, only the most probable macro-state with the highest degeneracy
$\Omega_{max}$ is considered. Then the normalization condition reads
$P\Omega_{max}=1$, giving
\begin{equation}
P=\dfrac{1}{\Omega_{max}}.\label{eq:19-4}
\end{equation}
On the other hand, when fluctuations are significant, all possible
degeneracies from $\Omega=1$ to $\Omega_{max}$ must be included.
From the normalization condition
\begin{equation}
\sum_{\Omega=1}^{\Omega_{max}}P\Omega=1\label{eq:19-5}
\end{equation}
we obtain
\begin{equation}
P=\dfrac{2}{\Omega_{max}(\Omega_{max}+1)},\label{eq:19-6}
\end{equation}
which for large systems can be approximated by $2/\Omega_{max}^{2}$.

Regardless of the specific value of $P$, the product $P\Omega$ is
the probability of a given macro-state and thus the probability 
\begin{equation}
W_{\lambda_{1},...,\lambda_{m}}=P\Omega\left(\lambda_{1},...,\lambda_{m}\right)\label{eq:19-7}
\end{equation}
of the corresponding set of internal parameters. Here, as in Sec.~\ref{sec:General-fluctations},
we simplify the notations by suppressing the fixed parameters $X_{1},...,X_{n}$
as arguments of $P$ and $\Omega$. The probability $W_{\lambda_{1},...,\lambda_{m}}$
reaches a maximum value, denoted $W_{m}$, at the most probable parameter
set $\lambda_{1}^{0},...,\lambda_{m}^{0}$, for which
\begin{equation}
W_{m}=P\Omega\left(\lambda_{1}^{0},...,\lambda_{m}^{0}\right)=P\Omega_{max}.\label{eq:19-8}
\end{equation}
Combining Eqs.(\ref{eq:19-7}) and (\ref{eq:19-8}),
\begin{equation}
W_{\lambda_{1},...,\lambda_{m}}=W_{m}\dfrac{\Omega\left(\lambda_{1},...,\lambda_{m}\right)}{\Omega_{max}}.\label{eq:19-9}
\end{equation}

We now define the non-equilibrium entropy of an isolated system by
\begin{equation}
\hat{S}\left(\lambda_{1},...,\lambda_{m}\right)\equiv k\ln\Omega\left(\lambda_{1},...,\lambda_{m}\right)\label{eq:19-10}
\end{equation}
and the equilibrium entropy by 
\begin{equation}
S\equiv k\ln\Omega_{max}.\label{eq:19-11}
\end{equation}
It immediately follows that the non-equilibrium entropy can never
exceed its equilibrium value $S$. It also follows that in equilibrium,
$\hat{S}$ can occasionally fluctuate down from $S$ but eventually
returns to $S$ as the most probable value. 

The probability of any given set of internal parameters given by Eq.(\ref{eq:19-9})
can now be rewritten in the form
\begin{equation}
W_{\lambda_{1},...,\lambda_{m}}=W_{m}\exp\left[\frac{\hat{S}(\lambda_{1},...,\lambda_{m})-S}{k}\right].\label{eq:19-12}
\end{equation}
This equation treats $\lambda_{1},...,\lambda_{m}$ as discrete parameters.
For continuous parameters, Eq.(\ref{eq:19-12}) remains valid but
gives the probability density of the internal parameters. The obtained
Eq.(\ref{eq:19-12}) is identical to Eq.(\ref{eq:16}) of the thermodynamic
theory of fluctuations (Sec.~\ref{sec:General-fluctations}).

\section{The Gaussian law of fluctuations\label{sec:Gaussian-law}}

In macroscopic systems, the peak of $W$ at the equilibrium parameter
set $\lambda_{1}^{0},...,\lambda_{m}^{0}$ is extremely sharp. In
most cases, one can safely replace the exponent in Eq.(\ref{eq:16})
by its Taylor expansion at $\lambda_{1}^{0},...,\lambda_{m}^{0}$
truncated at quadratic terms:
\begin{equation}
\hat{S}(\lambda_{1},...,\lambda_{m})-S=\sum_{i=1}^{m}\left(\dfrac{\partial\hat{S}}{\partial\lambda_{i}}\right)_{0}(\lambda_{i}-\lambda_{i}^{0})+\frac{1}{2}\sum_{i,j=1}^{m}\left(\dfrac{\partial^{2}\hat{S}}{\partial\lambda_{i}\partial\lambda_{j}}\right)_{0}(\lambda_{i}-\lambda_{i}^{0})(\lambda_{j}-\lambda_{j}^{0}),\label{eq:20}
\end{equation}
where the subscript 0 indicates that the derivatives are taken at
$\lambda_{1}^{0},...,\lambda_{m}^{0}$. The linear terms vanish by
the extremum condition (\ref{eq:6}). As a result, Eq.(\ref{eq:16})
reduces to the multi-variable Gaussian distribution
\begin{equation}
W(\lambda_{1},...,\lambda_{m})=W_{m}\exp\left[-\frac{1}{2}\sum_{i,j=1}^{m}\Lambda_{ij}(\lambda_{i}-\lambda_{i}^{0})(\lambda_{j}-\lambda_{j}^{0})\right]\label{eq:21}
\end{equation}
with the symmetrical matrix 
\begin{equation}
\Lambda_{ij}\equiv-\frac{1}{k}\left(\dfrac{\partial^{2}\hat{S}}{\partial\lambda_{i}\partial\lambda_{j}}\right)_{0}\label{eq:22}
\end{equation}
which we call the \emph{stability matrix} of the system. For normally
stable systems this matrix must be positive-definite.\footnote{We follow the classification \citep{Tisza:1961aa} dividing all systems
into normally stable, critically stable, unstable and metastable.
In normally stable systems, all fluctuations are finite.} Applying the normalization condition (\ref{eq:13}), the Gaussian
distribution can be rewritten as 
\begin{equation}
W(\lambda_{1},...,\lambda_{m})=\sqrt{\dfrac{|\Lambda|}{(2\pi)^{m}}}\exp\left[-\frac{1}{2}\sum_{i,j=1}^{m}\Lambda_{ij}(\lambda_{i}-\lambda_{i}^{0})(\lambda_{j}-\lambda_{j}^{0})\right],\label{eq:23}
\end{equation}
where $\Lambda$ is the determinant of $\Lambda_{ij}$.\footnote{The Gaussian distribution (\ref{eq:23}) can be interpreted as a consequence
of the central limit theorem for a system composed of many statistically
independent subsystems, see for example Chapter V of Khinchin \citep{Khinchin_1949}.}

We can now make use of the well-known mathematical properties of the
normal distribution. The average value of any parameter $\lambda_{i}$
coincides with its most probable value, 
\begin{equation}
\overline{\lambda}_{i}=\lambda_{i}^{0},\,\,\,\,\,i=1,...,m.\label{eq:24}
\end{equation}
For covariances of internal parameters we have 
\begin{equation}
\overline{\Delta\lambda_{i}\Delta\lambda_{j}}=\Lambda_{ij}^{-1},\,\,\,\,\,i,j=1,...,m,\label{eq:25}
\end{equation}
where $\Lambda_{ij}^{-1}$ is the inverse of the stability matrix
$\Lambda_{ij}$. In particular, the mean-square fluctuation of a parameter
$\lambda_{i}$ equals 
\begin{equation}
\overline{(\Delta\lambda_{i})^{2}}=\Lambda_{ii}^{-1},\,\,\,\,\,i=1,...,m.\label{eq:26}
\end{equation}

We can also introduce the following linear combinations of internal
parameters:
\begin{equation}
\varphi_{i}\equiv\sum_{j=1}^{m}\Lambda_{ij}\lambda_{j},\,\,\,\,\,i=1,...,m.\label{eq:27}
\end{equation}
It can be shown that
\begin{equation}
\overline{\Delta\varphi_{i}}=0,\,\,\,\,\,i=1,...,m,\label{eq:27-1}
\end{equation}
\begin{equation}
\overline{\Delta\varphi_{i}\Delta\lambda_{j}}=\delta_{ij},\,\,\,\,\,i,j=1,...,m,\label{eq:28}
\end{equation}
\begin{equation}
\overline{\Delta\varphi_{i}\Delta\varphi_{j}}=\Lambda_{ij},\,\,\,\,\,i,j=1,...,m,\label{eq:29}
\end{equation}
and therefore
\begin{equation}
\overline{(\Delta\varphi_{i})^{2}}=\Lambda_{ii},\,\,\,\,\,i=1,...,m.\label{eq:30}
\end{equation}
Thus $\Lambda_{ij}$ is the covariance matrix of the parameters $\varphi_{i}$. 

An important property of Eqs.(\ref{eq:25}), (\ref{eq:26}), (\ref{eq:29})
and (\ref{eq:30}) is that their left-hand sides contain fluctuations,
whereas the right-hand sides are computed in the state of equilibrium
and can be obtained by the standard formalism of equilibrium thermodynamics.
In other words, these equations relate fluctuations to equilibrium
properties of the system.

As a simple illustration, let us revisit the isolated system divided
in two compartments (Fig.~\ref{fig:cylinder}). Suppose the system
is in equilibrium. Consider a particular type of fluctuations in which
the compartments can only exchange energy at fixed volumes and numbers
of particles. Thus, the only internal parameter is $\lambda_{1}=E_{1}$.
The stability matrix (\ref{eq:22}) reduces to the scalar 
\begin{equation}
\Lambda=-\dfrac{1}{k}\left(\dfrac{\partial^{2}\hat{S}}{\partial E_{1}^{2}}\right)_{0}.\label{eq:34}
\end{equation}
From Eq.(\ref{eq:8}),
\begin{equation}
\dfrac{\partial\hat{S}}{\partial E_{1}}=\dfrac{1}{T_{1}}-\dfrac{1}{T_{2}}\label{eq:35}
\end{equation}
and thus
\begin{equation}
\Lambda=\dfrac{1}{k}\left(\dfrac{1}{T_{1}^{2}}\dfrac{\partial T_{1}}{\partial E_{1}}+\dfrac{1}{T_{2}^{2}}\dfrac{\partial T_{2}}{\partial E_{2}}\right)_{0}=\dfrac{1}{kT_{0}^{2}c_{v}}\left(\dfrac{1}{N_{1}}+\dfrac{1}{N_{2}}\right),\label{eq:36}
\end{equation}
where $T_{0}$ is the equilibrium temperature of both compartments
and 
\begin{equation}
c_{v}=\dfrac{1}{N}\left(\dfrac{\partial E}{\partial T}\right)_{V,N}\label{eq:37}
\end{equation}
is the specific heat of the substance per particle at the equilibrium
temperature $T_{0}$. Applying Eq.(\ref{eq:26}) we obtain the energy
fluctuation of compartment 1:
\begin{equation}
\left(\overline{(\Delta E_{1})^{2}}\right)_{V_{1},N_{1}}=\dfrac{1}{\Lambda}=kT_{0}^{2}c_{v}\dfrac{N_{1}N_{2}}{N_{1}+N_{2}}.\label{eq:38}
\end{equation}

When the second compartment is much larger than the first, $N_{2}\gg N_{1}$,
this equation reduces to 
\begin{equation}
\left(\overline{(\Delta E_{1})^{2}}\right)_{V_{1},N_{1}}=N_{1}kT_{0}^{2}c_{v}.\label{eq:39}
\end{equation}
In this limit, the second compartment serves as a heat bath (reservoir)
for the first. The relative fluctuation (variance) of $E_{1}$ becomes
\begin{equation}
\dfrac{\left[\left(\overline{(\Delta E_{1})^{2}}\right)_{V_{1},N_{1}}\right]^{1/2}}{\overline{E_{1}}}=\dfrac{T_{0}(c_{v}k)^{1/2}}{\varepsilon\sqrt{N_{1}}},\label{eq:40}
\end{equation}
where $\varepsilon=\overline{E_{1}}/N_{1}$ is the equilibrium energy
per particle. This relation shows that the relative fluctuation of
energy decreases with the system size as $1/\sqrt{N}$.

\section{Fluctuations in canonical systems\label{sec:Canonical}}

\subsection{General relations\label{sub:General-relations-canonical}}

We now return to the general fluctuation law formulated in Secs.~\ref{sec:General-fluctations}
and will apply it to a relatively small (but still macroscopic) subsystem
of an equilibrium isolated system. We will adopt the terminology in
which we refer to the selected subsystem as the ``canonical system'',
or simply ``system'', and to the remaining part of the isolated system
as the ``complementary system'' or ``reservoir''. 

\begin{figure}
\begin{centering}
\includegraphics[clip,width=0.4\columnwidth]{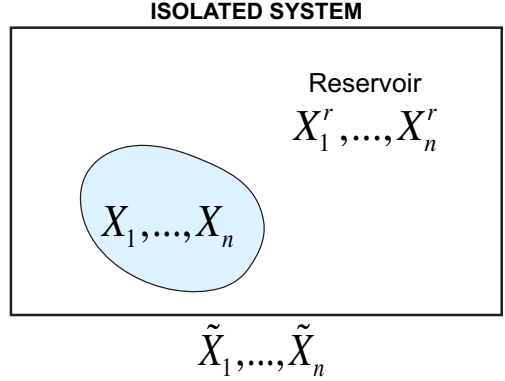}
\par\end{centering}

\protect\caption{A subsystem chosen inside an isolated system characterized by a set
extensive properties $\tilde{X}_{1},...,\tilde{X}_{n}$. These properties
are partitioned between the subsystem and the complementary system
(reservoir). The partitioning is specified by the parameter sets $X_{1},...,X_{n}$
and $X_{1}^{r},...,X_{n}^{r}$, respectively. \label{fig:canonical}}

\end{figure}

An example of the situation was already given by the system with two
compartments (Fig.~\ref{fig:cylinder}) when the second compartment
is much larger than the first. We will now consider a more general
case in which the entire isolated system is characterized by a set
of conserved extensive properties (additive invariants) $\tilde{X}_{1},...,\tilde{X}_{n}$.
These properties are partitioned between our system, $X_{1},...,X_{n}$,
and the reservoir, $X_{1}^{r},...,X_{n}^{r}$ (Fig.~\ref{fig:canonical}).
For some of these properties, the partitioning is fixed once and for
all, while $m$ remaining properties can ``flow'' between the system
and the reservoir as a result of equilibrium fluctuations. Such fluctuations
are subject to the conservation constraints
\begin{equation}
X_{i}+X_{i}^{r}=\tilde{X}_{i}=\textnormal{const},\,\,\,\,\,i=1,...,m.\label{eq:41}
\end{equation}

We assume that the exchange of the quantities $X_{i}$ occurs slowly
enough that we can consider both the system and the reservoir as quasi-equilibrium
systems. Accordingly, they obey the fundamental equations
\begin{equation}
S=S\left(X_{1},...,X_{n}\right)\label{eq:42}
\end{equation}
for the system and
\begin{equation}
S_{r}=S_{r}\left(X_{1}^{r},...,X_{n}^{r}\right)\label{eq:43}
\end{equation}
for the reservoir. Note that these two equations are generally different,
allowing the system and the reservoir to be composed of different
substances. 

In the subsequent analysis, a special role will be played by the entropy
derivatives
\begin{equation}
F_{i}\equiv\dfrac{\partial S}{\partial X_{i}},\label{eq:43-1}
\end{equation}
called the thermodynamic ``forces'' conjugate to the extensive variables
$X_{i}$. Similar derivatives $\partial S_{r}/\partial X_{i}^{r}$
define the forces $F_{i}^{r}$ in the reservoir. For example, for
a simple fluid with the fundamental equation $S(E,V,N)$, the thermodynamic
forces are $F_{1}=1/T$, $F_{2}=p/T$ and $F_{3}=-\mu/T$. Note that
the thermodynamic forces are intensive variables.

We define the non-equilibrium entropy of the entire isolated system
as
\begin{equation}
\hat{S}(\underbrace{X_{1},...,X_{m}}_{\lambda},X_{m+1},...,X_{n},\tilde{X}_{1},...,\tilde{X}_{n})\equiv S\left(X_{1},...,X_{n}\right)+S_{r}\left(\tilde{X}_{1}-X_{1},...,\tilde{X}_{n}-X_{n}\right),\label{eq:44}
\end{equation}
where $X_{1},...,X_{m}$ play the role of the fluctuating internal
parameters $\lambda_{i}$, whereas the remaining parameters $X_{m+1},...,X_{n},\tilde{X}_{1},...,\tilde{X}_{n}$
remain fixed. The equilibrium value of this entropy is 
\begin{eqnarray}
\tilde{S} & = & S\left(X_{1}^{0},...,X_{m}^{0},X_{m+1},...,X_{n}\right)\nonumber \\
 & + & S_{r}\left(\tilde{X}_{1}-X_{1}^{0},...,\tilde{X}_{m}-X_{m}^{0},\tilde{X}_{m+1}-X_{m+1},...,\tilde{X}_{n}-X_{n}\right),\label{eq:44-1}
\end{eqnarray}
where $X_{1}^{0},...,X_{m}^{0}$ are the equilibrium values of the
fluctuating parameters. To simplify the notations, from now on we
will suppress the fixed parameters in all equations, writing Eqs.(\ref{eq:44})
and (\ref{eq:44-1}) in the form
\begin{equation}
\hat{S}\left(X_{1},...,X_{m}\right)\equiv S\left(X_{1},...,X_{m}\right)+S_{r}\left(\tilde{X}_{1}-X_{1},...,\tilde{X}_{m}-X_{m}\right),\label{eq:44-2}
\end{equation}
\begin{equation}
\tilde{S}=S\left(X_{1}^{0},...,X_{m}^{0}\right)+S_{r}\left(\tilde{X}_{1}-X_{1}^{0},...,\tilde{X}_{m}-X_{m}^{0}\right).\label{eq:44-3}
\end{equation}

Before analyzing fluctuations, we will apply the equilibrium conditions
(\ref{eq:6}) to the internal parameters $\lambda_{i}=X_{i}$ ($i=1,...,m$):
\begin{equation}
\left(\dfrac{\partial\hat{S}}{\partial\lambda_{i}}\right)_{0}=\left(\dfrac{\partial S}{\partial X_{i}}\right)_{0}-\left(\dfrac{\partial S_{r}}{\partial X_{i}^{r}}\right)_{0}=F_{i}^{0}-\left(F_{i}^{r}\right)^{0}=0,\,\,\,\,\,i=1,...,m,\label{eq:45}
\end{equation}
where, as usual, the symbol 0 indicates that the derivatives are computed
for the equilibrium parameters $X_{1}^{0},...,X_{m}^{0}$. We thus
conclude that in equilibrium, the system and the reservoir have equal
values of the thermodynamic forces conjugate to the fluctuating parameters. 

We now turn to fluctuations. The probability density of the internal
parameters $X_{1},...,X_{m}$ is given by Eq.(\ref{eq:16}), which
takes the form
\begin{eqnarray}
W(X_{1},...,X_{m}) & = & W_{m}\exp\left[\frac{\hat{S}(X_{1},...,X_{m})-\tilde{S}}{k}\right]\nonumber \\
 & = & W_{m}\exp\left[\dfrac{S(X_{1},...,X_{m})-S(X_{1}^{0},...,X_{m}^{0})}{k}\right]\nonumber \\
 & \times & \exp\left[\dfrac{S_{r}(\tilde{X}_{1}-X_{1},...,\tilde{X}_{m}-X_{m})-S_{r}(\tilde{X}_{1}-X_{1}^{0},...,\tilde{X}_{m}-X_{m}^{0})}{k}\right].\label{eq:46}
\end{eqnarray}
In the second exponential factor, we will expand $S_{r}$ around the
equilibrium parameters $X_{1}^{0},...,X_{m}^{0}$:
\begin{eqnarray}
S_{r}(\tilde{X}_{1}-X_{1},...,\tilde{X}_{m}-X_{m})-S_{r}(\tilde{X}_{1}-X_{1}^{0},...,\tilde{X}_{m}-X_{m}^{0}) & = & -\sum_{i=1}^{m}F_{i}^{0}(X_{i}-X_{i}^{0})\nonumber \\
+\frac{1}{2}\sum_{i,j=1}^{m}\left(\dfrac{\partial^{2}S_{r}}{\partial X_{i}^{r}\partial X_{j}^{r}}\right)_{0}(X_{i}-X_{i}^{0})(X_{j}-X_{j}^{0}).\label{eq:47}
\end{eqnarray}

As will be shown later, the second derivatives appearing in the quadratic
terms scale as $1/N^{r}$, where $N^{r}$ is the number of particles
in the reservoir. Since the reservoir is assumed to be much larger
than our system, the quadratic terms in Eq.(\ref{eq:47}) can be neglected
in comparison with the linear terms. Neglecting the quadratic terms
constitutes what can be called the ``reservoir approximation'', in
which
\begin{equation}
S_{r}(\tilde{X}_{1}-X_{1},...,\tilde{X}_{m}-X_{m})-S_{r}(\tilde{X}_{1}-X_{1}^{0},...,\tilde{X}_{m}-X_{m}^{0})=-\sum_{i=1}^{m}F_{i}^{0}(X_{i}-X_{i}^{0}).\label{eq:48}
\end{equation}
Inserting this equation in Eq.(\ref{eq:46}), we obtain
\begin{equation}
W(X_{1},...,X_{m})=W_{m}\exp\left[\frac{S(X_{1},...,X_{m})-S(X_{1}^{0},...,X_{m}^{0})-{\displaystyle \sum_{i=1}^{m}F_{i}^{0}(X_{i}-X_{i}^{0})}}{k}\right].\label{eq:49}
\end{equation}
Note that this equation contains only properties of the canonical
system and does not depend on properties of the reservoir. The role
of the reservoir is only to impose the thermodynamic forces $F_{i}^{0}$. 

Equation (\ref{eq:49}) is an important relation that will serve as
the starting point for deriving all fluctuation properties of canonical
systems. 

In some canonical systems, there can be additional internal parameters,
say $Y_{1},Y_{2},...,Y_{l}$, that can fluctuate without affecting
the reservoir parameters $X_{1}^{r},...,X_{n}^{r}$. In such cases,
the set of internal parameters is $X_{1},...,X_{m},Y_{1},Y_{2},...,Y_{l}$
and Eq.(\ref{eq:44-2}) is replaced by

\begin{equation}
\hat{S}\left(X_{1},...,X_{m},Y_{1},Y_{2},...,Y_{l}\right)\equiv S\left(X_{1},...,X_{m},Y_{1},Y_{2},...,Y_{l}\right)+S_{r}\left(\tilde{X}_{1}-X_{1},...,\tilde{X}_{m}-X_{m}\right).\label{eq:44-2-1}
\end{equation}
The equilibrium conditions with respect to the parameters decoupled
from the reservoir is
\begin{equation}
\left(\dfrac{\partial\hat{S}}{\partial Y_{i}}\right)_{0}=\left(\dfrac{\partial S}{\partial Y_{i}}\right)_{0}\equiv F_{i+m}^{0}=0,\label{eq:49-4}
\end{equation}
where $i=1,2,...,l$. It is easy to see that the additional parameters
can be incorporated into the parameter list $X_{1},...,X_{m}$ by
simply assigning them zero thermodynamic forces. Accordingly, Eq.(\ref{eq:49})
remains valid, except that the sum with $F_{i}^{0}$ excludes the
terms corresponding to the decoupled parameters $Y_{1},Y_{2},...,Y_{l}$.

\subsection{The generalized canonical distribution\label{sub:Generalized-canonical-ensembles}}

Before proceeding with applications of Eq.(\ref{eq:49}), we return
for a moment to the statistical-mechanical interpretation of fluctuations
discussed in Sec.~\ref{sec:Stat-mech}. For discrete internal parameters
$X_{1},...,X_{m}$, Eq.(\ref{eq:49}) gives the probability $W_{X_{1},...,X_{m}}$
of a given parameter set $X_{1},...,X_{m}$. Recall that on the quasi-equilibrium
time scale $t_{q}$, our system can be treated as if it were equilibrium
and isolated with fixed values of $X_{1},...,X_{m}$. Thus, its fundamental
equation $S(X_{1},...,X_{m})$ gives the equilibrium entropy $S=k\ln\Omega_{max}$.
The probability $P$ of a single \emph{micro-state} can be obtained
from the relation $W_{X_{1},...,X_{m}}=P\Omega_{max}$, which gives
\begin{equation}
P=\dfrac{W_{X_{1},...,X_{m}}}{\Omega_{max}}=W_{X_{1},...,X_{m}}\exp\left[-\frac{S(X_{1},...,X_{m})}{k}\right].\label{eq:50}
\end{equation}
Combining this relation with Eq.(\ref{eq:49}) we obtain
\begin{equation}
P=W_{m}\exp\left[-\frac{S(X_{1}^{0},...,X_{m}^{0})+{\displaystyle \sum_{i=1}^{m}F_{i}^{0}(X_{i}-X_{i}^{0})}}{k}\right],\label{eq:51}
\end{equation}
which can be rewritten as 
\begin{equation}
P=A\exp\left(-\frac{1}{k}\sum_{i=1}^{m}F_{i}^{0}X_{i}\right),\label{eq:52}
\end{equation}
where 
\begin{equation}
A=W_{m}\exp\left[-\frac{S(X_{1}^{0},...,X_{m}^{0})-{\displaystyle \sum_{i=1}^{m}F_{i}^{0}X_{i}^{0}}}{k}\right]\label{eq:53}
\end{equation}
is a constant.

Equation (\ref{eq:52}) constitutes the \emph{generalized canonical
distribution} describing several statistical ensembles.\footnote{The way we arrived at Eq.(\ref{eq:52}) is not intended to be a rigorous
statistical-mechanical derivation of the canonical distribution. The
goal was only to show that the thermodynamic fluctuation relation
(\ref{eq:49}) is consistent with the distribution of micro-states
known from statistical mechanics.} 

Assuming that one of the extensive parameters, say $X_{1}$, is energy,
Eq.(\ref{eq:52}) can be rewritten in the form
\begin{equation}
P=A\exp\left(-\frac{{\displaystyle E+\sum_{i=2}^{m}T_{0}F_{i}^{0}X_{i}}}{kT_{0}}\right).\label{eq:52-2}
\end{equation}
As an illustration, suppose our system is a simple fluid. Either one
or two of its extensive parameters $E$, $V$ and $N$ can fluctuate
with the third parameter fixed (otherwise the system is undefined).
For example, suppose only energy can fluctuate while $V$ and $N$
remain fixed. From Eq.(\ref{eq:52-2}), the probability of a micro-state
is
\begin{equation}
P=A\exp\left(-\frac{E}{kT_{0}}\right),\;\;\textnormal{fixed }V,N,\label{eq:54}
\end{equation}
a relation which is known as the canonical, or $NVT$, distribution.
If both energy and volume are allowed to fluctuate, we have $X_{2}=V$
with $F_{2}=p/T$, giving the $NpT$ distribution 
\begin{equation}
P=A\exp\left(-\frac{E+p_{0}V}{kT_{0}}\right),\;\;\textnormal{fixed }N.\label{eq:55}
\end{equation}
If the energy and the number of particles can fluctuate at a fixed
volume (open system), we have $X_{2}=N$ and $F_{2}=-\mu/T$. Equation
(\ref{eq:52-2}) gives the grand-canonical, or $\mu VT$, distribution
\begin{equation}
P=A\exp\left(-\frac{E-\mu_{0}N}{kT_{0}}\right),\;\;\textnormal{fixed }V.\label{eq:56}
\end{equation}
Of course, the pre-exponential factor $A$ is different in all three
distributions and is related to the respective partition functions
through the probability normalization condition. 

Note that the distributions (\ref{eq:54}), (\ref{eq:55}) and (\ref{eq:56})
contain the temperature $T_{0}$, pressure $p_{0}$ and the chemical
potential $\mu_{0}$ corresponding to the exact equilibrium with the
reservoir. These properties are fixed parameters imposed by the reservoir,
by contrast to the fluctuating temperature, pressure and chemical
potential inside the system. 

We re-emphasize the important difference between Eq.(\ref{eq:52})
and the previously derived Eq.(\ref{eq:49}): Eq.(\ref{eq:52}) gives
the probability of a single \emph{micro-state} of the system, whereas
Eq.(\ref{eq:49}) is the probability distribution of the parameters
$X_{1},...,X_{m}$ characterizing different \emph{macro-states}.

\subsection{The Gaussian law of canonical fluctuations\label{sub:The-Gaussian-law-canonical}}

We now return to the probability distribution of the fluctuating parameters
given by Eq.(\ref{eq:49}). Let us expand the entropy of the system
around its equilibrium value, keeping only linear and quadratic terms:
\begin{equation}
S(X_{1},...,X_{m})-S(X_{1}^{0},...,X_{m}^{0})=\sum_{i=1}^{m}F_{i}^{0}(X_{i}-X_{i}^{0})+\frac{1}{2}\sum_{i,j=1}^{m}\left(\dfrac{\partial^{2}S}{\partial X_{i}\partial X_{j}}\right)_{0}(X_{i}-X_{i}^{0})(X_{j}-X_{j}^{0}).\label{eq:57}
\end{equation}
Inserting this expansion in Eq.(\ref{eq:49}), we arrive at the multi-variable
Gaussian distribution
\begin{equation}
W(X_{1},...,X_{m})=W_{m}\exp\left[-\frac{1}{2}\sum_{i,j=1}^{m}\Lambda_{ij}(X_{i}-X_{i}^{0})(X_{j}-X_{j}^{0})\right],\label{eq:58}
\end{equation}
with the stability matrix 
\begin{equation}
\Lambda_{ij}\equiv-\frac{1}{k}\left(\dfrac{\partial^{2}S}{\partial X_{i}\partial X_{j}}\right)_{0}=-\frac{1}{k}\left(\dfrac{\partial F_{i}}{\partial X_{j}}\right)_{0}.\label{eq:59}
\end{equation}

In this approximation, the average values $\overline{X}_{i}$ are
numerically equal to the most probable values $X_{i}^{0}$. Using
the standard properties of the Gaussian distribution (Sec.~\ref{sec:Gaussian-law}),
we obtain the fluctuation relations 
\begin{equation}
\overline{\Delta X_{i}\Delta X_{j}}=\Lambda_{ij}^{-1},\,\,\,\,\,i,j=1,...,m,\label{eq:60}
\end{equation}
and thus 
\begin{equation}
\overline{(\Delta X_{i})^{2}}=\Lambda_{ii}^{-1},\,\,\,\,\,i=1,...,m.\label{eq:61}
\end{equation}
We can also introduce the parameters 
\begin{equation}
\varphi_{i}\equiv\sum_{j=1}^{m}\Lambda_{ij}X_{j},\,\,\,\,\,i=1,...,m,\label{eq:62}
\end{equation}
for which 
\begin{equation}
\varphi_{i}-\overline{\varphi}_{i}=-\frac{1}{k}\sum_{j=1}^{m}\left(\dfrac{\partial F_{i}}{\partial X_{j}}\right)_{0}(X_{j}-X_{j}^{0})\equiv-\frac{1}{k}\Delta F_{i}.\label{eq:63}
\end{equation}
Here, $\Delta F_{i}$ is the deviation of $F_{i}$ from its equilibrium
value evaluated in the quadratic approximation (\ref{eq:57}). Using
the properties (\ref{eq:29}) and (\ref{eq:30}) of the Gaussian distribution,
we obtain the fluctuation relations for thermodynamic forces:
\begin{equation}
\overline{\Delta X_{i}\Delta F_{j}}=-k\delta_{ij},\,\,\,\,\,i,j=1,...,m,\label{eq:64}
\end{equation}
\begin{equation}
\overline{\Delta F_{i}\Delta F_{j}}=k^{2}\Lambda_{ij},\,\,\,\,\,i,j=1,...,m,\label{eq:65}
\end{equation}
and therefore 
\begin{equation}
\overline{(\Delta F_{i})^{2}}=k^{2}\Lambda_{ii},\,\,\,\,\,i=1,...,m.\label{eq:66}
\end{equation}

Equations (\ref{eq:61}) and (\ref{eq:66}) show that for normally
stable states, the derivatives $(\partial X_{i}/\partial F_{i})_{0}$
and $(\partial F_{i}/\partial X_{i})_{0}$ must be positive. At critical
points one of these derivatives tends to infinity and the respective
fluctuation diverges (unless the reservoir has a finite size, see
Sec.~\ref{sec:Finite-reservoir}). 

Another form of the foregoing equations is obtained by introducing
the thermodynamic potential 
\begin{equation}
\Theta(F_{1},...,F_{m})\equiv S-\sum_{i=1}^{m}F_{i}X_{i}\label{eq:66-10}
\end{equation}
as the Legendre transform of the entropy (Massieu function) \citep{Callen_book_1960,Munster:1970aa,Callen_book_1985}.
Then,
\begin{equation}
X_{i}=-\dfrac{\partial\Theta}{\partial F_{i}}\label{eq:66-11}
\end{equation}
and 
\begin{equation}
\Lambda_{ij}^{-1}=k\left(\dfrac{\partial^{2}\Theta}{\partial F_{i}\partial F_{j}}\right)_{0}.\label{eq:66-12}
\end{equation}
For normally stable states $(\partial^{2}\Theta/\partial F_{i}^{2})_{0}>0$
for all $i=1,...,m$. 

It is interesting to evaluate canonical fluctuations of entropy relative
to its equilibrium value $S(X_{1}^{0},...,X_{m}^{0})$. Using the
quadratic approximation (\ref{eq:57}), we have
\begin{equation}
\Delta S=\sum_{i=1}^{m}F_{i}^{0}\Delta X_{i}-\frac{1}{2}k\sum_{i,j=1}^{m}\Lambda_{ij}\Delta X_{i}\Delta X_{j}.\label{eq:57-1}
\end{equation}
Averaging this equation over the distribution (\ref{eq:58}) and taking
into account Eq.(\ref{eq:60}), we obtain
\begin{equation}
\overline{\Delta S}=\sum_{i=1}^{m}F_{i}^{0}\overline{\Delta X_{i}}-\frac{1}{2}k\sum_{i,j=1}^{m}\Lambda_{ij}\overline{\Delta X_{i}\Delta X_{j}}=-\frac{1}{2}k.\label{eq:66-1}
\end{equation}
Thus, in the Gaussian approximation the canonical entropy equals 
\begin{equation}
\bar{S}=S(X_{1}^{0},...,X_{m}^{0})-\frac{1}{2}k.\label{eq:66-2}
\end{equation}
Considering that entropy scales in proportion to the total of number
of particles $N$, the relative fluctuation $\overline{\Delta S}/\bar{S}\propto N^{-1}$.
Using the same Gaussian approximation, the mean-square fluctuation
of entropy scales in proportion to $N$,
\begin{equation}
\overline{(\Delta S)^{2}}=\sum_{i,j=1}^{m}F_{i}^{0}F_{j}^{0}\overline{\Delta X_{i}\Delta X_{j}}=\sum_{i,j=1}^{m}F_{i}^{0}F_{j}^{0}\Lambda_{ij}^{-1}\propto N.\label{eq:66-3}
\end{equation}
where we neglected higher-order moments of the distribution. Thus,
the relative fluctuation $\left(\overline{(\Delta S)^{2}}\right)^{1/2}/\bar{S}\propto N^{-1/2}$.
Both estimates show that, with increasing size of the system, the
canonical entropy $\bar{S}$ tends to the micro-canonical entropy
$S$ satisfying the fundamental equation (\ref{eq:1}). While the
two entropies are conceptually different, they converge to the same
numerical value in the thermodynamic limit.

\subsection{Canonical fluctuations in the energy scheme\label{sub:energy_scheme}}

The treatment of fluctuations based on the fundamental equation (\ref{eq:1})
is called the entropy scheme. In practical situations, energy is always
one of the extensive parameters fluctuating between the canonical
system and the reservoir.\footnote{While ``adiabatic ensembles'' in which the system is adiabatically
isolated but can still exchange particles with the reservoir are helpful
theoretical constructions, their practical implementation in experiments
or atomistic simulations is unfeasible.} Suppose $X_{1}=E$ and thus $F_{1}=1/T$. Equation (\ref{eq:1})
can be inverted to obtain 
\begin{equation}
E=E\left(Z_{1},...,Z_{n}\right),\label{eq:1-1}
\end{equation}
where $Z_{1}=S$ and $Z_{i}=X_{i}$, $i=2,...,n$. This equation is
called the fundamental equation in the energy representation, or the
energy scheme \citep{Tisza:1961aa}. In the energy scheme, the thermodynamic
forces $P_{i}=\partial E/\partial Z_{i}$ conjugate to the extensive
arguments $Z_{1},...,Z_{n}$ are $P_{1}=T$ and $P_{i}=-TF_{i}$,
$i=2,...,n$. For example, for a simple fluid $E=E(S,V,N)$ and thus
$P_{1}=T$, $P_{2}=-p$ and $P_{3}=\mu$.

In many applications, we are more interested in fluctuations of the
$P$-forces rather than $F$-forces. The formalism presented below
expresses canonical fluctuations in terms of the variable sets $Z_{1},...,Z_{n}$
and $P_{1},...,P_{n}$ corresponding to the energy scheme.

As before, we consider a system coupled to a reservoir and described
by $n$ extensive properties $X_{1},...,X_{n}$. Out of them, $m$
properties $X_{1},...,X_{m}$ can flow between the system and the
reservoir while the remaining properties $X_{m+1},...,X_{n}$ are
fixed. Assuming $X_{1}=E$, we transform the variable set $X_{1},...,X_{m}$
to $Z_{1},...,Z_{m}$ and perform a Legendre transformation to obtain
the thermodynamic potential appropriate for the the energy scheme:
\begin{equation}
\Phi(P_{1},...,P_{m})\equiv E-\sum_{i=1}^{m}P_{i}Z_{i}.\label{eq:1-2}
\end{equation}
This potential has the property 
\begin{equation}
\dfrac{\partial\Phi}{\partial P_{i}}=-Z_{i}.\label{eq:1-3}
\end{equation}

Returning to the parameter distribution (\ref{eq:49}), we can now
recast it in the form
\begin{equation}
W=W_{m}\exp\left[-\frac{E-E_{0}-{\displaystyle \sum_{i=1}^{m}P_{i}^{0}(Z_{i}-Z_{i}^{0})}}{kT_{0}}\right]\equiv W_{m}\exp\left(-\dfrac{\mathcal{R}}{kT_{0}}\right),\label{eq:49-1}
\end{equation}
where we denote
\begin{equation}
\mathcal{R}\equiv E-E_{0}-{\displaystyle \sum_{i=1}^{m}P_{i}^{0}(Z_{i}-Z_{i}^{0}).}\label{eq:49-2}
\end{equation}
It can be shown that $\mathcal{R}$ has the meaning of the reversible
work required for the creation of the fluctuated state of the canonical
system at a fixed value of the total entropy of the system and the
reservoir \citep{Landau-Lifshitz-Stat-phys,Willard_Gibbs}. 

Equation (\ref{eq:49-1}) is the canonical distribution function of
the extensive parameters $Z_{1},...,Z_{m}$. For small fluctuations,
we can expand $E-E_{0}$ in these parameters and neglect all terms
beyond quadratic to obtain the multi-variable Gaussian distribution

\begin{equation}
W(Z_{1},...,Z_{m})=W_{m}\exp\left[-\frac{1}{2}\sum_{i,j=1}^{m}K_{ij}(Z_{i}-Z_{i}^{0})(Z_{j}-Z_{j}^{0})\right],\label{eq:58-1}
\end{equation}
where the matrix 
\begin{equation}
K_{ij}\equiv\frac{1}{kT_{0}}\left(\dfrac{\partial^{2}E}{\partial Z_{i}\partial Z_{j}}\right)_{0}=\frac{1}{kT_{0}}\left(\dfrac{\partial P_{i}}{\partial Z_{j}}\right)_{0}\label{eq:59-1}
\end{equation}
can be called \citep{Tisza:1961aa} the generalized stiffness matrix.
Its inverse 
\begin{equation}
K_{ij}^{-1}=kT_{0}\left(\dfrac{\partial Z_{i}}{\partial P_{j}}\right)_{0}=-kT_{0}\dfrac{\partial^{2}\Phi}{\partial P_{i}\partial P_{j}}\label{eq:59-2}
\end{equation}
can be then called \citep{Tisza:1961aa} the generalized compliance
matrix. 

Using the standard properties of the Gaussian distribution (Sec.~\ref{sec:Gaussian-law}),
we obtain at once the covariances and mean-square fluctuations of
all extensive and intensive properties of the canonical system in
the energy scheme: 
\begin{equation}
\overline{\Delta Z_{i}\Delta Z_{j}}=K_{ij}^{-1},\,\,\,\,\,i,j=1,...,m,\label{eq:60-1}
\end{equation}
\begin{equation}
\overline{(\Delta Z_{i})^{2}}=K_{ii}^{-1},\,\,\,\,\,i=1,...,m,\label{eq:61-1}
\end{equation}
\begin{equation}
\overline{\Delta Z_{i}\Delta P_{j}}=kT_{0}\delta_{ij},\,\,\,\,\,i,j=1,...,m,\label{eq:64-1}
\end{equation}
\begin{equation}
\overline{\Delta P_{i}\Delta P_{j}}=(kT_{0})^{2}K_{ij},\,\,\,\,\,i,j=1,...,m,\label{eq:65-1}
\end{equation}
\begin{equation}
\overline{(\Delta P_{i})^{2}}=(kT_{0})^{2}K_{ii},\,\,\,\,\,i=1,...,m.\label{eq:66-6}
\end{equation}

Having these fluctuation relations, it is straightforward to find
the covariance of any two thermodynamic properties $y=y(Z_{1},...,Z_{m})$
and $z=z(Z_{1},...,Z_{m})$. Indeed, for small fluctuations,
\begin{equation}
\Delta y=\sum_{i=1}^{m}\left(\dfrac{\partial y}{\partial Z_{i}}\right)_{0}\Delta Z_{i},\label{eq:66-7}
\end{equation}
\begin{equation}
\Delta z=\sum_{i=1}^{m}\left(\dfrac{\partial z}{\partial Z_{i}}\right)_{0}\Delta Z_{i},\label{eq:66-8}
\end{equation}
from which
\begin{equation}
\overline{\Delta y\Delta z}=\sum_{i,j=1}^{m}\left(\dfrac{\partial y}{\partial Z_{i}}\right)_{0}\left(\dfrac{\partial z}{\partial Z_{j}}\right)_{0}\overline{\Delta Z_{i}\Delta Z_{j}}=\sum_{i,j=1}^{m}\left(\dfrac{\partial y}{\partial Z_{i}}\right)_{0}\left(\dfrac{\partial z}{\partial Z_{j}}\right)_{0}K_{ij}^{-1}.\label{eq:66-9}
\end{equation}
Similarly, for any two functions $y=y(P_{1},...,P_{m})$ and $z=z(P_{1},...,P_{m})$,
\begin{equation}
\overline{\Delta y\Delta z}=\sum_{i,j=1}^{m}\left(\dfrac{\partial y}{\partial P_{i}}\right)_{0}\left(\dfrac{\partial z}{\partial P_{j}}\right)_{0}\overline{\Delta P_{i}\Delta P_{j}}=(kT_{0})^{2}\sum_{i,j=1}^{m}\left(\dfrac{\partial y}{\partial P_{i}}\right)_{0}\left(\dfrac{\partial z}{\partial P_{j}}\right)_{0}K_{ij}.\label{eq:66-9-1}
\end{equation}
The same calculation method can be applied when $y$ is a function
$Z_{1},...,Z_{m}$ while $z$ is a function of $P_{1},...,P_{m}$,
or when both are functions of mixed sets of $Z$'s and $P$'s.

\subsection{Canonical fluctuations in a simple fluid\label{sub:Canonical-fluid}}

To demonstrate the foregoing formalism for computing canonical fluctuations,
we will apply it to a simple fluid coupled to a reservoir. We will
consider three different ensembles, deriving equations for the mean-square
fluctuations and covariances of  all thermodynamic properties in each
case. The calculations will be conducted in the energy scheme, although
the entropy scheme could be applied just as well. We are working with
the extensive and intensive parameter sets $(Z_{1},Z_{2},Z_{3})=(S,V,N)$
and $(P_{1},P_{2},P_{3})=(T,-p,\mu)$, respectively. To simplify the
notations, we will suppress the index 0 but it is implied that all
derivatives are taken at the equilibrium state.

While a number of fluctuation relations for fluids can be found in
the literature, in this paper we present a \emph{complete} set of
such relations for each ensemble. For the $NVT$ ensemble, there are
5 fluctuating parameters $(E,S,T,p,\mu)$. Out of 25 covariances that
can be formed among them, only 15 are distinct due to the symmetry
of the covariance. Similarly, in the $NpT$ and $\mu VT$ ensembles,
there are 6 fluctuating parameters $(E,S,V,T,p,\mu)$ and $(E,S,N,T,p,\mu)$,
respectively.\footnote{Recall that all intensive properties of a canonical system are allowed
to fluctuate, including those which are imposed by the reservoir.
For example, in the $\mu VT$ ensemble, the reservoir imposes a certain
temperature $T_{0}$ and chemical potential $\mu_{0}$; however, the
actual temperature and chemical potential inside the system fluctuate
around the imposed values.} Taking into account the symmetry of covariances, 21 distinct fluctuation
relations exist for each of these ensembles. All these fluctuation
relations will be derived below in a systematic manner enabled by
the proposed formalism.

The equilibrium properties appearing in right-hand sides of the fluctuation
relations will be presented in simplest possible form. They can be
further transformed to many other forms expressing them through specific
heats, moduli or other experimentally accessible properties. We do
not pursue such transformations as they lie outside the fluctuation
topic and depend on the problem at hand.

\subsubsection{$NVT$ ensemble\label{sub:NVT-ensemble}}

Suppose the volume and number of particles in the system are fixed
while the energy and entropy can fluctuate (canonical ensemble). The
stiffness and compliance matrices reduce to the scalars 
\begin{equation}
K_{11}=\frac{1}{kT_{0}}\left(\dfrac{\partial T}{\partial S}\right)_{V,N}=\dfrac{1}{Nkc_{v}}\label{eq:67}
\end{equation}
and
\begin{equation}
K_{11}^{-1}=Nkc_{v}.\label{eq:67-1}
\end{equation}
By Eq.(\ref{eq:61-1}), 
\begin{equation}
\overline{(\Delta S)^{2}}=K_{11}^{-1}=Nkc_{v}.\label{eq:67-2}
\end{equation}
Applying Eq.(\ref{eq:66-9}), we find the energy fluctuation
\begin{equation}
\overline{(\Delta E)^{2}}=\left(\dfrac{\partial E}{\partial S}\right)_{V,N}^{2}K_{11}^{-1}=NkT_{0}^{2}c_{v}\label{eq:68}
\end{equation}
and the energy-entropy covariance
\begin{equation}
\overline{\Delta E\Delta S}=\left(\dfrac{\partial E}{\partial S}\right)_{V,N}K_{11}^{-1}=NkT_{0}c_{v},\label{eq:68-1}
\end{equation}
where we used the relation $(\partial E/\partial S)_{V,N}=T_{0}$.
For temperature fluctuations, we follow Eq.(\ref{eq:66-6}) to obtain
\begin{equation}
\overline{(\Delta T)^{2}}=(kT_{0})^{2}K_{11}=\dfrac{kT_{0}^{2}}{Nc_{v}}.\label{eq:69}
\end{equation}

We next derive fluctuations of the chemical potential and pressure.
Both are dependent variables that can be expressed as functions of
temperature and the number density of particles, $\rho\equiv N/V$.
The latter is fixed, leaving only $T$ as an independent argument.
Applying Eq.(\ref{eq:66-9-1}) we obtain
\begin{equation}
\overline{(\Delta\mu)^{2}}=(kT_{0})^{2}\left(\dfrac{\partial\mu}{\partial T}\right)_{\rho}^{2}K_{11}=\dfrac{kT_{0}^{2}}{Nc_{v}}\left(\dfrac{\partial\mu}{\partial T}\right)_{\rho}^{2},\label{eq:69-4}
\end{equation}
\begin{equation}
\overline{(\Delta p)^{2}}=(kT_{0})^{2}\left(\dfrac{\partial p}{\partial T}\right)_{\rho}^{2}K_{11}=\dfrac{kT_{0}^{2}}{Nc_{v}}\left(\dfrac{\partial p}{\partial T}\right)_{\rho}^{2},\label{eq:69-5}
\end{equation}
\begin{equation}
\overline{\Delta\mu\Delta p}=(kT_{0})^{2}\left(\dfrac{\partial\mu}{\partial T}\right)_{\rho}\left(\dfrac{\partial p}{\partial T}\right)_{\rho}K_{11}=\dfrac{kT_{0}^{2}}{Nc_{v}}\left(\dfrac{\partial\mu}{\partial T}\right)_{\rho}\left(\dfrac{\partial p}{\partial T}\right)_{\rho},\label{eq:69-6}
\end{equation}
\begin{equation}
\overline{\Delta\mu\Delta T}=(kT_{0})^{2}\left(\dfrac{\partial\mu}{\partial T}\right)_{\rho}K_{11}=\dfrac{kT_{0}^{2}}{Nc_{v}}\left(\dfrac{\partial\mu}{\partial T}\right)_{\rho},\label{eq:69-7}
\end{equation}
\begin{equation}
\overline{\Delta p\Delta T}=(kT_{0})^{2}\left(\dfrac{\partial p}{\partial T}\right)_{\rho}K_{11}=\dfrac{kT_{0}^{2}}{Nc_{v}}\left(\dfrac{\partial p}{\partial T}\right)_{\rho}.\label{eq:69-8}
\end{equation}

For ``cross-fluctuations'' between extensive and intensive properties,
we first apply Eq.(\ref{eq:64-1}) to obtain
\begin{equation}
\overline{\Delta S\Delta T}=kT_{0}.\label{eq:69-10}
\end{equation}
The remaining ``cross-fluctuations'' are easily computed by the method
outlined by Eqs.(\ref{eq:66-7})-(\ref{eq:66-9-1}):
\begin{equation}
\overline{\Delta E\Delta T}=\left(\dfrac{\partial E}{\partial S}\right)_{V,N}\overline{\Delta S\Delta T}=kT_{0}^{2},\label{eq:69-11}
\end{equation}
\begin{equation}
\overline{\Delta S\Delta p}=\left(\dfrac{\partial p}{\partial T}\right)_{\rho}\overline{\Delta S\Delta T}=kT_{0}\left(\dfrac{\partial p}{\partial T}\right)_{\rho},\label{eq:69-12}
\end{equation}
\begin{equation}
\overline{\Delta E\Delta p}=\left(\dfrac{\partial E}{\partial S}\right)_{V,N}\left(\dfrac{\partial p}{\partial T}\right)_{\rho}\overline{\Delta S\Delta T}=kT_{0}^{2}\left(\dfrac{\partial p}{\partial T}\right)_{\rho},\label{eq:69-13}
\end{equation}
\begin{equation}
\overline{\Delta S\Delta\mu}=\left(\dfrac{\partial\mu}{\partial T}\right)_{\rho}\overline{\Delta S\Delta T}=kT_{0}\left(\dfrac{\partial\mu}{\partial T}\right)_{\rho},\label{eq:69-14}
\end{equation}
\begin{equation}
\overline{\Delta E\Delta\mu}=\left(\dfrac{\partial E}{\partial S}\right)_{V,N}\left(\dfrac{\partial\mu}{\partial T}\right)_{\rho}\overline{\Delta S\Delta T}=kT_{0}^{2}\left(\dfrac{\partial\mu}{\partial T}\right)_{\rho}.\label{eq:69-15}
\end{equation}

\begin{table}
\noindent \begin{centering}
\begin{tabular}{lccl}
\hline 
 &  &  & \tabularnewline
$\overline{(\Delta E)^{2}}=NkT_{0}^{2}c_{v}$ &  &  & $\overline{\Delta\mu\Delta p}=\dfrac{kT_{0}^{2}}{Nc_{v}}\left(\dfrac{\partial\mu}{\partial T}\right)_{\rho}\left(\dfrac{\partial p}{\partial T}\right)_{\rho}$\tabularnewline
 &  &  & \tabularnewline
$\overline{(\Delta S)^{2}}=Nkc_{v}$ &  &  & $\overline{\Delta S\Delta T}=kT_{0}$\tabularnewline
 &  &  & \tabularnewline
$\overline{(\Delta E)(\Delta S)}=NkT_{0}c_{v}$ &  &  & $\overline{\Delta E\Delta T}=kT_{0}^{2}$\tabularnewline
 &  &  & \tabularnewline
$\overline{(\Delta T)^{2}}=\dfrac{kT_{0}^{2}}{Nc_{v}}$ &  &  & $\overline{\Delta S\Delta p}=kT_{0}\left(\dfrac{\partial p}{\partial T}\right)_{\rho}$\tabularnewline
 &  &  & \tabularnewline
$\overline{(\Delta\mu)^{2}}=\dfrac{kT_{0}^{2}}{Nc_{v}}\left(\dfrac{\partial\mu}{\partial T}\right)_{\rho}^{2}$ &  &  & $\overline{\Delta E\Delta p}=kT_{0}^{2}\left(\dfrac{\partial p}{\partial T}\right)_{\rho}$\tabularnewline
 &  &  & \tabularnewline
$\overline{(\Delta p)^{2}}=\dfrac{kT_{0}^{2}}{Nc_{v}}\left(\dfrac{\partial p}{\partial T}\right)_{\rho}^{2}$ &  &  & $\overline{\Delta S\Delta\mu}=kT_{0}\left(\dfrac{\partial\mu}{\partial T}\right)_{\rho}$\tabularnewline
 &  &  & \tabularnewline
$\overline{\Delta T\Delta p}=\dfrac{kT_{0}^{2}}{Nc_{v}}\left(\dfrac{\partial p}{\partial T}\right)_{\rho}$ &  &  & $\overline{\Delta E\Delta\mu}=kT_{0}^{2}\left(\dfrac{\partial\mu}{\partial T}\right)_{\rho}$\tabularnewline
 &  &  & \tabularnewline
$\overline{\Delta\mu\Delta T}=\dfrac{kT_{0}^{2}}{Nc_{v}}\left(\dfrac{\partial\mu}{\partial T}\right)_{\rho}$ &  &  & \tabularnewline
 &  &  & \tabularnewline
\hline 
\end{tabular}
\par\end{centering}

\protect\caption{Complete set of fluctuation relations for a simple fluid in the $NVT$
ensemble.\label{tab:Fluctuations-NVT}}
\end{table}

Table \ref{tab:Fluctuations-NVT} summarizes all fluctuation relation
derived for the $NVT$ ensemble.

\subsubsection{$NpT$ ensemble}

Now suppose that the volume of the system can also fluctuate while
$N=$const. The stiffness matrix is now $2\times2$ and its $K_{11}$
component is still given by Eq.(\ref{eq:67}). For the remaining components
we have
\begin{equation}
K_{12}=\frac{1}{kT_{0}}\left(\dfrac{\partial T}{\partial V}\right)_{S,N}=-\frac{1}{kT_{0}}\dfrac{\left(\dfrac{\partial S}{\partial V}\right)_{T,N}}{\left(\dfrac{\partial S}{\partial T}\right)_{V,N}}=-\dfrac{\left(\dfrac{\partial p}{\partial T}\right)_{\rho}}{Nkc_{v}},\label{eq:70}
\end{equation}
where we used the Maxwell relation
\begin{equation}
\left(\dfrac{\partial S}{\partial V}\right)_{T,N}=\left(\dfrac{\partial p}{\partial T}\right)_{V,N},\label{eq:89}
\end{equation}
and 
\begin{equation}
K_{22}=-\frac{1}{kT_{0}}\left(\dfrac{\partial p}{\partial V}\right)_{S,N}=\frac{1}{Nkc_{v}}\left(\dfrac{\partial p}{\partial T}\right)_{\rho}^{2}-\frac{1}{kT_{0}}\left(\dfrac{\partial p}{\partial V}\right)_{T,N},\label{eq:71}
\end{equation}
where we used the identity
\begin{equation}
\left(\dfrac{\partial p}{\partial V}\right)_{S,N}=\left(\dfrac{\partial p}{\partial V}\right)_{T,N}-\dfrac{T}{Nc_{v}}\left(\dfrac{\partial p}{\partial T}\right)_{V,N}^{2}.\label{eq:88}
\end{equation}
Note that the derivatives $(\partial p/\partial V)_{T,N}$ and $(\partial p/\partial V)_{S,N}$
are related to the isothermal and adiabatic moduli, respectively. 

Thus, the stiffness matrix of the fluid takes the form
\begin{equation}
K={\displaystyle \frac{1}{Nkc_{v}}}\begin{pmatrix}1 &  & -\left(\dfrac{\partial p}{\partial T}\right)_{\rho}\\
\\
-\left(\dfrac{\partial p}{\partial T}\right)_{\rho} &  & {\displaystyle \left(\dfrac{\partial p}{\partial T}\right)_{\rho}^{2}-\frac{Nc_{v}}{T_{0}}\left(\dfrac{\partial p}{\partial V}\right)_{T,N}}
\end{pmatrix}\label{eq:74}
\end{equation}
with the compliance matrix 
\begin{equation}
K^{-1}=-kT_{0}{\displaystyle \left(\dfrac{\partial V}{\partial p}\right)_{T,N}}\begin{pmatrix}{\displaystyle \left(\dfrac{\partial p}{\partial T}\right)_{\rho}^{2}-\frac{Nc_{v}}{T_{0}}\left(\dfrac{\partial p}{\partial V}\right)_{T,N}} &  & \left(\dfrac{\partial p}{\partial T}\right)_{\rho}\\
\\
\left(\dfrac{\partial p}{\partial T}\right)_{\rho} &  & {\displaystyle 1}
\end{pmatrix}.\label{eq:76}
\end{equation}

We are ready to calculate the fluctuations. To find the entropy and
volume fluctuations, we apply Eq.(\ref{eq:60-1}):
\begin{equation}
\overline{(\Delta S)^{2}}=K_{11}^{-1}=Nkc_{v}-kT_{0}{\displaystyle \left(\dfrac{\partial V}{\partial p}\right)_{T,N}}\left(\dfrac{\partial p}{\partial T}\right)_{\rho}^{2}\equiv Nkc_{p},\label{eq:77}
\end{equation}
\begin{equation}
\overline{(\Delta V)^{2}}=K_{22}^{-1}=-kT_{0}\left(\dfrac{\partial V}{\partial p}\right)_{T,N},\label{eq:78}
\end{equation}
\begin{equation}
\overline{\Delta S\Delta V}=K_{12}^{-1}=-kT_{0}\left(\dfrac{\partial V}{\partial p}\right)_{T,N}\left(\dfrac{\partial p}{\partial T}\right)_{\rho}=kT_{0}\left(\dfrac{\partial V}{\partial T}\right)_{p,N},\label{eq:79}
\end{equation}
where $c_{p}$ is the specific heat at constant pressure. Fluctuation
relations involving energy are calculated by Eq.(\ref{eq:66-9}):
\begin{eqnarray}
\overline{(\Delta E)^{2}} & = & \left(\dfrac{\partial E}{\partial S}\right)_{V,N}^{2}K_{11}^{-1}+\left(\dfrac{\partial E}{\partial V}\right)_{S,N}^{2}K_{22}^{-1}+2\left(\dfrac{\partial E}{\partial S}\right)_{V,N}\left(\dfrac{\partial E}{\partial V}\right)_{S,N}K_{12}^{-1}\nonumber \\
 & = & NkT_{0}^{2}c_{v}-kT_{0}^{3}\left(\dfrac{\partial V}{\partial p}\right)_{T,N}\left[\left(\dfrac{\partial p}{\partial T}\right)_{\rho}-\dfrac{p_{0}}{T_{0}}\right]^{2}.\label{eq:79-1}
\end{eqnarray}
\begin{eqnarray}
\overline{\Delta E\Delta S} & = & \left(\dfrac{\partial E}{\partial S}\right)_{V,N}K_{11}^{-1}+\left(\dfrac{\partial E}{\partial V}\right)_{S,N}K_{12}^{-1}\nonumber \\
 & = & NkT_{0}c_{v}-kT_{0}^{2}\left(\dfrac{\partial V}{\partial p}\right)_{T,N}\left(\dfrac{\partial p}{\partial T}\right)_{\rho}\left[\left(\dfrac{\partial p}{\partial T}\right)_{\rho}-\dfrac{p_{0}}{T_{0}}\right].\label{eq:79-2}
\end{eqnarray}
\begin{eqnarray}
\overline{\Delta E\Delta V} & = & \left(\dfrac{\partial E}{\partial S}\right)_{V,N}K_{12}^{-1}+\left(\dfrac{\partial E}{\partial V}\right)_{S,N}K_{22}^{-1}\nonumber \\
 & = & -kT_{0}^{2}\left(\dfrac{\partial V}{\partial p}\right)_{T,N}\left[\left(\dfrac{\partial p}{\partial T}\right)_{\rho}-\dfrac{p_{0}}{T_{0}}\right].\label{eq:79-3}
\end{eqnarray}

Turning to fluctuations of intensive parameters, we use Eqs.(\ref{eq:65-1})
and (\ref{eq:66-6}) to obtain
\begin{equation}
\overline{(\Delta T)^{2}}=(kT_{0})^{2}K_{11}=\dfrac{kT_{0}^{2}}{Nc_{v}},\label{eq:79-4}
\end{equation}
\begin{equation}
\overline{(\Delta p)^{2}}=(kT_{0})^{2}K_{22}=-kT_{0}\left[\left(\dfrac{\partial p}{\partial V}\right)_{T,N}-\dfrac{T_{0}}{Nc_{v}}\left(\dfrac{\partial p}{\partial T}\right)_{\rho}^{2}\right],\label{eq:79-5}
\end{equation}
\begin{equation}
\overline{\Delta T\Delta p}=-(kT_{0})^{2}K_{12}=\dfrac{kT_{0}^{2}}{Nc_{v}}\left(\dfrac{\partial p}{\partial T}\right)_{\rho}.\label{eq:80}
\end{equation}
To derive fluctuation relations involving the chemical potential,
we treat it as a function of $T$ and $p$. Using Eq.(\ref{eq:66-9-1}),
\begin{eqnarray}
\overline{(\Delta\mu)^{2}} & =(kT_{0})^{2} & \left[\left(\dfrac{\partial\mu}{\partial T}\right)_{p}^{2}K_{11}+\left(\dfrac{\partial\mu}{\partial p}\right)_{T}^{2}K_{22}-2\left(\dfrac{\partial\mu}{\partial T}\right)_{p}\left(\dfrac{\partial\mu}{\partial p}\right)_{T}K_{12}\right]\nonumber \\
 & = & (kT_{0})^{2}\left[s_{0}^{2}K_{11}+v_{0}^{2}K_{22}+2s_{0}v_{0}K_{12}\right]\nonumber \\
 & = & \dfrac{kT_{0}^{2}}{Nc_{v}}\left[v_{0}\left(\dfrac{\partial p}{\partial T}\right)_{\rho}-s_{0}\right]^{2}-kT_{0}v_{0}^{2}\left(\dfrac{\partial p}{\partial V}\right)_{T,N},\label{eq:80-1}
\end{eqnarray}
where
\begin{equation}
s_{0}=-\left(\dfrac{\partial\mu}{\partial T}\right)_{p}\label{eq:69-9}
\end{equation}
is the equilibrium entropy per particle and 
\begin{equation}
v_{0}=\left(\dfrac{\partial\mu}{\partial p}\right)_{p}\label{eq:81}
\end{equation}
is the equilibrium volume per particle ($v_{0}=1/\rho_{0}$). Similarly,
\begin{equation}
\overline{\Delta\mu\Delta T}=(kT_{0})^{2}\left[\left(\dfrac{\partial\mu}{\partial T}\right)_{p}K_{11}-\left(\dfrac{\partial\mu}{\partial p}\right)_{T}K_{12}\right]=\dfrac{kT_{0}^{2}}{Nc_{v}}\left[v_{0}\left(\dfrac{\partial p}{\partial T}\right)_{\rho}-s_{0}\right],\label{eq:82}
\end{equation}
\begin{eqnarray}
\overline{\Delta\mu\Delta p} & = & (kT_{0})^{2}\left[-\left(\dfrac{\partial\mu}{\partial T}\right)_{p}K_{12}+\left(\dfrac{\partial\mu}{\partial p}\right)_{T}K_{22}\right]\label{eq:82-2}\\
 & = & -kT_{0}v_{0}\left(\dfrac{\partial p}{\partial V}\right)_{T,N}+\dfrac{kT_{0}^{2}}{Nc_{v}}\left(\dfrac{\partial p}{\partial T}\right)_{\rho}\left[v_{0}\left(\dfrac{\partial p}{\partial T}\right)_{\rho}-s_{0}\right].\label{eq:83}
\end{eqnarray}

Equation (\ref{eq:64-1}) gives the ``cross-fluctuations''
\begin{equation}
\overline{\Delta S\Delta T}=-\overline{\Delta V\Delta p}=kT_{0},\label{eq:84}
\end{equation}
\begin{equation}
\overline{\Delta S\Delta p}=0,\label{eq:84-1}
\end{equation}
\begin{equation}
\overline{\Delta V\Delta T}=0.\label{eq:84-6}
\end{equation}
The remaining ``cross-fluctuations'' are computed in a manner similar
to Eqs.(\ref{eq:66-7})-(\ref{eq:66-9-1}):
\begin{equation}
\overline{\Delta S\Delta\mu}=\left(\dfrac{\partial\mu}{\partial T}\right)_{p}\overline{\Delta S\Delta T}=-kT_{0}s_{0},\label{eq:84-2}
\end{equation}
 
\begin{equation}
\overline{\Delta E\Delta T}=\left(\dfrac{\partial E}{\partial S}\right)_{V,N}\overline{\Delta S\Delta T}=kT_{0}^{2},\label{eq:84-3}
\end{equation}
\begin{equation}
\overline{\Delta E\Delta p}=\left(\dfrac{\partial E}{\partial V}\right)_{S,N}\overline{\Delta V\Delta p}=kT_{0}p_{0},\label{eq:84-4}
\end{equation}
\begin{equation}
\overline{\Delta E\Delta\mu}=\left(\dfrac{\partial E}{\partial S}\right)_{V,N}\left(\dfrac{\partial\mu}{\partial T}\right)_{p}\overline{\Delta S\Delta T}+\left(\dfrac{\partial E}{\partial V}\right)_{S,N}\left(\dfrac{\partial\mu}{\partial p}\right)_{T}\overline{\Delta V\Delta p}=kT_{0}\left(p_{0}v_{0}-T_{0}s_{0}\right),\label{eq:84-5}
\end{equation}
\begin{equation}
\overline{\Delta V\Delta\mu}=\left(\dfrac{\partial\mu}{\partial p}\right)_{T}\overline{\Delta V\Delta p}=-kT_{0}v_{0}.\label{eq:84-7}
\end{equation}

\begin{table}
\noindent \begin{centering}
\begin{tabular}{lccl}
\hline 
 &  &  & \tabularnewline
$\overline{(\Delta S)^{2}}=Nkc_{v}-kT_{0}{\displaystyle \left(\dfrac{\partial V}{\partial p}\right)_{T,N}}\left(\dfrac{\partial p}{\partial T}\right)_{\rho}^{2}$ &  &  & $\overline{\Delta\mu\Delta p}={\displaystyle -kT_{0}v_{0}\left(\dfrac{\partial p}{\partial V}\right)_{T,N}}$\tabularnewline
 &  &  & $+\dfrac{kT_{0}^{2}}{Nc_{v}}\left(\dfrac{\partial p}{\partial T}\right)_{\rho}\left[v_{0}\left(\dfrac{\partial p}{\partial T}\right)_{\rho}-s_{0}\right]$\tabularnewline
 &  &  & \tabularnewline
$\overline{\Delta S\Delta V}=kT_{0}\left(\dfrac{\partial V}{\partial T}\right)_{p,N}$ &  &  & $\overline{\Delta S\Delta T}=kT_{0}$\tabularnewline
 &  &  & \tabularnewline
$\overline{(\Delta V)^{2}}=-kT_{0}\left(\dfrac{\partial V}{\partial p}\right)_{T,N}$ &  &  & $\overline{\Delta V\Delta p}=-kT_{0}$\tabularnewline
 &  &  & \tabularnewline
$\overline{(\Delta E)^{2}}=NkT_{0}^{2}c_{v}$ &  &  & $\overline{\Delta S\Delta p}=0$\tabularnewline
$-kT_{0}^{3}\left(\dfrac{\partial V}{\partial p}\right)_{T,N}\left[\left(\dfrac{\partial p}{\partial T}\right)_{\rho}-\dfrac{p_{0}}{T_{0}}\right]^{2}$ &  &  & \tabularnewline
 &  &  & \tabularnewline
\multirow{1}{*}{$\overline{\Delta E\Delta S}=NkT_{0}c_{v}$} &  &  & $\overline{\Delta V\Delta T}=0$\tabularnewline
$-kT_{0}^{2}\left(\dfrac{\partial V}{\partial p}\right)_{T,N}\left(\dfrac{\partial p}{\partial T}\right)_{\rho}\left[\left(\dfrac{\partial p}{\partial T}\right)_{\rho}-\dfrac{p_{0}}{T_{0}}\right]$ &  &  & \tabularnewline
 &  &  & \tabularnewline
$\overline{\Delta E\Delta V}=-kT_{0}^{2}\left(\dfrac{\partial V}{\partial p}\right)_{T,N}\left[\left(\dfrac{\partial p}{\partial T}\right)_{\rho}-\dfrac{p_{0}}{T_{0}}\right]$ &  &  & $\overline{\Delta S\Delta\mu}=-kT_{0}s_{0}$\tabularnewline
 &  &  & \tabularnewline
$\overline{(\Delta T)^{2}}=\dfrac{kT_{0}^{2}}{Nc_{v}}$ &  &  & $\overline{\Delta E\Delta T}=kT_{0}^{2}$\tabularnewline
 &  &  & \tabularnewline
$\overline{(\Delta p)^{2}}=-kT_{0}\left[\left(\dfrac{\partial p}{\partial V}\right)_{T,N}-\dfrac{T_{0}}{Nc_{v}}\left(\dfrac{\partial p}{\partial T}\right)_{\rho}^{2}\right]$ &  &  & $\overline{\Delta E\Delta p}=kT_{0}p_{0}$\tabularnewline
 &  &  & \tabularnewline
$\overline{\Delta T\Delta p}=\dfrac{kT_{0}^{2}}{Nc_{v}}\left(\dfrac{\partial p}{\partial T}\right)_{\rho}$ &  &  & $\overline{\Delta E\Delta\mu}=kT_{0}\left(p_{0}v_{0}-T_{0}s_{0}\right)$\tabularnewline
 &  &  & \tabularnewline
$\overline{(\Delta\mu)^{2}}={\displaystyle \dfrac{kT_{0}^{2}}{Nc_{v}}\left[v_{0}\left(\dfrac{\partial p}{\partial T}\right)_{\rho}-s_{0}\right]^{2}-kT_{0}v_{0}^{2}\left(\dfrac{\partial p}{\partial V}\right)_{T,N}}$ &  &  & $\overline{\Delta V\Delta\mu}=-kT_{0}v_{0}$\tabularnewline
 &  &  & \tabularnewline
$\overline{\Delta\mu\Delta T}=\dfrac{kT_{0}^{2}}{Nc_{v}}\left[v_{0}\left(\dfrac{\partial p}{\partial T}\right)_{\rho}-s_{0}\right]$ &  &  & \tabularnewline
 &  &  & \tabularnewline
\hline 
\end{tabular}
\par\end{centering}

\protect\caption{Complete set of fluctuation relations for a simple fluid in the $NpT$
ensemble.\label{tab:Fluctuations-NpT}}

\end{table}

The fluctuation relations for the $NpT$ ensemble are summarized in
Table \ref{tab:Fluctuations-NpT}. 

\begin{table}
\noindent \begin{centering}
\begin{tabular}{lccl}
\hline 
 &  &  & \tabularnewline
$\overline{(\Delta S)^{2}}=\rho_{0}kc_{v}V+kT_{0}{\displaystyle \left(\dfrac{\partial\mu}{\partial\rho}\right)_{T}^{-1}}\left(\dfrac{\partial\mu}{\partial T}\right)_{\rho}^{2}V$ &  &  & $\overline{\Delta\mu\Delta p}={\displaystyle \dfrac{kT_{0}}{\rho_{0}v_{0}V}\left(\dfrac{\partial\mu}{\partial\rho}\right)_{T}}$\tabularnewline
 &  &  & $+{\displaystyle \frac{kT_{0}^{2}}{\rho_{0}c_{v}v_{0}V}}\left(\dfrac{\partial\mu}{\partial T}\right)_{\rho}\left[s_{0}+\left(\dfrac{\partial\mu}{\partial T}\right)_{\rho}\right]$\tabularnewline
 &  &  & \tabularnewline
$\overline{\Delta S\Delta N}=-kT_{0}\left(\dfrac{\partial\mu}{\partial\rho}\right)_{T}^{-1}\left(\dfrac{\partial\mu}{\partial T}\right)_{\rho}V$ &  &  & $\overline{\Delta S\Delta T}=kT_{0}$\tabularnewline
 &  &  & \tabularnewline
$\overline{(\Delta N)^{2}}=kT_{0}{\displaystyle \left(\dfrac{\partial\mu}{\partial\rho}\right)_{T}^{-1}}V$ &  &  & $\overline{\Delta N\Delta\mu}=kT_{0}$\tabularnewline
 &  &  & \tabularnewline
$\overline{(\Delta E)^{2}}=\rho_{0}kT_{0}^{2}c_{v}V$ &  &  & $\overline{\Delta S\Delta\mu}=0$\tabularnewline
$+kT_{0}^{3}\left(\dfrac{\partial\mu}{\partial\rho}\right)_{T}^{-1}\left[\left(\dfrac{\partial\mu}{\partial T}\right)_{\rho}-\dfrac{\mu_{0}}{T_{0}}\right]^{2}V$ &  &  & \tabularnewline
 &  &  & \tabularnewline
\multirow{1}{*}{$\overline{\Delta E\Delta S}=\rho_{0}kT_{0}c_{v}V$} &  &  & $\overline{\Delta N\Delta T}=0$\tabularnewline
$-kT_{0}^{2}\left(\dfrac{\partial\mu}{\partial\rho}\right)_{T}^{-1}\left(\dfrac{\partial\mu}{\partial T}\right)_{\rho}\left[\left(\dfrac{\partial\mu}{\partial T}\right)_{\rho}-\dfrac{\mu_{0}}{T_{0}}\right]V$ &  &  & \tabularnewline
 &  &  & \tabularnewline
$\overline{\Delta E\Delta N}=-kT_{0}^{2}\left(\dfrac{\partial\mu}{\partial\rho}\right)_{T}^{-1}\left[\left(\dfrac{\partial\mu}{\partial T}\right)_{\rho}-\dfrac{\mu_{0}}{T_{0}}\right]V$ &  &  & $\overline{\Delta S\Delta p}=kT_{0}\dfrac{s_{0}}{v_{0}}$\tabularnewline
 &  &  & \tabularnewline
$\overline{(\Delta T)^{2}}=\dfrac{kT_{0}^{2}}{\rho_{0}c_{v}V}$ &  &  & $\overline{\Delta E\Delta T}=kT_{0}^{2}$\tabularnewline
 &  &  & \tabularnewline
$\overline{(\Delta\mu)^{2}}=\dfrac{kT_{0}}{\rho_{0}V}\left[\rho_{0}\left(\dfrac{\partial\mu}{\partial\rho}\right)_{T}+\dfrac{T_{0}}{c_{v}}\left(\dfrac{\partial\mu}{\partial T}\right)_{\rho}^{2}\right]$ &  &  & $\overline{\Delta E\Delta\mu}=kT_{0}\mu_{0}$\tabularnewline
 &  &  & \tabularnewline
$\overline{\Delta T\Delta\mu}=\dfrac{kT_{0}^{2}}{\rho_{0}c_{v}V}\left(\dfrac{\partial\mu}{\partial T}\right)_{\rho}$ &  &  & $\overline{\Delta E\Delta p}=\dfrac{kT_{0}}{v_{0}}\left(\mu_{0}+T_{0}s_{0}\right)$\tabularnewline
 &  &  & \tabularnewline
$\overline{(\Delta p)^{2}}={\displaystyle \dfrac{kT_{0}}{v_{0}^{2}V}\left(\dfrac{\partial\mu}{\partial\rho}\right)_{T}+\frac{kT_{0}^{2}}{\rho_{0}c_{v}v_{0}^{2}V}\left[s_{0}+\left(\dfrac{\partial\mu}{\partial T}\right)_{\rho}\right]^{2}}$ &  &  & $\overline{\Delta N\Delta p}=\dfrac{kT_{0}}{v_{0}}$\tabularnewline
 &  &  & \tabularnewline
$\overline{\Delta p\Delta T}={\displaystyle \frac{kT_{0}^{2}}{\rho_{0}c_{v}v_{0}V}}\left[s_{0}+\left(\dfrac{\partial\mu}{\partial T}\right)_{\rho}\right]$ &  &  & \tabularnewline
 &  &  & \tabularnewline
\hline 
\end{tabular}
\par\end{centering}

\protect\caption{Complete set of fluctuation relations for a simple fluid in the $\mu VT$
ensemble.\label{tab:Fluctuations-muVT}}
\end{table}

\subsubsection{$\mu VT$ ensemble\label{sub:grand-canonic-ensemble}}

We now turn to the grand-canonical ensemble, in which the volume of
the system is fixed while the energy and the number of particles can
fluctuate. The equilibrium temperature $T_{0}$ and the chemical potential
$\mu_{0}$ are fixed by the reservoir. The independent fluctuating
variables are $S$ and $N$, with the conjugate thermodynamic forces
$P_{1}=T$, and $P_{2}=\mu$. Energy is a dependent extensive parameter.
The chemical potential and pressure are also dependent parameters
treated as functions of $T$ and $\rho$.

All fluctuation relations can be obtained from those for the $NpT$
ensemble by simply swapping the variables $V\rightarrow N$, $p\rightarrow-\mu$.
In particular, the derivative $(\partial p/\partial T)_{\rho}$ is
replaced by $-(\partial\mu/\partial T)_{\rho}$ and $(\partial p/\partial V)_{T,N}$
by $-V^{-1}(\partial\mu/\partial\rho)_{T}$. Furthermore, it easy
to verify that $s_{0}$ is replaced by $s_{0}/v_{0}$ and $v_{0}$
by $1/v_{0}$. The stiffness matrix becomes
\begin{equation}
K={\displaystyle \frac{1}{N_{0}kc_{v}}}\begin{pmatrix}1 &  & \left(\dfrac{\partial\mu}{\partial T}\right)_{\rho}\\
\\
\left(\dfrac{\partial\mu}{\partial T}\right)_{\rho} &  & {\displaystyle \left(\dfrac{\partial\mu}{\partial T}\right)_{\rho}^{2}+\frac{\rho_{0}c_{v}}{T_{0}}\left(\dfrac{\partial\mu}{\partial\rho}\right)_{T}}
\end{pmatrix}\label{eq:74-1}
\end{equation}
 with the inverse
\begin{equation}
K^{-1}=kT_{0}{\displaystyle V\left(\dfrac{\partial\mu}{\partial\rho}\right)_{T}^{-1}}\begin{pmatrix}{\displaystyle \left(\dfrac{\partial\mu}{\partial T}\right)_{\rho}^{2}+\frac{\rho_{0}c_{v}}{T_{0}}\left(\dfrac{\partial\mu}{\partial\rho}\right)_{T}} &  & -\left(\dfrac{\partial\mu}{\partial T}\right)_{\rho}\\
\\
-\left(\dfrac{\partial\mu}{\partial T}\right)_{\rho} &  & {\displaystyle 1}
\end{pmatrix}.\label{eq:76-1}
\end{equation}
The results of the calculations are summarized in Table \ref{tab:Fluctuations-muVT}. 

The following scaling properties of fluctuations are evident from
Tables \ref{tab:Fluctuations-NVT}, \ref{tab:Fluctuations-NpT} and
\ref{tab:Fluctuations-muVT}:
\begin{itemize}
\item Covariances of extensive properties scale in proportion to the system
size ($N$ or $V$). 
\item Covariances of intensive properties scale as inverse of the system
size ($N^{-1}$ or $V^{-1}$).
\item ``Cross-fluctuations'' between extensive and intensive properties
are independent of the system size.
\end{itemize}
The first of these properties explains why we could neglect the quadratic
terms in the expansion for the reservoir entropy, justifying the reservoir
approximation (\ref{eq:48}) discussed in Sec.~\ref{sub:General-relations-canonical}.

\section{Finite-reservoir ensembles\label{sec:Finite-reservoir}}

\subsection{General considerations}

We now revisit the system coupled to a reservoir discussed in Sec.~\ref{sub:General-relations-canonical}
(Fig.~\ref{fig:canonical}). The system and the reservoir are described
by extensive parameters $X_{1},...,X_{n}$ and $X_{1}^{r},...,X_{n}^{r}$,
respectively, with fixed total amounts $\tilde{X}_{i}=X_{i}+X_{i}^{r}$.
Suppose $m$ of these parameters are allowed to flow back and forth
between the system and the reservoir, whereas the remaining $(n-m)$
parameters are fixed. Both the system and the reservoir are assumed
to be single-phase substances with different fundamental equations
$S=S(X_{1},...,X_{n})$ and $S_{r}=S_{r}(X_{1}^{r},...,X_{n}^{r})$,
respectively. The probability distribution of the fluctuating parameters
of the system is given by Eq.(\ref{eq:46}), which is repeated here
for convenience:
\begin{eqnarray}
W(X_{1},...,X_{m}) & = & W_{m}\exp\left[\dfrac{S(X_{1},...,X_{m})-S(X_{1}^{0},...,X_{m}^{0})}{k}\right]\nonumber \\
 & \times & \exp\left[\dfrac{S_{r}(\tilde{X}_{1}-X_{1},...,\tilde{X}_{m}-X_{m})-S_{r}(\tilde{X}_{1}-X_{1}^{0},...,\tilde{X}_{m}-X_{m}^{0})}{k}\right].\label{eq:100}
\end{eqnarray}

In Sec.~\ref{sub:General-relations-canonical}, the entropy of the
reservoir was approximated by the linear expansion (\ref{eq:48})
at the equilibrium parameter set $X_{1}^{0},...,X_{m}^{0}$. This
linear approximation was justified by the large size of the reservoir
in comparison with the system. We will now lift this assumption and
adopt a more general expansion, Eq.(\ref{eq:47}), which includes
quadratic terms: 
\begin{eqnarray}
S_{r}(\tilde{X}_{1}-X_{1},...,\tilde{X}_{m}-X_{m})-S_{r}(\tilde{X}_{1}-X_{1}^{0},...,\tilde{X}_{m}-X_{m}^{0}) & = & -\sum_{i=1}^{m}F_{i}^{0}(X_{i}-X_{i}^{0})\nonumber \\
-\frac{1}{2}k\sum_{i,j=1}^{m}\Lambda_{ij}^{r}(X_{i}-X_{i}^{0})(X_{j}-X_{j}^{0}),\label{eq:101}
\end{eqnarray}
where 
\begin{equation}
\Lambda_{ij}^{r}\equiv-\frac{1}{k}\left(\dfrac{\partial^{2}S_{r}}{\partial X_{i}^{r}\partial X_{j}^{r}}\right)_{0}\label{eq:102}
\end{equation}
is the stability matrix of the reservoir. As usual, the derivatives
are taken at the equilibrium parameter set $X_{1}^{0},...,X_{m}^{0}$. 

Before analyzing fluctuations, let us examine the micro-state probability
distribution $P(X_{1},...,X_{m})$. Recall that the latter is obtained
from $W(X_{1},...,X_{m})$ by dropping $S(X_{1},...,X_{m})$ in the
first line of Eq.(\ref{eq:100}) {[}see Eq.(\ref{eq:50}) in Sec.~\ref{sub:Generalized-canonical-ensembles}{]}.
Thus, 
\begin{eqnarray}
P & = & W_{m}\exp\left[\dfrac{-S(X_{1}^{0},...,X_{m}^{0})+S_{r}(\tilde{X}_{1}-X_{1},...,\tilde{X}_{m}-X_{m})-S_{r}(\tilde{X}_{1}-X_{1}^{0},...,\tilde{X}_{m}-X_{m}^{0})}{k}\right]\nonumber \\
 & = & A\exp\left[-\frac{1}{k}\sum_{i=1}^{m}F_{i}^{0}X_{i}-\frac{1}{2}\sum_{i,j=1}^{m}\Lambda_{ij}^{r}(X_{i}-X_{i}^{0})(X_{j}-X_{j}^{0})\right],\label{eq:103}
\end{eqnarray}
where $A$ is a constant that can be found from the probability normalization
condition. Thus, in addition to the usual linear terms $F_{i}^{0}X_{i}$
appearing in the generalized canonical distribution (\ref{eq:52}),
we now have a quadratic form representing the effect of the finite
reservoir on the micro-state probability of the system. 

The matrix elements $\Lambda_{ij}^{r}$ scale as $1/N_{r}$. When
the size of the reservoir increases to infinity, the quadratic terms
in Eq.(\ref{eq:103}) vanish and we return to the generalized canonical
distribution (\ref{eq:52}). In the opposite limit when the reservoir
becomes extremely small in comparison with our system, the probability
distribution is dominated by the quadratic terms. The distribution
becomes a very sharp Gaussian peak. Accordingly, all micro-states
whose $X_{i}$ are even slightly deviated from $X_{i}^{0}$ become
extremely improbable. The system behaves as if isolated and Eq.(\ref{eq:103})
approaches the micro-canonical distribution. In other words, Eq.(\ref{eq:103})
interpolates between the generalized canonical and micro-canonical
distributions; we can smoothly transition from one distribution to
the other by varying the reservoir size and thus scaling the reservoir
stability matrix (\ref{eq:102}).

We now return to the parameter distribution function $W(X_{1},...,X_{m})$
and expand the entropy of our system in a manner similar to Eq.(\ref{eq:101}):
\begin{equation}
S(X_{1},...,X_{m})-S(X_{1}^{0},...,X_{m}^{0})=\sum_{i=1}^{m}F_{i}^{0}(X_{i}-X_{i}^{0})-\frac{1}{2}k\sum_{i,j=1}^{m}\Lambda_{ij}(X_{i}-X_{i}^{0})(X_{j}-X_{j}^{0}),\label{eq:104}
\end{equation}
with the usual stability matrix 
\begin{equation}
\Lambda_{ij}\equiv-\frac{1}{k}\left(\dfrac{\partial^{2}S}{\partial X_{i}\partial X_{j}}\right)_{0}.\label{eq:105}
\end{equation}
Inserting the expansions (\ref{eq:101}) and (\ref{eq:104}) in Eq.(\ref{eq:100}),
we obtain the multi-variable Gaussian distribution
\begin{equation}
W(X_{1},...,X_{m})=W_{m}\exp\left[-\frac{1}{2}\sum_{i,j=1}^{m}\tilde{\Lambda}_{ij}(X_{i}-X_{i}^{0})(X_{j}-X_{j}^{0})\right],\label{eq:106}
\end{equation}
with the stability matrix
\begin{equation}
\tilde{\Lambda}_{ij}=\Lambda_{ij}+\Lambda_{ij}^{r}.\label{eq:107}
\end{equation}

We thus arrive at the remarkable result that the finite size of the
reservoir does not change the Gaussian form of the parameter distribution
of the system. However, the reservoir does contribute to the matrix
elements of the Gaussian distribution in an additive manner. We can,
therefore, easily compute the mean-square fluctuations and covariances
of the parameters $X_{1},...,X_{m}$ using the known Eqs.(\ref{eq:60})
and (\ref{eq:61}) with $\Lambda_{ij}$ replaced by the full stability
matrix $\tilde{\Lambda}_{ij}$:
\begin{equation}
\overline{\Delta X_{i}\Delta X_{j}}=\tilde{\Lambda}_{ij}^{-1},\,\,\,\,\,i,j=1,...,m,\label{eq:108}
\end{equation}
\begin{equation}
\overline{(\Delta X_{i})^{2}}=\tilde{\Lambda}_{ii}^{-1},\,\,\,\,\,i=1,...,m.\label{eq:109}
\end{equation}

A number of other interesting relations can be derived. It follows
from Eqs.(\ref{eq:104}) and (\ref{eq:101}) that
\begin{equation}
F_{i}=\dfrac{\partial S}{\partial X_{i}}=F_{i}^{0}-k\sum_{j=1}^{m}\Lambda_{ij}(X_{j}-X_{j}^{0}),\label{eq:110}
\end{equation}
\begin{equation}
F_{i}^{r}=\dfrac{\partial S_{r}}{\partial X_{i}^{r}}=F_{i}^{0}+k\sum_{j=1}^{m}\Lambda_{ij}^{r}(X_{j}-X_{j}^{0}).\label{eq:111}
\end{equation}
Averaging these equations over the distribution and using the fact
that for a Gaussian distribution $\overline{X_{j}}=X_{j}^{0}$, we
obtain
\begin{equation}
\overline{F_{i}^{r}}=\overline{F_{i}}=F_{i}^{0}.\label{eq:112}
\end{equation}
In other words, not only the most probable but also average values
of the thermodynamic forces are the same in the system and in the
reservoir. Furthermore, subtracting Eq.(\ref{eq:110}) from Eq.(\ref{eq:111})
we have
\begin{equation}
F_{i}^{r}-F_{i}=k\sum_{l=1}^{m}\tilde{\Lambda}_{il}(X_{l}-X_{l}^{0}).\label{eq:113}
\end{equation}
Starting from this equation, we can generate a set of fluctuation
relations involving the difference $(F_{i}^{r}-F_{i})$. For example,
multiplying Eq.(\ref{eq:113}) by $\Delta X_{j}$ and averaging over
the parameter distribution, we have
\begin{equation}
\overline{(F_{i}^{r}-F_{i})\Delta X_{j}}=k\delta_{ij},\,\,\,\,\,i,j=1,...,m,\label{eq:114}
\end{equation}
where we used Eq.(\ref{eq:108}). Likewise, multiplying Eq.(\ref{eq:113})
by $(F_{j}^{r}-F_{j})$ and averaging over the distribution, we obtain
\begin{equation}
\overline{(F_{i}^{r}-F_{i})(F_{j}^{r}-F_{j})}=k^{2}\tilde{\Lambda}_{ij},\,\,\,\,\,i,j=1,...,m,\label{eq:115}
\end{equation}
where we used Eq.(\ref{eq:114}). In particular,
\begin{equation}
\overline{(F_{i}^{r}-F_{i})^{2}}=k^{2}\tilde{\Lambda}_{ii},\,\,\,\,\,i=1,...,m.\label{eq:116}
\end{equation}
These equations describe fluctuations of the differences between the
values of the intensive parameters inside our system and in the reservoir.

We can also compute fluctuations inside our system. To this end, we
rewrite Eq.(\ref{eq:110}) as 
\begin{equation}
\Delta F_{i}=-k\sum_{p=1}^{m}\Lambda_{ip}\Delta X_{p}.\label{eq:117}
\end{equation}
Multiplying this equation by $\Delta X_{j}$ and averaging over the
distribution, we obtain the covariances
\begin{equation}
\overline{\Delta F_{i}\Delta X_{j}}=-k\sum_{p=1}^{m}\Lambda_{ip}\tilde{\Lambda}_{pj}^{-1},\,\,\,\,\,i,j=1,...,m.\label{eq:118}
\end{equation}
Similarly, multiplying Eq.(\ref{eq:117}) by $\Delta F_{j}$ and averaging
over the distribution, we have 
\begin{equation}
\overline{\Delta F_{i}\Delta F_{j}}=k^{2}\sum_{p,q=1}^{m}\Lambda_{ip}\Lambda_{pq}\tilde{\Lambda}_{qj}^{-1},\,\,\,\,\,i,j=1,...,m,\label{eq:119}
\end{equation}
\begin{equation}
\overline{(\Delta F_{i})^{2}}=k^{2}\sum_{p,q=1}^{m}\Lambda_{ip}\Lambda_{pq}\tilde{\Lambda}_{qi}^{-1},\,\,\,\,\,i=1,...,m.\label{eq:120}
\end{equation}
As expected, in the limit of an infinitely large reservoir when $\tilde{\Lambda}_{ij}$
becomes identical to $\Lambda_{ij}$, Eqs.(\ref{eq:118}), (\ref{eq:119})
and (\ref{eq:120}) reduce to Eqs.(\ref{eq:64}) (\ref{eq:65}) and
(\ref{eq:66}) for the generalized canonical ensemble.

\subsection{Application to a simple fluid}

To demonstrate the finite-ensemble formalism, consider a fixed volume
$V$ of a simple fluid embedded in another (generally, different)
simple fluid serving as a reservoir. This situation was already discussed
in Sec.~\ref{sub:grand-canonic-ensemble} in the context of the grand-canonical
ensemble. This time, however, the reservoir will be treated as an
infinite source of heat (thermostat) but a \emph{finite} source of
particles. We thus have two fluctuating extensive properties, $X_{1}=E$
and $X_{2}=N$, with the conjugate thermodynamic forces $F_{1}=1/T$
and $F_{2}=-\mu/T$.

Because the reservoir has an infinite heat capacity, we have $\Lambda_{11}^{r}=0$.
Assuming for simplicity that the chemical potential of the reservoir
$\mu_{r}$ is temperature-independent, we additionally have $\Lambda_{12}^{r}=\Lambda_{21}^{r}=0$.
Thus, the reservoir stability matrix is
\begin{equation}
\Lambda_{ij}^{r}=\begin{pmatrix}0 & 0\\
0 & \Lambda_{22}^{r}
\end{pmatrix}\label{eq:121}
\end{equation}
with a single nonzero parameter $\Lambda_{22}^{r}$.

The micro-state probability distribution for our system can be found
from Eq.(\ref{eq:103}):
\begin{equation}
P=A\exp\left[-\frac{E-\mu_{0}N}{kT_{0}}-\frac{1}{2}\Lambda_{22}^{r}(N-N_{0})^{2}\right].\label{eq:130}
\end{equation}
As a side note, a similar equation can be obtained for a closed system
coupled to a thermostat with a finite heat capacity. In this case,
the role of the number of particles is played by the energy, so that
the additional term contributed by the thermostat is quadratic in
$(E-E_{0})$:
\begin{equation}
P=A\exp\left[-\frac{E}{kT_{0}}-\frac{1}{2}\Lambda_{11}^{r}(E-E_{0})^{2}\right].\label{eq:131}
\end{equation}
This situation is known in the literature as the Gaussian ensemble
\citep{Challa88,Johal2003}. By analogy, formula (\ref{eq:130}) can
also be referred to as the Gaussian ensemble distribution.

Equation (\ref{eq:130}) can be rewritten the form 
\begin{equation}
P=A\exp\left(-\frac{E-\hat{\mu}N}{kT_{0}}\right),\label{eq:132}
\end{equation}
where
\begin{equation}
\hat{\mu}\equiv a-bN\label{eq:133}
\end{equation}
with 
\begin{equation}
a=\mu_{0}+\Lambda_{22}^{r}kT_{0}N_{0},\label{eq:134}
\end{equation}
\begin{equation}
b=\frac{1}{2}\Lambda_{22}^{r}kT_{0}.\label{eq:135}
\end{equation}
Equation (\ref{eq:132}) looks similar to the grand-canonical distribution,
except that the fixed chemical potential of the reservoir $\mu_{0}$
is replaced an ``effective'' chemical potential $\hat{\mu}$ linearly
dependent on the current number of particles $N$. In the limit of
$\Lambda_{22}^{r}\rightarrow0$ (infinitely large reservoir), this
dependence disappears and we return to the standard grand-canonical
distribution with the chemical potential $\mu_{0}$.

Equations (\ref{eq:132})-(\ref{eq:135}) can be implemented in Monte
Carlo computer simulations using an algorithm similar to that of the
grand-canonical ensemble \citep{FrenkelS02} but with a chemical potential
adjustable ``on the fly''. For binary mixtures, this approach was
implemented in the semi-grand canonical mode as ``variance-constrained''
Monte Carlo \citep{Sadigh2012,Sadigh2012a}. A similar semi-grand
canonical Monte Carlo method, called feedback Monte Carlo, was recently
employed for interface free energy calculations by the capillary fluctuation
approach \citep{Mishin2014}. Related Monte Carlo algorithms were
earlier developed for closed systems coupled to a finite-capacity
thermostat. The initially proposed ``dynamical ensemble'' method \citep{Huller1993b}
developed for such systems later evolved into more sophisticated and
computationally more efficient iterative Monte Carlo schemes based
on Eq.(\ref{eq:131}) \citep{Velazquez2010a,Velazquez2010b}. 

We now return to the analysis of fluctuations. To simplify the calculations,
we will neglect the temperature dependence of the chemical potential,
treating it solely as a function of the particle density $\rho$.
Following the calculation methods presented in Sec.~\ref{sub:Canonical-fluid},
it is easy to obtain the stability matrix of the fluid in the entropy
scheme,

\begin{equation}
\Lambda_{ij}={\displaystyle \frac{1}{\rho_{0}c_{v}kT_{0}^{2}V}}\begin{pmatrix}1 &  & -\mu_{0}\\
\\
-\mu_{0} &  & \rho_{0}c_{v}T_{0}\left(\dfrac{\partial\mu}{\partial\rho}\right)_{T}+\mu_{0}^{2}
\end{pmatrix}\label{eq:122}
\end{equation}
and its inverse
\begin{equation}
\Lambda_{ij}^{-1}=kT_{0}V\left(\dfrac{\partial\mu}{\partial\rho}\right)_{T}^{-1}\begin{pmatrix}\rho_{0}c_{v}T_{0}\left(\dfrac{\partial\mu}{\partial\rho}\right)_{T}+\mu_{0}^{2} &  & \mu_{0}\\
\\
\mu_{0} &  & 1
\end{pmatrix}.\label{eq:123}
\end{equation}
Accordingly,
\begin{equation}
\tilde{\Lambda}_{ij}={\displaystyle \frac{1}{\rho_{0}c_{v}kT_{0}^{2}V}}\begin{pmatrix}1 &  & -\mu_{0}\\
\\
-\mu_{0} &  & \rho_{0}c_{v}T_{0}\left(\dfrac{\partial\mu}{\partial\rho}\right)_{T}+\mu_{0}^{2}+\rho_{0}c_{v}kT_{0}^{2}\Lambda_{22}^{r}V
\end{pmatrix}\label{eq:124}
\end{equation}
and
\begin{equation}
\tilde{\Lambda}_{ij}^{-1}=\dfrac{kT_{0}V}{\left(\dfrac{\partial\mu}{\partial\rho}\right)_{T}+kT_{0}V\Lambda_{22}^{r}}\begin{pmatrix}\rho_{0}c_{v}T_{0}\left(\dfrac{\partial\mu}{\partial\rho}\right)_{T}+\mu_{0}^{2}+\rho_{0}c_{v}kT_{0}^{2}\Lambda_{22}^{r}V &  & \mu_{0}\\
\\
\mu_{0} &  & 1
\end{pmatrix}.\label{eq:125}
\end{equation}

We can now compute the mean-square fluctuations of the energy and
the number of particles: 
\begin{equation}
\overline{(\Delta E)^{2}}=\tilde{\Lambda}_{11}^{-1}=\dfrac{\rho_{0}c_{v}kT_{0}^{2}V\left(\dfrac{\partial\mu}{\partial\rho}\right)_{T}+kT_{0}V\left(\mu_{0}^{2}+\rho_{0}c_{v}kT_{0}^{2}\Lambda_{22}^{r}V\right)}{\left(\dfrac{\partial\mu}{\partial\rho}\right)_{T}+kT_{0}V\Lambda_{22}^{r}},\label{eq:126}
\end{equation}
\begin{equation}
\overline{(\Delta N)^{2}}=\tilde{\Lambda}_{22}^{-1}=\dfrac{kT_{0}V}{\left(\dfrac{\partial\mu}{\partial\rho}\right)_{T}+kT_{0}V\Lambda_{22}^{r}}.\label{eq:127}
\end{equation}
Similarly, for the covariance $\overline{\Delta E\Delta N}$ we have
\begin{equation}
\overline{\Delta E\Delta N}=\tilde{\Lambda}_{12}^{-1}=\dfrac{kT_{0}\mu_{0}V}{\left(\dfrac{\partial\mu}{\partial\rho}\right)_{T}+kT_{0}V\Lambda_{22}^{r}}.\label{eq:128}
\end{equation}

Fluctuations of the chemical potential could be found directly from
Eq.(\ref{eq:120}). Instead, we will take a shortcut by taking advantage
of the fact that $\mu=\mu(\rho)$ and therefore 
\begin{equation}
\Delta\mu=\left(\dfrac{\partial\mu}{\partial\rho}\right)_{T}\dfrac{\Delta N}{V}.\label{eq:129}
\end{equation}
Taking a square of this equation and averaging over the distribution,
we obtain
\begin{equation}
\overline{(\Delta\mu)^{2}}=\left(\dfrac{\partial\mu}{\partial\rho}\right)_{T}^{2}\dfrac{\overline{(\Delta N)^{2}}}{V^{2}}=\dfrac{\dfrac{kT_{0}}{V}\left(\dfrac{\partial\mu}{\partial\rho}\right)_{T}^{2}}{\left(\dfrac{\partial\mu}{\partial\rho}\right)_{T}+kT_{0}V\Lambda_{22}^{r}},\label{eq:137}
\end{equation}
where we used Eq.(\ref{eq:127}). Applying the same method,
\begin{equation}
\overline{\Delta\mu\Delta N}=\left(\dfrac{\partial\mu}{\partial\rho}\right)_{T}\dfrac{\overline{(\Delta N)^{2}}}{V}=\dfrac{kT_{0}\left(\dfrac{\partial\mu}{\partial\rho}\right)_{T}}{\left(\dfrac{\partial\mu}{\partial\rho}\right)_{T}+kT_{0}V\Lambda_{22}^{r}},\label{eq:138}
\end{equation}
\begin{equation}
\overline{\Delta\mu\Delta E}=\left(\dfrac{\partial\mu}{\partial\rho}\right)_{T}\dfrac{\overline{\Delta E\Delta N}}{V}=\dfrac{\mu_{0}kT_{0}\left(\dfrac{\partial\mu}{\partial\rho}\right)_{T}}{\left(\dfrac{\partial\mu}{\partial\rho}\right)_{T}+kT_{0}V\Lambda_{22}^{r}}.\label{eq:139}
\end{equation}
Continuing along this line, all other covariances can be readily calculated.

Consider limiting cases of the above relations. When $\Lambda_{22}^{r}\rightarrow0$
(the reservoir is an infinite source of particles), we exactly recover
all results for the grand-canonical ensemble listed in Table \ref{tab:Fluctuations-muVT}.
When $\Lambda_{22}^{r}\rightarrow\infty$ (the reservoir cannot supply
or absorb particles), the system behaves as virtually closed ($NVT$).
Accordingly, the energy fluctuation (\ref{eq:126}) reduces to 
\begin{equation}
\overline{(\Delta E)^{2}}=\rho_{0}kT_{0}^{2}c_{v}V\label{eq:140}
\end{equation}
in agreement with the $NVT$ result (Table \ref{tab:Fluctuations-NVT}).
All other mean-square fluctuations and covariances appearing in Eqs.(\ref{eq:127})-(\ref{eq:139})
tend to zero as $1/\Lambda_{22}^{r}$. As evident from Table \ref{tab:Fluctuations-NVT},
in the $NVT$ ensemble they are proportional to $(\partial\mu/\partial T)_{\rho}$
and thus generally not zero. In the present calculations, they vanished
because we neglected the temperature dependence of $\mu$ for the
sake of simplicity. More accurate but rather tedious calculations
taking into account the temperature dependence of $\mu$ exactly recover
the fluctuation relations from Table \ref{tab:Fluctuations-NVT} when
$\Lambda_{22}^{r}\rightarrow\infty$.

It is interesting to note that at the limit of normal thermodynamic
stability when $(\partial\mu/\partial\rho)_{T}\rightarrow0$, the
canonical fluctuation $\overline{(\Delta N)^{2}}$ diverges to infinity
(Table \ref{tab:Fluctuations-muVT}). However, when the system is
coupled to a finite-size reservoir with a positive coefficient $\Lambda_{22}^{r}>0$,
the latter stabilizes the fluid and its compositional fluctuations
remain finite and equal to $1/\Lambda_{22}^{r}$, see Eq.(\ref{eq:127}).
The finite size of the reservoir smooths the singularity.

\section{Fluctuations of grain boundary properties\label{sec:Fluctuations-in-GBs}}

GBs are interfaces separating regions of the same crystalline solid
phase (called grains) with different crystallographic orientations.
GBs can strongly impact physical and mechanical properties of crystalline
materials, especially in nano-structured systems. While thermodynamics
and kinetics of GBs has been studied extensively for several decades
\citep{Balluffi95}, it is only recently that fluctuations of GB properties
have become the subject of dedicated research, primarily by atomistic
computer simulations \citep{Foiles06a,Hoyt09a,Song10,Fensin2010,Karma-2012a,Adland2013a,Mishin2014,Schmitz_2014}. 

In preparation for the discussion of GB fluctuations, we will give
a briefly account of GB thermodynamics focusing on a plane GB in a
binary substitutional solid solution.

\begin{figure}
\noindent \begin{centering}
\includegraphics[clip,width=0.7\textwidth]{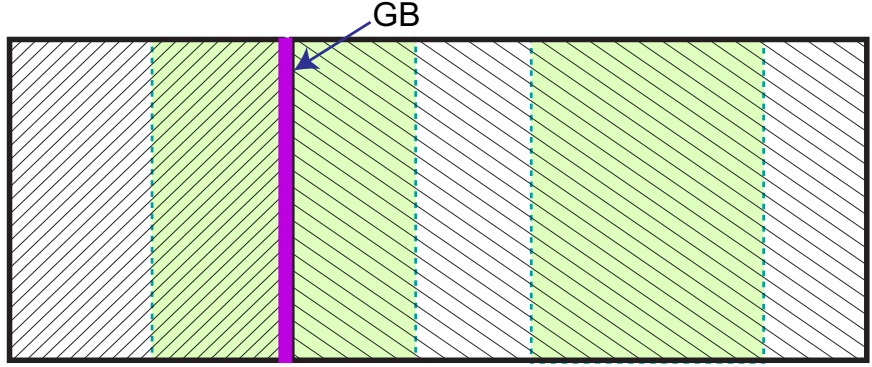}
\par\end{centering}

\protect\caption{Two grains with different crystallographic orientations separated
by a plane grain boundary (GB). The shaded areas bounded by dashed
lines mark a region containing the GB and a comparison region inside
one of the grains containing the same total number of atoms. These
two regions are selected for calculations of GB excess quantities.\label{fig:GB1}}

\end{figure}

\subsection{Grain boundary thermodynamics}

Consider a bicrystal (i.e., a system of two grains) enclosed in a
rectangular box with rigid impermeable walls (Fig.~\ref{fig:GB1}).
Since the system is closed, the total number of particles $N$ is
fixed. We will assume that the grains possess appropriate crystallographic
symmetries such that the mechanical stresses existing in the bicrystal
do not create a difference in strain energy densities in the grans
that would cause GB migration. 

Before discussing GB thermodynamics, we will first specify thermodynamic
properties of the homogeneous single-crystalline solid solution forming
the grains. We postulate the following fundamental equation of this
solution in the energy representation:\footnote{Generally, the fundamental equation of a stressed solid depends on
the deformation gradient relative to a chosen reference state \citep{Willard_Gibbs,Voorhees2004,Frolov2012a}.
In the particular case considered here, the cross-section of the solid
is fixed and only deformation in the normal direction is permitted.
Under such conditions, only the normal component of the deformation
gradient is relevant and varies as a linear function of the volume.
It is convenient to represent this component by the volume itself.}
\begin{equation}
E_{g}=E_{g}(S_{g},V_{g},N_{s}^{g},N_{2}^{g}).\label{eq:150}
\end{equation}
Here, the suffix $g$ indicates that the properties refer to a region
in one of the grains, such as the shaded layer shown in Fig.~\ref{fig:GB1},
$N_{2}^{g}$ is the number of atoms of species 2 and $N_{s}^{g}$
is the number of lattice sites in the region. Vacancies are neglected,
so that the number of atoms of species 1 is $N_{1}^{g}=N_{s}^{g}-N_{2}^{g}$.
The differential of Eq.(\ref{eq:150}) is
\begin{equation}
dE_{g}=TdS_{g}-pdV_{g}+\varphi dN_{s}^{g}+MdN_{2}^{g},\label{eq:151}
\end{equation}
where $T=\partial E_{g}/\partial S_{g}$ is temperature, $p=-\partial E_{g}/\partial V_{g}$
is pressure (negative of the stress component normal to the GB plane),
$M=\partial E_{g}/\partial N_{2}^{g}$ is the diffusion potential
of species 2 relative to species 1 \citep{Larche73,Larche_Cahn_78a,Larche1985},
and $\varphi=\partial E_{g}/\partial N_{s}^{g}$ is a thermodynamic
potential conjugate to the number of sites $N_{s}^{g}$. On the other
hand, the energy defined by Eq.(\ref{eq:150}) is a homogeneous function
of first degree in all four arguments. Using the Euler theorem for
homogeneous functions \citep{Courant:1989aa} we have
\begin{equation}
E_{g}=TE_{g}-pV_{g}+\varphi N_{s}^{g}+MN_{2}^{g}.\label{eq:152}
\end{equation}
Combining Eqs.(\ref{eq:151}) and (\ref{eq:152}) we obtain the Gibbs-Duhem
equation
\begin{equation}
-S_{g}dT+V_{g}dp-N_{s}^{g}d\varphi-N_{2}^{g}dM=0.\label{eq:153}
\end{equation}

To develop GB thermodynamics, we define the GB excess of any extensive
property $X$ by
\begin{equation}
\tilde{X}\equiv X-X_{g},\label{eq:154}
\end{equation}
where $X$ is the value of the property for a bicrystalline layer
containing the GB and $X_{g}$ is the value of the same property for
a homogeneous grain layer containing the same total number of atoms.
Examples of such layers are shown by shaded regions in Fig.~\ref{fig:GB1}.
The choice of these layers is totally arbitrary and does not affect
$\tilde{X}$ as long as their boundaries are unaffected by the GB
or the walls. According to this definition of excess, a bicrystalline
layer can be thought of as composed of two subsystems: (1) the reference
grain layer containing the same total number of atoms and (2) the
excess system representing the GB and described by the excess properties
$\tilde{S}$, $\tilde{E}$, $\tilde{V}$, etc.

Just as the grain thermodynamics has been formulated above starting
from a postulated fundamental equation (\ref{eq:150}), we will postulate
a fundamental equation of the GB in the form
\begin{equation}
\tilde{E}=\tilde{E}(\tilde{S},\tilde{V},\tilde{N}_{2},A).\label{eq:155}
\end{equation}
Note that $\tilde{N}_{1}$ is not an independent variable because
\begin{equation}
\tilde{N}_{1}+\tilde{N}_{2}=0\label{eq:155-1}
\end{equation}
by our definition of excesses. The excess energy (\ref{eq:155}) is
a homogeneous function of first degree in all four arguments. Applying
the Euler theorem \citep{Courant:1989aa} we have
\begin{equation}
\tilde{E}=\dfrac{\partial\tilde{E}}{\partial\tilde{S}}\tilde{S}+\dfrac{\partial\tilde{E}}{\partial\tilde{V}}\tilde{V}+\dfrac{\partial\tilde{E}}{\partial\tilde{N}_{2}}\tilde{N}_{2}+\dfrac{\partial\tilde{E}}{\partial A}A.\label{eq:155-2}
\end{equation}
On the other hand, the differential of Eq.(\ref{eq:155}) is
\begin{equation}
d\tilde{E}=\dfrac{\partial\tilde{E}}{\partial\tilde{S}}d\tilde{S}+\dfrac{\partial\tilde{E}}{\partial\tilde{V}}d\tilde{V}+\dfrac{\partial\tilde{E}}{\partial\tilde{N}_{2}}d\tilde{N}_{2}+\dfrac{\partial\tilde{E}}{\partial A}dA.\label{eq:155-3}
\end{equation}
Combining Eqs.(\ref{eq:155-2}) and (\ref{eq:155-3}) we obtain the
GB version of the Gibbs-Duhem equation:
\begin{equation}
\tilde{S}d\dfrac{\partial\tilde{E}}{\partial\tilde{S}}+\tilde{V}d\dfrac{\partial\tilde{E}}{\partial\tilde{V}}+\tilde{N}_{2}d\dfrac{\partial\tilde{E}}{\partial\tilde{N}_{2}}+Ad\dfrac{\partial\tilde{E}}{\partial A}=0.\label{eq:155-4}
\end{equation}

Next, we find the conditions of thermodynamic equilibrium between
the GB and the grains. To this end, we consider a variation of the
total energy $(\tilde{E}+E_{g})$ of a bicrystalline layer at ax fixed
composition, volume and entropy. In equilibrium, this variation $(d\tilde{E}+dE_{g})$
must be zero \citep{Willard_Gibbs}. Here, $dE_{g}$ is given by Eq.(\ref{eq:151})
and $d\tilde{E}$ by Eq.(\ref{eq:155-3}). The variation is subject
to the following constraints:
\begin{equation}
d\tilde{N}_{1}+dN_{1}^{g}=0,\label{eq:157}
\end{equation}
\begin{equation}
d\tilde{N}_{2}+dN_{2}^{g}=0,\label{eq:162}
\end{equation}
\begin{equation}
d\tilde{V}+dV_{g}=0,\label{eq:160}
\end{equation}
\begin{equation}
d\tilde{S}+dS_{g}=0,\label{eq:159}
\end{equation}
as well as $dA=0$. Furthermore, by adding Eqs.(\ref{eq:157}) and
(\ref{eq:162}) and taking into account Eq.(\ref{eq:155-1}), it follows
that the number of sites in the reference grain region is conserved:
$dN_{s}^{g}=dN_{1}^{g}+dN_{2}^{g}=0$. Inserting these constraints
into the total energy variation, we arrive at the equilibrium condition
\begin{equation}
\left(\dfrac{\partial\tilde{E}}{\partial\tilde{S}}-T\right)d\tilde{S}+\left(\dfrac{\partial\tilde{E}}{\partial\tilde{V}}+p\right)d\tilde{V}+\left(\dfrac{\partial\tilde{E}}{\partial\tilde{N}_{2}}-M\right)d\tilde{N}_{2}=0.\label{eq:163}
\end{equation}
Because the variations $d\tilde{S}$, $d\tilde{V}$ and $d\tilde{N}_{2}$
are independent and can be positive or negative, the coefficients
multiplying them must be zero. We thus obtain the equilibrium values
of the derivatives of $\tilde{E}$: 
\begin{equation}
\left(\dfrac{\partial\tilde{E}}{\partial\tilde{S}}\right)_{0}=T_{0},\label{eq:165}
\end{equation}
\begin{equation}
\left(\dfrac{\partial\tilde{E}}{\partial\tilde{N}_{2}}\right)_{0}=M_{0},\label{eq:166}
\end{equation}
\begin{equation}
\left(\dfrac{\partial\tilde{E}}{\partial\tilde{V}}\right)_{0}=-p_{0}.\label{eq:167}
\end{equation}
As usual, the subscript 0 labels the equilibrium state. 

The remaining derivative 
\begin{equation}
\gamma_{0}\equiv\left(\dfrac{\partial\tilde{E}}{\partial A}\right)_{0}\label{eq:164}
\end{equation}
is called the equilibrium GB free energy. The precise thermodynamic
meaning of this property will be discussed later.

\subsection{Quasi-equilibrium grain boundary states}

So far, we have only specified the derivatives of $\tilde{E}$ with
respect to $\tilde{S}$, $\tilde{N}_{2}$ , $\tilde{V}$ and $A$
for equilibrium states of the system. Namely, such derivatives are
equal to the equilibrium temperature, diffusion potential, negative
of pressure in the grains and the equilibrium GB free energy, respectively.
For the analysis of GB fluctuations, we must consider these derivatives
without assuming equilibrium between the GB and the grains. Such derivatives
form a new set of variables $T$, $p$, $M$ and $\gamma$ defined
as follows:
\begin{equation}
\dfrac{\partial\tilde{E}}{\partial\tilde{S}}\equiv T,\label{eq:165-1}
\end{equation}
\begin{equation}
\dfrac{\partial\tilde{E}}{\partial\tilde{N}_{2}}\equiv M,\label{eq:166-1}
\end{equation}
\begin{equation}
\dfrac{\partial\tilde{E}}{\partial\tilde{V}}\equiv-p,\label{eq:167-1}
\end{equation}
\begin{equation}
\dfrac{\partial\tilde{E}}{\partial A}\equiv\gamma.\label{eq:169-4}
\end{equation}
These new variables can be interpreted as, respectively, the local
temperature, negative of local pressure, local diffusion potential
and the GB free energy when the GB is \emph{not} in equilibrium with
the grains. Indeed, by postulating the fundamental equation (\ref{eq:155})
we implicitly assumed that the GB follows this equation even in the
absence of equilibrium with the grains. One can think of such GB states
as being disconnected from the actual grains and equilibrated with
imaginary grains with a different set of intensive parameters $T$,
$M$ and $p$. In terms of the variables (\ref{eq:165-1})-(\ref{eq:169-4}),
the differential form of the fundamental equation (\ref{eq:155})
becomes
\begin{equation}
d\tilde{E}=Td\tilde{S}-pd\tilde{V}+Md\tilde{N}_{2}+\gamma dA.\label{eq:155-3-1}
\end{equation}

The existence of quasi-equilibrium states in which the GB obeys a
fundamental equation without being in equilibrium with the grains
is an important supposition of the interface fluctuation theory. As
discussed in Secs.~\ref{sec:Non-equilibrium-entropy} and \ref{sec:Second-law},
the entire thermodynamic theory of equilibrium fluctuations is built
on the assumption that all fluctuated states are states of quasi-equilibrium.
The theory discussed here is an application of this principle to GBs. 

Suppose the bicrystal has been initially equilibrated and then the
GB deviates from this initial equilibrium to a nearby quasi-equilibrium
state where it is no longer in equilibrium with the grains. Denoting
the equilibrium excess entropy, volume and number of atoms $\tilde{S}_{0}$,
$\tilde{V}_{0}$ and $\tilde{N}_{2}^{0}$, respectively, the Gibss-Duhem
equation (\ref{eq:155-4}) for this variations becomes 
\begin{eqnarray}
 &  & \tilde{S}_{0}d\dfrac{\partial\tilde{E}}{\partial\tilde{S}}+\tilde{V}_{0}d\dfrac{\partial\tilde{E}}{\partial\tilde{V}}+\tilde{N}_{2}^{0}d\dfrac{\partial\tilde{E}}{\partial\tilde{N}_{2}}+Ad\dfrac{\partial\tilde{E}}{\partial A}\nonumber \\
 & = & \tilde{S}_{0}dT-\tilde{V}_{0}dp+\tilde{N}_{2}^{0}dM+Ad\gamma=0.\label{eq:190-1}
\end{eqnarray}
This equation can be rewritten in the form 
\begin{equation}
d\gamma=-\dfrac{\tilde{S}_{0}}{A}dT+\dfrac{\tilde{V}_{0}}{A}dp-\dfrac{\tilde{N}_{2}^{0}}{A}dM,\label{eq:191-1}
\end{equation}
which will be used later. 

In the particular case when the states of the grains also vary in
such a manner that the entire system undergoes an equilibrium process,
Eq.(\ref{eq:191-1}) becomes
\begin{equation}
d\gamma_{0}=-\dfrac{\tilde{S}_{0}}{A}dT_{0}+\dfrac{\tilde{V}_{0}}{A}dp_{0}-\dfrac{\tilde{N}_{2}^{0}}{A}dM_{0}.\label{eq:191-2}
\end{equation}
This is one of the forms of the Gibbs adsorption equation for a fixed
GB area. This equilibrium equation should not be confused with Eq.(\ref{eq:191-1})
describing fluctuations in which the GB deviates from equilibrium
with the grains.

We can now understand the meaning of the parameter $\gamma$. Let
us rewrite Eq.(\ref{eq:155-2}) in the form
\begin{equation}
\tilde{E}=T\tilde{S}-p\tilde{V}+M\tilde{N}_{2}+\gamma A,\label{eq:191-3}
\end{equation}
from which 
\begin{equation}
\gamma=\dfrac{\tilde{E}-T\tilde{S}+p\tilde{V}-M\tilde{N}_{2}}{A}.\label{eq:191-4}
\end{equation}
As was shown by Gibbs \citep{Willard_Gibbs}, an expression of the
form $E-TS+pV-\Sigma_{i}\mu_{i}N_{i}$ is the reversible work required
for the creation of a new system with extensive properties $E$, $S$,
$V$ and $N_{i}$ by drawing the energy and matter from an infinite
reservoir with a temperature $T$, pressure $p$ and chemical potentials
$\mu_{i}$. Thus, the numerator in Eq.(\ref{eq:191-4}) can be interpreted
as the reversible work of GB formation from an imaginary infinite
reservoir with the intensive parameters $T$, $p$ and $M$. In the
present case, the term $\mu_{1}\tilde{N}_{1}+\mu_{2}\tilde{N}_{2}$
reduces to $M\tilde{N}_{2}$ due to the constraint imposed by Eq.(\ref{eq:155-1}).
In short, $\gamma$ is the reversible work of GB formation per unit
area. Accordingly, the equilibrium GB free energy $\gamma_{0}$ is
the reversible work of GB formation per unit area from a reservoir
with the same intensive parameters $T_{0}$, $p_{0}$ and $M_{0}$
as in the actual grains: 
\begin{equation}
\gamma_{0}=\dfrac{\tilde{E}-T_{0}\tilde{S}+p_{0}\tilde{V}-M_{0}\tilde{N}_{2}}{A}.\label{eq:191-4-1}
\end{equation}
In other words, this is the reversible work of GB formation between
two infinitely large grains.

\subsection{Grain boundary fluctuations}

We are now in a position to analyze GB fluctuations. We will compute
generalized canonical fluctuations of GB properties treating the grains
as an infinite reservoir coupled to the GB. Although the foregoing
discussion utilized the energy scheme, the latter is impractical for
GB fluctuations.\footnote{The energy scheme involves fluctuations of the excess entropy $\tilde{S}$,
which is not accessible by experiments or atomistic simulations. In
the entropy scheme, $\tilde{S}$ is replaced by the more accessible
excess energy $\tilde{E}$.} Instead, we will now switch to the entropy scheme in which the independent
additive invariants fluctuating between the GB and the grains are
$X_{1}=\tilde{E}$, $X_{2}=\tilde{V}$ and $X_{3}=\tilde{N}_{2}$.
The conjugate thermodynamic forces are readily found from Eq.(\ref{eq:155-3-1}):
$F_{1}=1/T$, $F_{2}=p/T$ and $F_{3}=-M/T$. We can also rewrite
Eq.(\ref{eq:191-1}) in terms of the new variables:
\begin{equation}
d\gamma=\sum_{i=1}^{3}B_{i}dF_{i},\label{eq:191-1-1}
\end{equation}
with the coefficients
\begin{equation}
B_{1}\equiv\dfrac{T_{0}(\tilde{E}_{0}-\gamma_{0}A)}{A},\label{eq:195-1}
\end{equation}
\begin{equation}
B_{2}\equiv\dfrac{T_{0}\tilde{V}_{0}}{A},\label{eq:196-1}
\end{equation}
\begin{equation}
B_{3}\equiv-\dfrac{T_{0}\tilde{N}_{2}^{0}}{A}.\label{eq:197-1}
\end{equation}

In the Gaussian approximation to fluctuations, the distribution of
the excess parameters $X_{i}$ is given by Eq.(\ref{eq:58}) with
the stability matrix
\begin{equation}
\Lambda_{ij}=-\dfrac{1}{k}\begin{pmatrix}\left(\dfrac{\partial^{2}\tilde{S}}{\partial\tilde{E}^{2}}\right)_{0} &  & \left(\dfrac{\partial^{2}\tilde{S}}{\partial\tilde{E}\partial\tilde{V}}\right)_{0} &  & \left(\dfrac{\partial^{2}\tilde{S}}{\partial\tilde{E}\partial\tilde{N}_{2}}\right)_{0}\\
\\
\left(\dfrac{\partial^{2}\tilde{S}}{\partial\tilde{E}\partial\tilde{V}}\right)_{0} &  & \left(\dfrac{\partial^{2}\tilde{S}}{\partial\tilde{V}^{2}}\right)_{0} &  & \left(\dfrac{\partial^{2}\tilde{S}}{\partial\tilde{V}\partial\tilde{N}_{2}}\right)_{0}\\
\\
\left(\dfrac{\partial^{2}\tilde{S}}{\partial\tilde{E}\partial\tilde{N}_{2}}\right)_{0} &  & \left(\dfrac{\partial^{2}\tilde{S}}{\partial\tilde{V}\partial\tilde{N}_{2}}\right)_{0} &  & \left(\dfrac{\partial^{2}\tilde{S}}{\partial\tilde{N}_{2}^{2}}\right)_{0}
\end{pmatrix}.\label{eq:182}
\end{equation}
The elements of this matrix represent equilibrium excess properties
of the GB. For example,
\begin{equation}
\left(\dfrac{\partial^{2}\tilde{S}}{\partial\tilde{E}^{2}}\right)_{0}=\left(\dfrac{\partial}{\partial\tilde{E}}\left(\dfrac{1}{T}\right)\right)_{0}=-\dfrac{1}{T_{0}^{2}A\tilde{c}_{v}},\label{eq:183}
\end{equation}
where
\begin{equation}
\tilde{c}_{v}\equiv\dfrac{1}{A}\left(\dfrac{\partial\tilde{E}}{\partial T}\right)_{0}\label{eq:184-3}
\end{equation}
is the GB heat capacity (more accurately, the excess heat capacity
per unit GB area computed at fixed $\tilde{V}/A$ and $\tilde{N}_{2}/A$).
Similarly, 
\begin{equation}
\left(\dfrac{\partial^{2}\tilde{S}}{\partial\tilde{V}^{2}}\right)_{0}=\left(\dfrac{\partial}{\partial\tilde{V}}\left(\dfrac{p}{T}\right)\right)_{0}=-\dfrac{1}{T_{0}\tilde{V}_{0}\tilde{\kappa}_{T}}-\dfrac{p_{0}}{T_{0}^{2}\tilde{V}_{0}\tilde{\alpha}},\label{eq:185}
\end{equation}
where 
\begin{equation}
\tilde{\kappa}_{T}\equiv-\dfrac{1}{\tilde{V}_{0}}\left(\dfrac{\partial\tilde{V}}{\partial p}\right)_{0}\label{eq:186}
\end{equation}
is the isothermal compressibility of the GB and
\begin{equation}
\tilde{\alpha}\equiv\dfrac{1}{\tilde{V}_{0}}\left(\dfrac{\partial\tilde{V}}{\partial T}\right)_{0}\label{eq:187-1}
\end{equation}
is the GB thermal expansion coefficient.

Equilibrium GB properties can be computed directly by atomistic simulations.
For example, $\tilde{\kappa}_{T}$ can be found by computing the excess
GB volume as a function of pressure at a fixed temperature followed
by numerical differentiation \citep{Frolov2012b}. However, analysis
of GB fluctuations offers a more efficient approach in which the entire
set of equilibrium GB properties can be extracted from a single simulation
run using fluctuation relations. Indeed, the excess properties $\tilde{E}$,
$\tilde{V}$ and $\tilde{N}_{2}$ are readily accessible by atomistic
simulations. By computing their covariances, the inverse stability
matrix can be found from the fluctuation relation

\begin{equation}
\overline{\Delta X_{i}\Delta X_{j}}=\Lambda_{ij}^{-1},\enskip\enskip i,j=1,2,3.\label{eq:181}
\end{equation}
The latter can be then inverted to obtain the stability matrix $\Lambda_{ij}$
and thus the equilibrium GB properties such as $\tilde{c}_{v}$, $\tilde{\kappa}_{T}$,
$\tilde{\alpha}$, etc. Furthermore, knowing the stability matrix,
fluctuations of thermodynamic forces can be predicted from the fluctuation
relations
\begin{equation}
\overline{\Delta F_{i}\Delta F_{j}}=k^{2}\Lambda_{ij},\,\,\,\,\,i,j=1,...,3.\label{eq:190}
\end{equation}
Knowing these covariances, straightforward algebraic manipulations
can be applied to find $\overline{(\Delta M)^{2}}$, $\overline{(\Delta T)^{2}}$
and $\overline{(\Delta p)^{2}}$, as well as the covariances $\overline{\Delta M\Delta T}$,
$\overline{\Delta M\Delta p}$ and $\overline{\Delta p\Delta T}$.
As a general rule,
\begin{equation}
\overline{\Delta P_{i}\Delta P_{j}}=\sum_{m,n=1}^{3}k^{2}\left(\dfrac{\partial P_{i}}{\partial F_{m}}\right)_{0}\left(\dfrac{\partial P_{j}}{\partial F_{n}}\right)_{0}\Lambda_{nm},\,\,\,\,\,i,j=1,...,3,\label{eq:191}
\end{equation}
where the transformation matrix $\partial P_{i}/\partial F_{m}$ is
\begin{equation}
\left(\dfrac{\partial P_{i}}{\partial F_{m}}\right)_{0}=\begin{pmatrix}-T_{0}^{2} &  & p_{0}T_{0} &  & -M_{0}T_{0}\\
\\
0 &  & -T_{0} &  & 0\\
\\
0 &  & 0 &  & -T_{0}
\end{pmatrix}.\label{eq:190-2}
\end{equation}
To our knowledge, the proposed calculation scheme has not been implemented.

As an interesting application, one can predict the equilibrium mean-square
fluctuation $\overline{(\Delta\gamma)^{2}}$ of the GB free energy,
where $\Delta\gamma\equiv\gamma-\gamma_{0}$. The fluctuation form
of Eq.(\ref{eq:191-1-1}) is
\begin{equation}
\Delta\gamma=\sum_{i=1}^{3}B_{i}\Delta F_{i}.\label{eq:198}
\end{equation}
Taking a square of this equation and averaging over the distribution,
we obtain
\begin{equation}
\overline{(\Delta\gamma)^{2}}=\sum_{i,j=1}^{3}B_{i}B_{j}\overline{\Delta F_{i}\Delta F_{j}}.\label{eq:199}
\end{equation}
Applying Eq.(\ref{eq:190}) for $\overline{\Delta F_{i}\Delta F_{j}}$,
this equation finally becomes
\begin{equation}
\overline{(\Delta\gamma)^{2}}=k^{2}\sum_{i,j=1}^{3}B_{i}B_{j}\Lambda_{ij}.\label{eq:200}
\end{equation}
This calculation requires the knowledge of $\Lambda_{ij}$ and the
equilibrium values of the excess GB energy, volume and segregation
as ingredients for the parameters $B_{i}$. The GB free energy is
usually treated as a static property. An estimate of its fluctuation
$\overline{(\Delta\gamma)^{2}}$ would provide a critical assessment
of this approximation.

\section{Fluctuations in pre-melted grain boundaries\label{sec:Pre-melting}}

In this section we address another fluctuation topic associated with
GBs. At high temperatures approaching the melting point of the solid,
many GBs become atomically disordered and can develop a liquid-like
structure reminiscent of a thin liquid film \citep{Balluffi95}. Our
goal is to describe equilibrium thermodynamic properties of this film
and derive the distribution function of its width. The analysis is
based on the sharp interface model of pre-melting (Fig.~\ref{fig:Sharp-interface-model}).
The grains are formed by the same binary solid solution phase as discussed
in Sec.~\ref{sec:Fluctuations-in-GBs}.

\subsection{Thermodynamic background}

\subsubsection{Equilibrium state of pre-melted grain boundary}

\begin{figure}
\noindent \begin{centering}
\includegraphics[clip,width=0.6\textwidth]{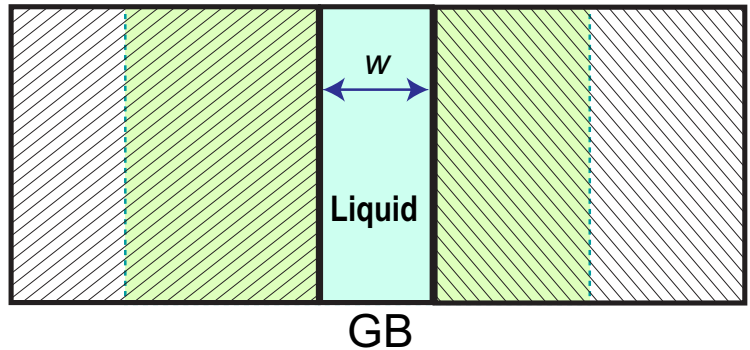}
\par\end{centering}

\protect\caption{Sharp interface model of GB pre-melting. The pre-melted GB is modeled
by a liquid layer of a width $w$ bounded by two sharp interfaces
endowed with thermodynamic properties of actual solid-liquid interfaces
between bulk phases.\label{fig:Sharp-interface-model} }

\end{figure}

We will consider the same bicrystal in a rigid box as discussed in
Sec.~\ref{sec:Fluctuations-in-GBs}, but this time we will adopt
a particular model of GB structure, namely, a liquid layer of a certain
width $w$ bounded by two solid-liquid interfaces (Fig.~\ref{fig:Sharp-interface-model}).
We start by describing equilibrium thermodynamic properties of this
boundary. 

As in Sec.~\ref{sec:Fluctuations-in-GBs}, thermodynamic properties
of the solid solution forming the grains are fully described by the
fundamental equation (\ref{eq:150}) in the energy representation.
The binary liquid solution (index $L$) is described by its own fundamental
equation
\begin{equation}
E_{L}=E_{L}(S_{L},V_{L},N_{1}^{L},N_{2}^{L})\label{eq:203}
\end{equation}
with the differential form
\begin{equation}
dE_{L}=T_{L}dS_{L}-p_{L}dV_{L}+\mu_{1}dN_{1}^{L}+\mu_{2}dN_{2}^{L}.\label{eq:204}
\end{equation}
Here, $T_{L}$ and $p_{L}$ are the temperature and pressure of the
liquid and $\mu_{1}$ and $\mu_{2}$ are the chemical potentials of
the components. 

To describe thermodynamics of the solid-liquid interfaces bounding
the liquid layer, we will first take a detour and consider a single
plane interface between the bulk solid and liquid phases. We can choose
an arbitrary geometric dividing surface partitioning our system in
two parts with volumes $V_{g}$ and $V_{L}$. We then compute the
excess of any extensive property $X$ by
\begin{equation}
\tilde{X}=X-X_{g}-X_{L},\label{eq:205}
\end{equation}
where $X$ is the value of the property for the solid-liquid system
with a volume $(V_{g}+V_{L})$ and $X_{g}$ and $X_{L}$ are properties
of the phases computed as if the volumes $V_{g}$ and $V_{L}$ remained
homogeneous all the way to the dividing surface \citep{Willard_Gibbs}.
By this definition, $\tilde{V}=0$. Because the densities of all properties
are generally different on either side of the interface, the excess
defined by Eq.(\ref{eq:205}) depends on the exact position of the
dividing surface within the interface region. We will adopt the rule
by which we always place the dividing surface in such a manner that
the excess of the total number of atoms $N$ vanishes: $\tilde{N}=0$.
The fundamental equation of the solid-liquid interface is formulated
in the form
\begin{equation}
\tilde{E}=\tilde{E}(\tilde{S},\tilde{N}_{2},A),\label{eq:206}
\end{equation}
with the differential form 
\begin{equation}
d\tilde{E}=\dfrac{\partial\tilde{E}}{\partial\tilde{S}}d\tilde{S}+\dfrac{\partial\tilde{E}}{\partial\tilde{N}_{2}}d\tilde{N}_{2}+\dfrac{\partial\tilde{E}}{\partial A}dA.\label{eq:207}
\end{equation}

Returning to the pre-melted GB modeled by a thin layer of the liquid
phase (Fig.~\ref{fig:Sharp-interface-model}), we will assume that
the solid-liquid interfaces bounding this layer follow the same fundamental
equation (\ref{eq:206}) as if they were interfaces between two bulk
phases. This is, of course, an approximations. In reality, the liquid
layer width is on the order of a nanometer and its thermodynamic properties
in this narrow confinement are different from properties of bulk liquid.
In addition, the solid-liquid interfaces separated by such a short
distance can interact with each other due to their finite thickness
comparable with the width of the layer. The difference between the
actual GB structure and this highly idealized liquid-layer structure
is accounted for by introducing an interaction between the two solid-liquid
interfaces called the ``disjoining'' interaction. The disjoining interaction
(index $d$) is included in the model by adding an extra energy term
$E_{d}$ and an associated entropy $S_{d}$, and postulating the following
fundamental equation of the disjoining interaction:
\begin{equation}
E_{d}=E_{d}(S_{d},V_{L}).\label{eq:208}
\end{equation}
The dependence on $V_{L}$ has been added to make the disjoining interaction
a function of the distance between the solid-liquid interfaces, which
for a fixed cross-section considered here, is proportional to $V_{L}$.
Both $E_{d}$ and $S_{d}$ are proportional to the GB area and have
the same dimensions as the excess quantities $\tilde{E}$ and $\tilde{S}$,
respectively. Both $E_{d}$ and $S_{d}$ tend to zero if the width
of the liquid layer increases to infinity (bulk solid-liquid system).

To summarize the model, the GB is represented by a homogeneous liquid
layer separated from the grains by two geometrically sharp solid-liquid
interfaces to which we assign thermodynamic properties of actual solid-liquid
interfaces between bulk phases (Fig.~\ref{fig:Sharp-interface-model}).
These sharp interfaces coincide with the dividing surfaces separated
by a distance $w$ and interact with each other by disjoining forces
that follow the fundamental equation (\ref{eq:208}).

We will next derive the conditions of thermodynamic equilibrium by
considering a variation of state of a bicrystalline layer indicated
in Fig.~\ref{fig:Sharp-interface-model} by a shaded region with
dashed boundaries. During this variation, the layer remains isolated
from the rest of the system and keeps a constant value of its entropy.
Under these constraints, the variation of the total energy 
\begin{equation}
\hat{E}=E_{g}+E_{L}+\tilde{E}+E_{d}\label{eq:208-1}
\end{equation}
must be zero \citep{Willard_Gibbs}. The terms of this equation are
represented by the fundamental equations (\ref{eq:150}), (\ref{eq:203}),
(\ref{eq:206}) and (\ref{eq:208}), respectively. Accordingly, 
\begin{eqnarray}
d\hat{E} & = & TdS_{g}-p_{g}dV_{g}+\varphi dN_{s}^{g}+MdN_{2}^{g}\nonumber \\
 & + & T_{L}dS_{L}-p_{L}dV_{L}+\mu_{1}dN_{1}^{L}+\mu_{2}dN_{2}^{L}\nonumber \\
 & + & \dfrac{\partial\tilde{E}}{\partial\tilde{S}}d\tilde{S}+\dfrac{\partial\tilde{E}}{\partial\tilde{N}_{2}}d\tilde{N}_{2}+\dfrac{\partial\tilde{E}}{\partial A}dA\nonumber \\
 & + & \dfrac{\partial E_{d}}{\partial S_{d}}dS_{d}+\dfrac{\partial E_{d}}{\partial V_{L}}dV_{L}=0\label{eq:209}
\end{eqnarray}
subject to the following constraints:
\begin{equation}
dS_{g}+dS_{L}+d\tilde{S}+dS_{d}=0,\label{eq:210}
\end{equation}
\begin{equation}
dV_{g}+dV_{L}=0,\label{eq:211}
\end{equation}
\begin{equation}
dN_{1}^{g}+dN_{1}^{L}+d\tilde{N}_{1}=0,\label{eq:212}
\end{equation}
\begin{equation}
dN_{2}^{g}+dN_{2}^{L}+d\tilde{N}_{2}=0,\label{eq:213}
\end{equation}
\begin{equation}
d\tilde{N}_{1}+d\tilde{N}_{2}=0,\label{eq:213-1}
\end{equation}
\begin{equation}
dN_{1}^{g}+dN_{2}^{g}-dN_{s}^{g}=0,\label{eq:214}
\end{equation}
as well as $dA=0$. Here, $-p_{g}$ is the component of the mechanical
stress in the grains normal to the GB. 

Out of 14 differentials appearing in the right-hand sides of the Eqs.(\ref{eq:209})-(\ref{eq:214}),
7 are eliminated by the constraints, leaving 7 independent variations:
\begin{eqnarray}
d\hat{E} & = & \left(T_{L}-T\right)dS_{L}+\left(\dfrac{\partial\tilde{E}}{\partial\tilde{S}}-T\right)d\tilde{S}+\left(\dfrac{\partial E_{d}}{\partial S_{d}}-T\right)dS_{d}\nonumber \\
 & + & \left(\dfrac{\partial E_{d}}{\partial V_{L}}-p_{L}+p_{g}\right)dV_{L}+\left(\varphi-\mu_{1}\right)dN_{1}^{g}\nonumber \\
 & + & \left(\varphi-\mu_{2}+M\right)dN_{2}^{g}+\left(\dfrac{\partial\tilde{E}}{\partial\tilde{N}_{2}}+\mu_{1}-\mu_{2}\right)d\tilde{N}_{2}=0.\label{eq:215}
\end{eqnarray}
Equating the differential coefficients to zero, we obtain the following
equilibrium conditions:
\begin{equation}
T=T_{L}=\left(\dfrac{\partial\tilde{E}}{\partial\tilde{S}}\right)_{0}=\left(\dfrac{\partial E_{d}}{\partial S_{d}}\right)_{0}\equiv T_{0},\label{eq:216}
\end{equation}
 
\begin{equation}
p_{L}^{0}=p_{g}^{0}+\left(\dfrac{\partial E_{d}}{\partial V_{L}}\right)_{0},\label{eq:217}
\end{equation}
\begin{equation}
\left(\dfrac{\partial\tilde{E}}{\partial\tilde{N}_{2}}\right)_{0}=M_{0}=\mu_{2}^{0}-\mu_{1}^{0},\label{eq:218}
\end{equation}
\begin{equation}
\varphi_{0}=\mu_{1}^{0},\label{eq:219}
\end{equation}
where $T_{0}$ is the uniform temperature throughout the equilibrium
system. As usual, the superscript 0 marks the equilibrium state. 

Equation (\ref{eq:217}) is the mechanical equilibrium condition,
which can be written in the form
\begin{equation}
p_{L}^{0}=p_{g}^{0}+p_{d}^{0},\label{eq:221}
\end{equation}
where $p_{d}^{0}$ is the equilibrium value of the \emph{disjoining
pressure} defined by
\begin{equation}
p_{d}\equiv\dfrac{\partial E_{d}}{\partial V_{L}}.\label{eq:222}
\end{equation}
Thus, the liquid phase within the GB layer is subject to an additional
pressure $p_{d}$ on top of the pressure $p_{g}$ existing in the
grains.

\subsubsection{Calculation of the disjoining pressure}

Calculation of the equilibrium disjoining pressure $p_{d}^{0}$ can
be simplified by reformulation of the equilibrium conditions in the
density form. The fundamental equation (\ref{eq:150}) of the solid
solution can be rewritten in the density form 
\begin{equation}
\varepsilon_{g}=\varepsilon_{g}(s_{g},v_{g},c_{g}),\label{eq:201-1}
\end{equation}
where $\varepsilon_{g}$, $s_{g}$ and $v_{g}$ are, respectively,
the energy, entropy and volume per lattice site (which is same as
per atom since we neglect vacancies), and $c_{g}$ is the site (atomic)
fraction of species 2. Accordingly, $T=\partial\varepsilon_{g}/\partial s_{g}$,
$p_{g}=-\partial\varepsilon_{g}/\partial v_{g}$ and $M=\partial\varepsilon_{g}/\partial c_{g}$.
From Eq.(\ref{eq:152}) we have
\begin{equation}
\varphi=\varepsilon_{g}-Ts_{g}+p_{g}v_{g}-Mc_{g},\label{eq:202-3}
\end{equation}
while the Gibbs-Duhem equation (\ref{eq:153}) becomes
\begin{equation}
s_{g}dT-v_{g}dp_{g}+d\varphi+c_{g}dM=0.\label{eq:202-7-1}
\end{equation}

It is more convenient to express all properties of the grain as functions
of the variable set $(T,p_{g},c_{g})$. The thermodynamic potential
appropriate for these variables is 
\begin{equation}
\omega(T,p_{g},c_{g})\equiv\varphi+Mc_{g}=\varepsilon_{g}-T_{g}s_{g}+p_{g}v_{g}.\label{eq:220-3}
\end{equation}
Its differential form is
\begin{equation}
d\omega=-s_{g}dT+v_{g}dp_{g}+Mdc_{g},\label{eq:220-8}
\end{equation}
which is easily derivable from Eq.(\ref{eq:202-7-1}). Similarly,
for the liquid solution we will use the variable set $(T_{L},p_{L},c_{L})$
and, accordingly, the Gibbs free energy
\begin{equation}
g(T_{L},p_{L},c_{L})\equiv\varepsilon_{L}-T_{L}s_{L}+p_{L}v_{L}=(1-c_{L})\mu_{1}+c_{L}\mu_{2}\label{eq:220-9}
\end{equation}
with the differential
\begin{equation}
dg=-s_{L}dT_{L}+v_{L}dp_{L}+(\mu_{2}-\mu_{1})dc_{L},\label{eq:220-10}
\end{equation}
where all densities are counted per atom.

We now reformulate the equilibrium conditions (\ref{eq:218}) and
(\ref{eq:219}) in the form
\begin{equation}
M(T_{0},p_{g}^{0},c_{g}^{0})=\mu_{2}(T_{0},p_{L}^{0},c_{L}^{0})-\mu_{1}(T_{0},p_{L}^{0},c_{L}^{0}),\label{eq:220-1}
\end{equation}
\begin{equation}
\omega(T_{0},p_{g}^{0},c_{g}^{0})=g(T_{0},p_{L}^{0},c_{L}^{0})+M(T_{0},p_{g}^{0},c_{g}^{0})(c_{g}^{0}-c_{L}^{0}),\label{eq:219-1}
\end{equation}
respectively. This is a system of two non-linear equations with respect
to five equilibrium parameters $(T_{0},p_{g}^{0},p_{L}^{0},c_{g}^{0},c_{L}^{0})$.
It follows that our solid-liquid system has three degrees of freedom.
Let us choose the independent variable set $(T_{0},p_{g}^{0},c_{g}^{0})$.
For any given state of the grains defined by these parameters, we
can solve Eqs.(\ref{eq:220-1}) and (\ref{eq:219-1}) to find the
composition $c_{L}^{0}$ and pressure $p_{L}^{0}$ in the liquid layer,
and thus the disjoining pressure $p_{d}^{0}=p_{L}^{0}-p_{g}^{0}$. 

For practical applications, we will replace this exact calculation
scheme by a simple approximation. At given temperature $T_{0}$ and
pressure $p_{g}^{0}$, let $c_{g}^{*}$ and $c_{L}^{*}$ be the solid
and liquid compositions corresponding to the bulk solid-liquid coexistence
$(p_{d}^{0}=0)$. These two compositions must satisfy the equations\footnote{We will not go into details, but Eqs.(\ref{eq:220-2}) and (\ref{eq:219-2})
can be given a geometric interpretation of a common tangent construction,
$M$ being the slope of the common tangent to the curves $\omega(T_{0},p_{g}^{0},c)$
and $g(T_{0},p_{g}^{0},c)$ as functions of composition $c$ at fixed
$T_{0}$ and $p_{g}^{0}$. }
\begin{equation}
M(T_{0},p_{g}^{0},c_{g}^{*})=\mu_{2}(T_{0},p_{g}^{0},c_{L}^{*})-\mu_{1}(T_{0},p_{g}^{0},c_{L}^{*}),\label{eq:220-2}
\end{equation}
\begin{equation}
\omega(T_{0},p_{g}^{0},c_{g}^{*})=g(T_{0},p_{g}^{0},c_{L}^{*})+M(T_{0},p_{g}^{0},c_{g}^{*})(c_{g}^{*}-c_{L}^{*}).\label{eq:219-2}
\end{equation}
Note that the pressure continuity condition $p_{L}^{0}=p_{g}^{0}$
valid for a two-phase bulk system with a plane interface eliminates
one degree of freedom, leaving our system with two degrees of freedom
in compliance with the Gibbs phase rule. For a fixed pressure $p_{g}^{0}$,
Eqs.(\ref{eq:220-2}) and (\ref{eq:219-2}) define the solidus and
liquidus lines, $c_{g}^{*}(T_{0})$ and $c_{L}^{*}(T_{0})$, on the
phase diagram of the system. 

We seek an approximate form of Eqs.(\ref{eq:220-1}) and (\ref{eq:219-1})
in the vicinity of the solidus line. As a measure of proximity of
the grains to the solidus line at given $T_{0}$ and $p_{g}^{0}$,
we choose the diffusion potential deviation 
\begin{equation}
\Delta M\equiv M(T_{0},p_{g}^{0},c_{g}^{0})-M(T_{0},p_{g}^{0},c_{g}^{*})\label{eq:240}
\end{equation}
from its solidus value $M(T_{0},p_{g}^{0},c_{g}^{*})$. All functions
appearing in Eqs.(\ref{eq:220-1}) and (\ref{eq:219-1}) can be then
approximated by linear expansions in the small parameters $\Delta M$,
$(c_{g}^{0}-c_{g}^{*})$, $(c_{L}^{0}-c_{L}^{*})$ and $(p_{L}^{0}-p_{g}^{0})$.
The linearized form of these equations becomes
\begin{equation}
\Delta M=\left(\dfrac{\partial(\mu_{2}-\mu_{1})}{\partial c_{L}}\right)_{*}(c_{L}^{0}-c_{L}^{*})+\left(\dfrac{\partial(\mu_{2}-\mu_{1})}{\partial p_{L}}\right)_{*}p_{d}^{0},\label{eq:242}
\end{equation}
\begin{eqnarray}
\left(\dfrac{\partial\omega}{\partial c_{g}}\right)_{*}(c_{g}^{0}-c_{g}^{*}) & = & \left(\dfrac{\partial g}{\partial c_{L}}\right)_{*}(c_{L}^{0}-c_{L}^{*})+\left(\dfrac{\partial g}{\partial p_{L}}\right)_{*}p_{d}^{0}\nonumber \\
 & + & M(T_{0},p_{g}^{0},c_{g}^{*})(c_{g}^{0}-c_{L}^{0}-c_{g}^{*}+c_{L}^{*})+(c_{g}^{*}-c_{L}^{*})\Delta M,\label{eq:241}
\end{eqnarray}
where the asterisk indicates that the derivatives are taken in the
bulk-coexistence state ($\Delta M=0$). All zero-order terms have
canceled out by Eqs.(\ref{eq:220-2}) and (\ref{eq:219-2}). The equations
obtained are simplified by utilizing the relations
\begin{equation}
\left(\dfrac{\partial\omega}{\partial c_{g}}\right)_{*}=\left(\dfrac{\partial g}{\partial c_{L}}\right)_{*}=M(T_{0},p_{g}^{0},c_{g}^{*}),\label{eq:243}
\end{equation}
\begin{equation}
\left(\dfrac{\partial g}{\partial p_{L}}\right)_{*}=v_{L}^{*},\label{eq:244}
\end{equation}
\begin{equation}
\left(\dfrac{\partial(\mu_{2}-\mu_{1})}{\partial p_{L}}\right)_{*}=\bar{v}_{L}^{2*}-\bar{v}_{L}^{1*},\label{eq:245}
\end{equation}
where $\bar{v}_{L}^{i}=\partial\mu_{i}/\partial p_{L}$ are partial
molar volumes of the components in the liquid phase, with the asterisk
indicating that these volumes refer to the bulk-coexistence state.
Using these relations, Eqs.(\ref{eq:242}) and (\ref{eq:241}) become,
respectively,
\begin{equation}
\Delta M=\left(\dfrac{\partial(\mu_{2}-\mu_{1})}{\partial c_{L}}\right)_{*}(c_{L}^{0}-c_{L}^{*})+(\bar{v}_{L}^{2*}-\bar{v}_{L}^{1*})p_{d}^{0},\label{eq:247}
\end{equation}
\begin{equation}
0=v_{L}^{*}p_{d}^{0}+(c_{g}^{*}-c_{L}^{*})\Delta M.\label{eq:246}
\end{equation}
Equation (\ref{eq:246}) immediately gives us the equilibrium disjoining
pressure,
\begin{equation}
p_{d}^{0}=\dfrac{c_{L}^{*}-c_{g}^{*}}{v_{L}^{*}}\Delta M,\label{eq:248}
\end{equation}
while Eq.(\ref{eq:247}) can be solved for the deviation of the liquid-layer
composition $c_{L}^{0}$ from the liquidus composition $c_{L}^{*}$
:
\begin{equation}
c_{L}^{0}-c_{L}^{*}=\dfrac{\Delta M-(v_{L}^{2*}-v_{L}^{1*})p_{d}^{0}}{\left(\dfrac{\partial(\mu_{2}-\mu_{1})}{\partial c_{L}}\right)_{*}}=\dfrac{1-\dfrac{v_{L}^{2*}-v_{L}^{1*}}{v_{L}^{*}}(c_{L}^{*}-c_{g}^{*})}{\left(\dfrac{\partial(\mu_{2}-\mu_{1})}{\partial c_{L}}\right)_{*}}\Delta M.\label{eq:249}
\end{equation}
In many systems, the difference between the partial molar volumes
of the components is small in comparison with $v_{L}^{*}$ and the
latter equation can be simplified to
\begin{equation}
c_{L}^{0}-c_{L}^{*}=\dfrac{1}{\left(\dfrac{\partial(\mu_{2}-\mu_{1})}{\partial c_{L}}\right)_{*}}\Delta M.\label{eq:250}
\end{equation}

For a single-component system, $\Delta M$ loses its significance.
The appropriate measure of proximity of the system to bulk melting
is then the difference $(T_{0}-T_{m})$ between the temperature of
the bicrystal and the bulk melting point $T_{m}$ at a given pressure
$p_{g}^{0}$. The equilibrium conditions (\ref{eq:220-1}) and (\ref{eq:219-1})
reduce to one equation,
\begin{equation}
\omega(T_{0},p_{g}^{0},0)=g(T_{0},p_{L}^{0},0),\label{eq:251}
\end{equation}
which for bulk equilibrium becomes
\begin{equation}
\omega(T_{m},p_{g}^{0},0)=g(T_{m},p_{g}^{0},0).\label{eq:252}
\end{equation}
Linearizing Eq.(\ref{eq:251}) in the small parameters $(T_{0}-T_{m})$
and $p_{d}^{0}$ we have
\begin{equation}
\left(\dfrac{\partial\omega}{\partial T}\right)_{*}(T_{0}-T_{m})=\left(\dfrac{\partial g}{\partial T}\right)_{*}(T_{0}-T_{m})+\left(\dfrac{\partial g}{\partial p_{L}}\right)_{*}p_{d}^{0},\label{eq:252-1}
\end{equation}
from which
\begin{equation}
p_{d}^{0}=\dfrac{s_{L}^{*}-s_{g}^{*}}{v_{L}^{*}}(T_{0}-T_{m})=\dfrac{H_{m}}{T_{m}v_{L}^{*}}(T_{0}-T_{m}),\label{eq:253}
\end{equation}
where $H_{m}=(s_{L}^{*}-s_{g}^{*})T_{m}$ is the enthalpy of melting
per atom.

\subsection{Analysis of fluctuations}

To describe fluctuations in the pre-melted GB, we treat the grains
as an infinite reservoir and apply the canonical fluctuation relation
(\ref{eq:49-1}). The system coupled to the reservoir is the GB possessing
the energy 
\begin{equation}
E_{GB}\equiv E_{L}+\tilde{E}+E_{d}.\label{eq:254}
\end{equation}

We will initially consider the full set of fluctuating parameters,
which will later be reduced to a smaller set. Namely, we will initially
treat all arguments of the fundamental equations (\ref{eq:203}),
(\ref{eq:206}) and (\ref{eq:208}), except for the fixed GB area,
as the fluctuating parameters. In the energy scheme, the fluctuating
parameter set is
\begin{equation}
Z=(S_{L},\tilde{S},S_{d},V_{L},N_{1}^{L},N_{2}^{L},\tilde{N}_{2}).\label{eq:255}
\end{equation}
The thermodynamic forces conjugate to these parameters are
\begin{equation}
P=(T_{L},\dfrac{\partial\tilde{E}}{\partial\tilde{S}},\dfrac{\partial E_{d}}{\partial S_{d}},-p_{L}+\dfrac{\partial E_{d}}{\partial V_{L}},\mu_{1},\mu_{2},\dfrac{\partial\tilde{E}}{\partial\tilde{N}_{2}}).\label{eq:256}
\end{equation}
The reversible work of GB formation (\ref{eq:49-2}) takes the form
\begin{equation}
\mathcal{R}=E_{GB}-E_{GB}^{0}-{\displaystyle \sum_{i=1}^{7}P_{i}^{0}(Z_{i}-Z_{i}^{0}),}\label{eq:49-2-1}
\end{equation}
where the thermodynamic forces are taken in the state of equilibrium.
Using the equilibrium conditions (\ref{eq:216})-(\ref{eq:218}),
we obtain
\begin{eqnarray}
\mathcal{R} & = & (E_{L}-E_{L}^{0})+(\tilde{E}-\tilde{E}_{0})+(E_{d}-E_{d}^{0})\nonumber \\
 &  & -T_{0}(S_{L}-S_{L}^{0})-T_{0}(\tilde{S}-\tilde{S}_{0})-T_{0}(S_{d}-S_{d}^{0})\nonumber \\
 &  & +p_{L}^{0}(V_{L}-V_{L}^{0})-p_{d}^{0}(V_{L}-V_{L}^{0})\nonumber \\
 &  & -\mu_{1}^{0}(N_{1}^{L}-N_{1}^{L0})-\mu_{2}^{0}(N_{2}^{L}-N_{2}^{L0})-M_{0}(\tilde{N}_{2}-\tilde{N}_{2}^{0}).\label{eq:257}
\end{eqnarray}

Next, we will make some approximations that will reduce the number
of fluctuating parameters. The difference $(E_{L}-E_{L}^{0})$ appearing
in Eq.(\ref{eq:257}) will be approximated by its linear expansion
around equilibrium:
\begin{equation}
E_{L}-E_{L}^{0}=T_{0}(S_{L}-S_{L}^{0})-p_{L}^{0}(V_{L}-V_{L}^{0})+\mu_{1}^{0}(N_{1}^{L}-N_{1}^{L0})+\mu_{2}^{0}(N_{2}^{L}-N_{2}^{L0}).\label{eq:258}
\end{equation}
Likewise, $(\tilde{E}-\tilde{E}_{0})$ will also be evaluated in the
linear approximation:
\begin{equation}
\tilde{E}-\tilde{E}_{0}=T_{0}(\tilde{S}-\tilde{S}_{0})+M_{0}(\tilde{N}_{2}-\tilde{N}_{2}^{0}).\label{eq:259}
\end{equation}
The terms related to the disjoining interaction will remain intact.
Inserting the linear expansions (\ref{eq:258}) and (\ref{eq:259})
in Eq.(\ref{eq:257}), several terms cancel out and we are left with
\begin{equation}
\mathcal{R}=(E_{d}-E_{d}^{0})-T_{0}(S_{d}-S_{d}^{0})-p_{d}^{0}(V_{L}-V_{L}^{0}).\label{eq:260-1}
\end{equation}

Although Eq.(\ref{eq:260-1}) contains variations of three variables,
only one fluctuating parameter is independent. To see this, we apply
the Legendre transformation with respect to the disjoining entropy
to obtain the thermodynamic potential

\begin{equation}
\Psi(T,V_{L})=E_{d}(S_{d},V_{L})-TS_{d}.\label{eq:227}
\end{equation}
This thermodynamic potential is called the \emph{disjoining potential}
and its differential form is
\begin{equation}
d\Psi=-S_{d}dT+p_{d}dV_{L}.\label{eq:226}
\end{equation}
The latter equation suggests another definition of the disjoining
pressure:
\begin{equation}
p_{d}=\left(\dfrac{\partial\Psi}{\partial V_{L}}\right)_{T}.\label{eq:228}
\end{equation}
Thus, Eq.(\ref{eq:260-1}) can be rewritten in the form 
\begin{equation}
\mathcal{R}=\Psi(T_{0},V_{L})-\Psi(T_{0},V_{L}^{0})-p_{d}^{0}(V_{L}-V_{L}^{0}),\label{eq:228-1}
\end{equation}
showing that, at a given temperature $T_{0}$, $\mathcal{R}$ is solely
a function of $V_{L}$. In other words, $V_{L}$ is the only independent
fluctuating parameter. By neglecting higher-order terms in Eqs.(\ref{eq:258})
and (\ref{eq:259}), we have suppressed fluctuations of all other
independent parameters. 

The probability distribution of the liquid volume $V_{L}$ is given
by Eq.(\ref{eq:49-1}), which becomes

\begin{equation}
W(V_{L})=W_{m}\exp\left(-\dfrac{\mathcal{R}}{kT_{0}}\right)=W_{m}\exp\left[-\dfrac{\Psi(T_{0},V_{L})-\Psi(T_{0},V_{L}^{0})-p_{d}^{0}(V_{L}-V_{L}^{0})}{kT_{0}}\right].\label{eq:225}
\end{equation}
Remembering that the cross-section of the bicrystal is fixed, the
finally arrive at the distribution function of the liquid-layer width
$w$:
\begin{equation}
W(w)=W_{m}\exp\left[-A\dfrac{\psi(T_{0},w)-\psi(T_{0},w_{0})-p_{d}^{0}(w-w_{0})}{kT_{0}}\right],\label{eq:230}
\end{equation}
where $w_{0}=V_{L}^{0}/A$ is the equilibrium width of the layer and
$\psi=\Psi/A$ is the disjoining potential per unit area. Note that
this distribution depends on the GB area $A$: the larger the area,
the sharper is the peak of the distribution around $w_{0}$. The disjoining
pressure can be rewritten as $p_{d}=(\partial\psi/\partial w)_{T}$.
Its equilibrium value $p_{d}^{0}$ appearing in Eq.(\ref{eq:230})
can be found from the bulk thermodynamic properties using Eq.(\ref{eq:248})
for a binary alloy and Eq.(\ref{eq:253}) for a single-component system.
Knowing $p_{d}^{0}$ and the distribution of $w$ at a given temperature
$T_{0}$, one can invert Eq.(\ref{eq:230}) and extract the disjoining
potential $\psi(T_{0},w)$. 

The GB width distribution (\ref{eq:230}) in conjunction with Eqs.(\ref{eq:248})
and (\ref{eq:253}) for the equilibrium disjoining pressure answers
the goal of this section. For single-component systems, an equation
similar to (\ref{eq:230}) was derived in previous work \citep{Hoyt09a,Fensin2010,Song10,Spatschek2013}
and utilized for calculations of disjoining interactions in Ni GBs
by atomistic simulations \citep{Hoyt09a,Fensin2010,Song10}. In the
present paper, this distribution has been extended to binary systems.

As a further approximation, we can expand $\psi(T_{0},w)$ around
the equilibrium width keeping only linear and quadratic terms:
\begin{equation}
\psi(T_{0},w)=\psi(T_{0},w_{0})+p_{d}^{0}(w-w_{0})+\dfrac{1}{2}\left(\dfrac{\partial^{2}\psi}{\partial w^{2}}\right)_{0}(w-w_{0})^{2},\label{eq:260}
\end{equation}
where the second derivative is taken at equilibrium. This results
in the Gaussian distribution
\begin{equation}
W(w)=W_{m}\exp\left[-\dfrac{A}{2kT_{0}}\left(\dfrac{\partial^{2}\psi}{\partial w^{2}}\right)_{0}(w-w_{0})^{2}\right],\label{eq:261}
\end{equation}
from which the mean-square fluctuation of the GB width is
\begin{equation}
\overline{(w-w_{0})^{2}}=\dfrac{kT_{0}}{A(\partial^{2}\psi/\partial w^{2})_{0}}.\label{eq:263}
\end{equation}
This estimate shows that the root-mean-square fluctuation of the liquid-layer
width scales with the GB area as $1/\sqrt{A}$.

According to the foregoing derivation, the GB width $w$ appearing
in the sharp-interface model of GB premelting (Fig.~\ref{fig:Sharp-interface-model})
has the meaning of the distance between two dividing surfaces of the
solid-liquid interfaces. Such dividing surfaces are characterized
by a zero excess of the total number of particles {[}$\tilde{V}=0$,
$\tilde{N}=0$, see excess definition (\ref{eq:205}){]}. It should
be, therefore, understood that the alternate definitions of $w$ adopted
in atomistic simulations \citep{Hoyt09a,Fensin2010,Song10} constitute
approximations to the strict definition.

\section{Concluding remarks\label{sec:Conclusions}}

The thermodynamic theory of fluctuations presented here expands the
postulational basis of the traditional classical thermodynamics by
incorporating equilibrium fluctuations. The main additional element
of the expanded theory is the fluctuation postulate expressed by Eq.(\ref{eq:12})
{[}or equivalently, Eq.(\ref{eq:16}){]}. Due to its statistical nature,
this postulate signifies a break with the determinism of classical
thermodynamics. The statistical component introduced by this postulate
propagates through the entire logical structure of thermodynamics.
The equilibrium state of a system is no longer a state but a distribution
over states. Thermodynamic properties become distributions around
the average values which are measured by macroscopic experiments.
Similar expansions of classical thermodynamics were previously discussed
by Tisza and Quay \citep{Tisza:1963} and Callen \citep{Callen_book_1960,Callen_book_1985,Callen:1965aa}
and can be traced back to Einstein \citep{Einstein_1910}. The fluctuation
postulate formulated in this paper describes fluctuations in an isolated
system and stems from the microcanonical distribution in statistical
mechanics. In this sense, our approach is different from that of Tisza
and Quay \citep{Tisza:1963} and Callen \citep{Callen_book_1960,Callen_book_1985,Callen:1965aa}
and is more aligned with Einstein's proposal \citep{Einstein_1910}.

The fluctuation postulate relies on the concept of non-equilibrium
entropy. Without a proper definition of non-equilibrium entropy, we
cannot talk about its increase during equilibration or behavior during
fluctuations. In some sense, non-equilibrium entropy is defined already
in classical thermodynamics. The thermodynamic equilibrium conditions
of classical thermodynamics are formulated through the entropy maximum
principle \citep{Willard_Gibbs,Tisza:1961aa,Callen_book_1960,Callen_book_1985}.
In this principle, the actual equilibrium state of the system is compared
with possible \emph{virtual} states in which the system is \emph{imagined}
to be partitioned into isolated equilibrium subsystems; their entropies
are then summed up and compared with the equilibrium entropy of the
actual system. The expanded thermodynamics goes one step further:
it assumes that such states are \emph{actually implemented} during
spontaneous fluctuations away from the equilibrium state. The term
``quasi-equilibrium'' introduced in this paper captures three defining
properties of such fluctuated states: (1) non-equilibrium, (2) real
(as opposed to virtual states of classical thermodynamics), and (3)
composed of subsystems that can be treated as isolated (on a certain
time scale) and each described by a fundamental equation. The quasi-equilibrium
states are characterized by a set of internal parameters $\lambda_{i}$.
The fluctuation postulate specifies the distribution function of these
parameters for an isolated system.

The second law of thermodynamics is re-formulated to include fluctuations.
In the refined formulation, it is only the entropy $\bar{S}$ averaged
over the fluctuations that monotonically increases and reaches a maximum
at equilibrium. On shorter time scales, the non-equilibrium entropy
$\hat{S}$ incessantly fluctuates around its average value and can
never accede a certain maximum value $S$ (Figs.~\ref{fig:time-scales}
and \ref{fig:Fluctuations_1}). Thus, the average entropy $\bar{S}$
obtained by macroscopic measurements (performed on the time scale
$t_{TD}$) is always somewhat smaller than $S$.

In classical thermodynamics, a central role is played by the fundamental
equation which encapsulates all equilibrium thermodynamic properties
of the system. This equation, first discovered by Gibbs \citep{Willard_Gibbs},
looks exactly like our Eq.(\ref{eq:1}) but has a subtly different
meaning. In the classical fundamental equation, the entropy $S$ is
a static property defined through the maximum-entropy principle (e.g.,
Callen's Postulate II \citep{Callen_book_1960,Callen_book_1985} or
Tisza's Postulate \emph{Pb1} \citep{Tisza:1961aa}) and measured experimentally.
In the expanded theory, $S$ appearing in Eq.(\ref{eq:1}) has the
statistical-mechanical meaning of the maximum micro-canonical entropy
of an isolated system. As noted in the previous paragraph, this entropy
is slightly larger than the measured entropy $\bar{S}$. In other
words, the measured entropy $\bar{S}$ does not exactly satisfy the
fundamental equation (\ref{eq:1}) of the present theory. This fact,
of course, does not create any problems. It only reflects the possibility
of different definitions of the fundamental equation for a fluctuating
system. Since both $S$ and $\bar{S}$ are well-defined for any given
set of conserved parameters $X_{1},...,X_{n}$, $\bar{S}(X_{1},...,X_{n})$
could also be taken as the fundamental equation \citep{Green:1951aa,Callen_book_1960,Callen:1965aa,Callen_book_1985}.
We would then have to re-define the temperature, pressure and all
other intensities, and to modify many other equations of the theory.
In the end, taking the fundamental equation $\bar{S}(X_{1},...,X_{n})$
as the point of departure would not lead to a logically consistent
thermodynamic fluctuation theory consistent with statistical mechanics.\footnote{As already mentioned, Greene and Callen \citep{Green:1951aa} resolved
these difficulties by making an additional approximation that essentially
identifies the canonical and micro-canonical entropies. This identification
is not satisfying for a rigorous fluctuation theory. The distinction
between the two entropies was discussed in Sec.~\ref{sub:The-Gaussian-law-canonical}} The advantage our Eq.(\ref{eq:1}) with the entropy defined as above
is that it leads to a simple and complete logical structure fully
consistent with statistical mechanics. In the limit of an infinite
large system, the fluctuations vanish and $S$ and $\bar{S}$ become
numerically equal. Accordingly, Eq.(\ref{eq:1}) becomes the fundamental
equation in the sense of classical thermodynamics.

Similar considerations apply to canonical fluctuations. If the average
values $\bar{X}_{1},...,\bar{X}_{m}$ of the fluctuating extensive
parameters $X_{1},...,X_{m}$ are formally inserted in the fundamental
equation (\ref{eq:1}), we do not obtain the canonical entropy $\bar{S}$.
However, in the limit of a large system, all canonical averages do
satisfy (\ref{eq:1}) and the latter becomes the classical fundamental
equation.

For the sake of simplicity, we have only considered a particular class
of thermodynamic systems for which all arguments of the fundamental
equation are additive invariants. In the future, the theory can be
generalized to include pseudo-thermodynamic and quasi-thermodynamic
variables \citep{Tisza:1961aa}. Pseudo-thermodynamic variables are
neither additive nor conserved; an example is furnished by long-range
order parameters in atomically ordered compounds. Quasi-thermodynamic
variables can be additive but need not be conserved. They give rise
to additional work terms in the total energy variation. However, the
conjugate thermodynamic forces do not become spatially homogeneous
when the system reaches equilibrium and thus cannot be considered
as intensities. Examples include the total magnetic or electric-dipole
moment of the system, the conjugate variables for which are, respectively,
the magnetic and electric field. Elastic strains and stresses in inhomogeneously
deformed crystalline solids belong to the same category. All such
cases require a separate treatment.

To facilitate practical applications of the theory, we have presented
a regular procedure for calculations of equilibrium fluctuations of
extensive parameters, intensive parameters and densities in systems
with any number of parameters. The proposed formalism has been demonstrated
by deriving the complete set of fluctuation relations for a simple
fluid in three different ensembles. The results are summarized in
Tables \ref{tab:Fluctuations-NVT}, \ref{tab:Fluctuations-NpT} and
\ref{tab:Fluctuations-muVT} and may present a reference value. It
should be noted that these calculations treat all fluctuations in
the Gaussian approximation. Greene and Callen \citep{Green:1951aa,Callen_book_1960}
proposed a more general method that does not rely on the Gaussian
approximation and enables calculations of higher-order correlation
moments such as $\overline{\Delta X_{i}\Delta X_{j}\Delta X_{k}}$,
$\overline{\Delta X_{i}\Delta X_{j}\Delta X_{k}\Delta X_{l}}$, etc.
Such moments are rarely important and were not discussed in this paper.

The proposed fluctuation formalism has been applied to solve two problems
related to GBs in binary solid solutions. First, we developed a set
of equations relating the fluctuations of excess GB properties, readily
accessible by atomistic simulations, to equilibrium GB characteristics
such as the excess elastic modulus, excess heat capacity and others.
These equations offer a computationally efficient approach to computing
the entire set of such properties from a single simulation run. Furthermore,
the covariances between the fluctuations generate a set of thermodynamic
relations that can be tested by simulations to verify the fundamental
understanding of GB thermodynamics. Implementation of this program
is presently underway. 

Second, we have extended the sharp interface model of GB pre-melting
to binary alloys and derived the distribution of the fluctuating liquid-layer
width $w$. This distribution has recently been applied to calculate
the disjoining potential for several GBs in Cu-Ag alloys \citep{Hickman_to_be_published_2015}.
The disjoining potential is a key GB property allowing predictions
of different pre-melting scenaria, such as the continuous melting,
thin-to-thick transitions or abrupt melting after over-heating \citep{Mishin09c}.
In the version of the pre-melting model considered here, we essentially
treat the grains, the liquid layer and the solid-liquid interfaces
as parts of the same infinite reservoir. We only account for fluctuations
of $w$ due disjoining interaction. All thermodynamic relations are
in place to develop a more general model that would include fluctuations
of intensive properties of the liquid layer and/or the solid-liquid
interface properties. This generalization could be the subject of
future work.

Finally, the approach demonstrated here for GBs can be readily extended
to other interfaces. It would be interesting to study the fluctuations
of excess properties of solid-liquid and especially solid-solid interfaces
to determine their excess elastic and thermal properties and test
some of the basic equations of interface thermodynamics.

\vspace{0.15in}

\textbf{Acknowledgements.} This work was supported by the National
Science Foundation, Division of Materials Research, the Metals and
Metallic Nanostructures Program.


\end{document}